\documentclass[11pt,nofootinbib]{article}
\usepackage{amsfonts,amssymb,amsmath,mathtools}
\usepackage{graphicx}
\usepackage{geometry}
\usepackage[normalem]{ulem}
\usepackage{array}
\usepackage[small,bf]{caption}
\usepackage[usenames,dvipsnames]{xcolor}
\usepackage[linktocpage]{hyperref}
\hypersetup{
	colorlinks=true,
	linkcolor=purple,
	citecolor=blue,
	anchorcolor = yellow
}
\usepackage{cite}
\usepackage{multirow}
\usepackage[utf8]{inputenc}
\newcommand*{\email}[1]{\href{mailto:#1}{\nolinkurl{#1}} }
\usepackage[symbol]{footmisc}

\newcommand{\fsl}[1]{\ensuremath{\mathrlap{\!\not{\phantom{#1}}}#1}}
\newcommand{\ajk}{\alpha_{jk}}

\newcommand{\mjk}{\Delta m_{jk}^2}
\newcommand{\nab}{\nu_\alpha \rightarrow\nu_\beta}
\newcommand{\ab}{\alpha \rightarrow \beta}

\newcommand{\sumU}{\sum_{j,k}U_{\alpha j}^*U_{\beta j}U_{\alpha k}U_{\beta k}^*}

\newcommand{\vx}{\bold{x}}
\newcommand{\vL}{\bold{L}}
\newcommand{\vp}{\bold{p}}
\newcommand{\vP}{\bold{P}}
\newcommand{\vj}{\bold{v}_j}
\newcommand{\vk}{\bold{v}_k}
\newcommand{\vPj}{\bold{P}_j}
\newcommand{\vPk}{\bold{P}_k}
\newcommand{\bx}{\bar{x}}
\newcommand{\bp}{\bar{p}}
\newcommand{\vbx}{\bar{\vx}}
\newcommand{\vbp}{\bar{\vp}}

\newcommand{\sigbx}{\sigma_{\bar{x}}}
\newcommand{\sigbp}{\sigma_{\bar{p}}}

\definecolor{Violet}{rgb}{0.58, 0.4, 1}

\usepackage[utf8]{inputenc}
\usepackage[acronym]{glossaries}

\begin{document}
	\begin{titlepage}
		
		\begin{center}
			{ \bf\LARGE  Microscopic and Macroscopic Effects in the Decoherence of Neutrino Oscillations} 
			\\[8mm]
			Ting Cheng\footnote{E-mail: \email{ting.cheng@mpi-hd.mpg.de}},
			Manfred Lindner\footnote{E-mail: \email{manfred.lindner@mpi-hd.mpg.de}},
			Werner Rodejohann\footnote{E-mail: \email{werner.rodejohann@mpi-hd.mpg.de}}
			\\[1mm]
		\end{center}
		\vspace*{0.50cm}
		\centerline{\it Max-Planck-Institut f\"ur Kernphysik, Saupfercheckweg 1, 69117 Heidelberg, Germany}
		\vspace*{0.2cm}
		
		\vspace*{1.20cm}

		\begin{abstract}
			\noindent
			We present a generic structure (the layer structure) for decoherence effects in neutrino oscillations, which includes decoherence from quantum mechanical and classical uncertainties. The calculation is done by combining the concept of open quantum system and quantum field theory, forming a structure composed of phase spaces from microscopic to macroscopic level. 
			Having information loss at different levels, quantum mechanical uncertainties parameterize decoherence by an intrinsic mass eigenstate separation effect, while decoherence for classical uncertainties is typically dominated by a statistical averaging effect. 
			With the help of the layer structure, we classify the former as state decoherence (SD) and the latter as phase decoherence (PD), then further conclude that both SD and PD result from phase wash-out effects of different phase structures on different layers. 
			Such effects admit for simple numerical calculations of decoherence for a given width and shape of  uncertainties. 
			While our structure is generic, so are the uncertainties, nonetheless, a few notable ones are: the wavepacket size of the external particles, the effective interaction volume at production and detection, the energy reconstruction model and the neutrino  production profile. Furthermore, we estimate the experimental sensitivities for SD and PD parameterized by the uncertainty parameters, for reactor neutrinos and decay-at-rest neutrinos, using a traditional rate measuring method and a novel phase measuring method.
		\end{abstract}
	\end{titlepage}
	
	\setcounter{footnote}{0}

	{
		\hypersetup{linkcolor=black}
		\tableofcontents
	}
	\newpage

	%%%%%%%%%%%%%%%%%%%%%%%%%%%%%%%%%%%%%%%%%%%%%%%%%%%%%%%%%%%%%%%%%%%%%%%%%%%%%%%%%%%
	\section{Introduction}

	By virtue of the more and more precisely measured phenomenon of neutrino oscillation \cite{SajjadAthar:2021prg}, the quantum coherence of neutrino mass eigenstates can routinely be observed  on a macroscopic level. There are two common bases that are utilized to express the quantum state of neutrinos, viz.\ the mass eigenstates and the flavor eigenstates. While the former determines how neutrinos propagate, the charged-current interaction, for the production and detection of neutrinos, is characterized by the latter. Evolution of neutrinos is effectively encapsulated by the flavor transition probability (FTP), $P_{\nab}$, which represents the probability that a neutrino produced as flavor $\nu_\alpha$ is detected as flavor $\nu_\beta$. The non-trivial mixing between these two bases, described by the Pontecorvo–Maki–Nakagawa–Sakata (PMNS) matrix \cite{Gribov:1968kq,Maki:1962mu,Pontecorvo:1967fh}, implies that $P_{\nab}$ is not diagonal. 
	Consequently, after a neutrino is produced as flavor eigenstate, it propagates in a superposition of mass eigenstates. 
	Such superposition in the Hilbert space describes quantum coherence, which could lead to observational interference patterns. 
	In fact, the loss of coherence, decoherence, represents a transition from the quantum to classical level, for it describes the loss of interference pattern from the correlation between quantum states \cite{Schlosshauer:2003zy,Zeh}.	
	In addition, the smallness of the neutrino masses and of their difference admits that quantum coherence can be observed macroscopically through the FTP. 
	Moreover, with increasing precision of neutrino oscillation experiments, the degree of the quantum correlation may be measured more accurately, therefore, decoherence effects warrant further investigation.

	As all observable effects of mixed quantum states, neutrino coherence is expected to be lost at some stage. 
	Theories for calculating quantum decoherence in neutrino oscillation (or neutrino decoherence) have been widely discussed in the literature. 
	These theories include the degree of wavepacket (WP) separation calculated through quantum mechanics (QM) \cite{PhysRevD.48.4318,Giunti:1997wq,Akhmedov:2010ms,Kiers:1995zj,Kayser:1981ye} and quantum field theory (QFT) \cite{Beuthe:2001rc,Giunti:2002xg,Akhmedov:2010ms,Akhmedov:2010ua,Naumov,Grimus:2019hlq}.
	Another approach deals with the effect of losing information to an open quantum system from Liouville dynamics calculated with density matrices {(e.g.\ through the Lindblad equation)} \cite{Lind,Lisi:2000zt,Hansen:2016klk,Benatti:2000ph,Stirner:2018ojk,Coelho:2017byq,Gomes:2020muc,Farzan:2008zv, Jones:2014sfa}, or through the Wigner quasi-probability distribution \cite{Wigner:1932eb,Stirner:2018ojk,Vlasenko:2013fja}. There is also literature comparing one approach with another, for instance, QM vs.\ QFT approach for WP separation in \cite{Akhmedov:2010ms}, Lindblad equation vs.\ the WP format in \cite{Hansen:2016klk}, and Lindblad equation vs.\ Wigner quasi-probability distribution in \cite{Stirner:2018ojk}.
	Among these theories, QFT is able to describe the situation on the most fundamental level by considering neutrino oscillation as the propagator of a full process described by a Feynman diagram. However, the open quantum system method is more tailored for quantum decoherence effect in a generic way by considering a system of interest in an environment. In this sense, decoherence in a system reflects loosing information to the environment, and is calculated by tracing out states of the environment entangled to the system.
	Nonetheless, regardless of the way one chooses to formulate the decoherence effect, it will result in additional terms to the coherent interference patterns caused by quantum correlation. For neutrino oscillation, the decoherence effect appears as a complex function $\Psi_{jk}$ in the FTP as
	\begin{equation}
	    P_{\nab} = \sumU e^{i\psi_{jk}} \Psi_{jk},
	\end{equation}
	where $\psi_{jk}$ is the coherent phase, usually estimated as $\mjk L_0/(2E_0)$ for some traveling distance $L_0$ and energy $E_0$. The decoherence term $\Psi_{jk}$ would, in general, erase the interference pattern, hence, $|\Psi_{jk}| \leq 1$. However, since it could also be complex, is might also cause a phase shift w.r.t.\ $\psi_{jk}$.

	In this work, we introduce the concept of the open quantum system method to the QFT calculations by considering the propagator describing neutrino oscillation as the system of interest, and everything else in the diagram as the environment, which is to be integrated out. 
	Furthermore, since neutrino oscillation is considered as a phenomenon resulting from the coherence of kinematics between mass eigenstates, the states of the environment we integrate out are in the phase space (PS). 
	In particular, if a state is described by creation and annihilation operators represented in the coordinate and/or the momentum space, we call it the ``Fock-PS"; on the other hand, if a state is described by occupation numbers on a PS forming a Wigner quasi-probability distribution \cite{Wigner:1932eb}, we call it the ``Wigner-PS".
	Notably, since the Fock space representations for mass basis and flavor basis are unitarily inequivalent with each other \cite{Blasone:1995zc,PhysRevD.45.2414}, at least one of them must be unphysical, and the debate on which of them is unphysical is still on-going,  e.g.\ \cite{Giunti:2003dg,Tureanu:2020odo,Torres:2020gzm,Smaldone:2021mii,Blasone:2020wer}.  
	As both representations approximately agree with each other in the relativistic limit,
	we choose to build the Fock-PS for mass states here, with flavor states represented by a superposition of mass states. Nonetheless, one can also build a flavor based Fock-PS for the layer structure, for it does not specify the representation we choose. 
		
	Concretely, we start from following the QFT description for calculating neutrino oscillation in \cite{Beuthe:2001rc}, which already includes integrating out the momentum space of the external particles, so we complete the picture by also integrating out the space-time components of the PS
	(although the PS does not include a temporal dimension, the kinematic of the states is time-dependent, therefore when we say ``PS variables", a temporal component is also included).
	Furthermore, we focus on the structure of the PS while calculating the FTP, which will be called the ``layer structure", composed of three layers. 
	Vertical-wise, the layer structure includes three layers, namely the microscopic layer, the physical layer and the measurement layer; and horizontal-wise, each layer represents a PS composed of space-time variables and momentum variables, which are able to determine the kinematics of the neutrinos fully. 
	The microscopic layer relates to the theories in the literature \cite{PhysRevD.48.4318,Giunti:1997wq,Akhmedov:2010ms,Beuthe:2001rc,Giunti:2002xg,Akhmedov:2010ua,Naumov,Lind,Lisi:2000zt,Hansen:2016klk,Benatti:2000ph,Wigner:1932eb,Stirner:2018ojk,Vlasenko:2013fja}, which can be represented by either the Fock-PS or the Wigner-PS on which Fock states and the Wigner quasi-probability distribution are represented, respectively. 
	With the layer structure we are able to account for the decoherence effect from information loss to the environment, from a microscopic to macroscopic level.
	Different from the existing literature in which one also calculates neutrino decoherence on a macroscopic level \cite{Akhmedov:2020vua,Naumov}, we focus on classifying and understanding decoherence in a generic picture. At the end, neutrino decoherence for continuously emitted neutrinos is mainly parameterized by four uncertainties appearing on the layer structure. Those are the coordinate/momentum uncertainty on the microscopic layer (quantum effects quantified by $\sigma_x$/$\sigma_p$), and that on physical layer (macroscopic effects such as energy resolution or neutrino production profile $\sigma_L$/$\sigma_E$), providing an interface between microscopic mechanisms and the macroscopic experiments. As for non-continuously emitted neutrinos, we would simply have an additional temporal uncertainty on the physical layer, $\sigma_{T}$.  
	
    As for the phenomenology part of neutrino decoherence, roughly speaking,  measurements of neutrinos produced in both long and short baseline experiments and in the atmosphere are best fitted with neutrinos considered as fully coherent, see for instance \cite{Esteban:2020cvm} for an updated global fit; as for neutrinos produced outside the Earth, such as solar and supernova neutrinos, it is best described as fully incoherent. 
    There are also many discussions on neutrino decoherence phenomenons, such as general neutrino decoherence for reactor experiments in \cite{deGouvea:2020hfl,deGouvea:2021uvg,JUNO:2021ydg};
    gravitation fluctuation or cosmological effects causing atmospheric neutrino decoherence in \cite{Lisi:2000zt,Stuttard:2020qfv,Boriero:2017tkh,DeGouvea:2020ang};
    matter effect responsible for accelerator or atmospheric neutrino decoherence in \cite{Carpio:2017nui,Carpio:2018gum,BalieiroGomes:2018gtd,Coelho:2017byq,Coloma:2018idr}.
    Our structure carries the potential of including all the mechanisms above and more, since both the QFT approach and the open quantum system concept is included. Hence, we do not go into the details of these theories but give a generic picture on  what measurable parameters it could reflect on. 
    Furthermore, we also include decoherence signatures caused by classical uncertainties or the ignorance of the observer, such as the production profile (for instance, the exact shape  of the neutrino source) of the neutrino and the energy reconstruction model.  

    Notably, we introduce the phase wash-out (PWO) effect,  which is an averaging effect with respect to some phase structure, washing-out the oscillation signatures. In this paper, we will show that decoherence signatures of neutrino oscillation can all be described by some PWO effects, and the distinction between difference decoherence parameters comes from different dependence on the phase structure(s). In other words, while quantum coherence reflects on the oscillation signature, decoherence can be described by some wash-out of such signature,  and the PS (e.g.\ the traveling distance and the energy of neutrinos) dependence of each decoherence effect comes from the formalism of the phase structure being washed-out. 
    Therefore, the PWO effect arising from neutrino decoherence results in a damping and/or phase shift signature to the oscillation.   
    We further analyze the damping/phase shift signatures w.r.t.\ both classical and QM uncertainty parameters for reactor/decay-at-rest (DAR) neutrinos. In particular, for damping signatures which are expected in all literature mentioned above, we estimate the sensitivity of the parameters by directly analysing the neutrino count rate in a conventional way. However, the phase shift signals are estimated to be not as suitable for such method, so instead, we evaluate the possibility of measuring the distance dependence of the oscillation phase for the phase shift signals, considering that a moving detector is possible.

    The paper is organised as follows. In Sec.~\ref{sec.2}, we introduce the layer structure and calculate neutrino FTP for neutrinos propagating in vacuum throughout the layers, as a demonstration for decoherence coming from only the production and detection site. In Sec.~\ref{sec.3}, with the help of the layer structure, we classify neutrino decoherence into ``state decoherence" and ``phase decoherence", where the former is related to the WP separation mechanism and the latter to the information loss mechanism.
    We also show how both types of decoherence are the result of the PWO effect. In Sec.~\ref{sec.4}, we discuss two kinds of phenomenological analyses for the decoherence parameters, namely, the ``rate measuring method" (RMM) and ``phase measuring method" (PMM). In particular, we estimate the sensitivity to the parameters for reactor and DAR neutrinos.  We summarize our results and conclude with some remarks in Sec.~\ref{sec.5}. Technical details are delegated to appendices. 
    Before we discuss the physics in detail, we provide a glossary of terms and definitions needed in this paper.

	\section*{Terminology introduced in this paper}
	
	\begin{description}
		\item \textbf{Phase space (PS) variables} 
		\newline Three coordinate variables, three momentum variables (composing the six-dimension PS) and a temporal variable. 
		
		\item \textbf{Layer structure}  
		\newline Composed of three layers (layer 1-3) from the microscopic Hilbert space to the macroscopic measurement space, where all spaces are represented by PS variables. The structure is illustrated in Sec.~\ref{sec.2}, and its value of providing a simple and generic picture of decoherence effects is shown in Sec.~\ref{sec.3}. 
		
		\item \textbf{Layer-Moving-Operator (LMO)} 
		\newline Operators moving some physical quantity up one layer, characterised by some weighting functions on the lower layer. The definition is given in Eq.~\eqref{LMO}.
		
		\item \textbf{Microscopic layer (layer 1)}
		\newline Configurations of fundamental theories are described on this layer, such as the Feynman diagram, and intrinsic quantum mechanical uncertainties. More explanations are given in Sec.~\ref{Sec.2-1} and Sec.~\ref{Sec.2-2}.    
		
		\item \textbf{Physical layer (layer 2)}
		\newline As an intermediate layer between the fundamental theories and experimental measurements, this layer describe the statistical ensemble. On top of quantum uncertainties brought up from the first layer, this layer also include uncertainties due to a lack of knowledge.  More explanations are given in Sec.~\ref{Sec.2-3}.      
		
		\item \textbf{Measurement layer (layer 3)}   
		\newline This layer describes realistic experimental measurements including effects  such as energy resolution. Examples are given in Sec.~\ref{Sec.2-4}.
		
		\item \textbf{Fock phase space (Fock-PS)}
		\newline A representation of layer 1 where the occupation of the PS is written in terms of Fock states. The case for neutrino oscillations calculated by QFT is demonstrated in Sec.~\ref{Sec.2-1}.  
		
		\item \textbf{Wigner phase space (Wigner-PS)}
		\newline A representation of layer 1 where the occupation of the PS is written in terms of Wigner quasi-probability distributions. More explanations are given in Sec.~\ref{Sec.2-2}.
		
		\item \textbf{Relativistic phase space (Relativistic-PS)}
		\newline A representation of layer 2, by taking the expectation values of the PS variables on the first layer assuming a  relativistic system (e.g.\ massless neutrinos). The case for neutrino oscillation is demonstrated in Sec.~\ref{Sec.2-3}. 	
		
		\item \textbf{Measurement phase space (Measurement-PS)}
		\newline A representation of layer 3, given by PS variables from experimental measurement.			
		
		\item \textbf{Weighting function}
		\newline Localized distributions that characterise uncertainties included in the LMO. Examples of uncertainties from layer 1 and layer 2 for neutrino oscillation are summerized in Sec.~\ref{Sec.3-2} and Sec.~\ref{Sec.3-3}, respectively.   
		
		\item \textbf{Phase wash-out (PWO) effect}
		\newline An averaging effect which washes out oscillation signatures by introducing a damping term and a phase shift term.   
		Mathematical formalism and properties are given in Appendix~\ref{Sec.AppA}. 
		
		\item \textbf{Uncertainty parameters ($\sigma_n$)}
		\newline  Widths of the weighting functions w.r.t.\ some PS variable $n$ which parameterize decoherence signatures. Some analysing methods, as well as its sensitivity estimation of three relevant uncertainty parameters are shown in Sec.~\ref{sec.4} for neutrino oscillation experiments.

		\item \textbf{State decoherence (SD)}
		\newline Decoherence by the separation of superposition states on the physical layer, which is equivalent to a PWO effect on the Wigner-PS under a factorisation condition (see Appendix~\ref{Sec.AppC}) and is dominated by uncertainties on layer 1 (see Sec.~\ref{Sec.3-2}).      
		
		\item \textbf{Phase decoherence (PD)}
		\newline Decoherence by the PWO effect on the physical layer dominated by the macroscopic uncertainties on layer 2 (see Sec.~\ref{Sec.3-3}). 
		
	\end{description}

	%2%%%%%%%%%%%%%%%%%%%%%%%%%%%%%%%%%%%%%%%%%%%%%%%%%%%%%%%%%%%%%%
	\section{The Layer Structure} \label{sec.2}
\begin{figure}
		\centering
		\includegraphics[width=0.8 \textwidth]{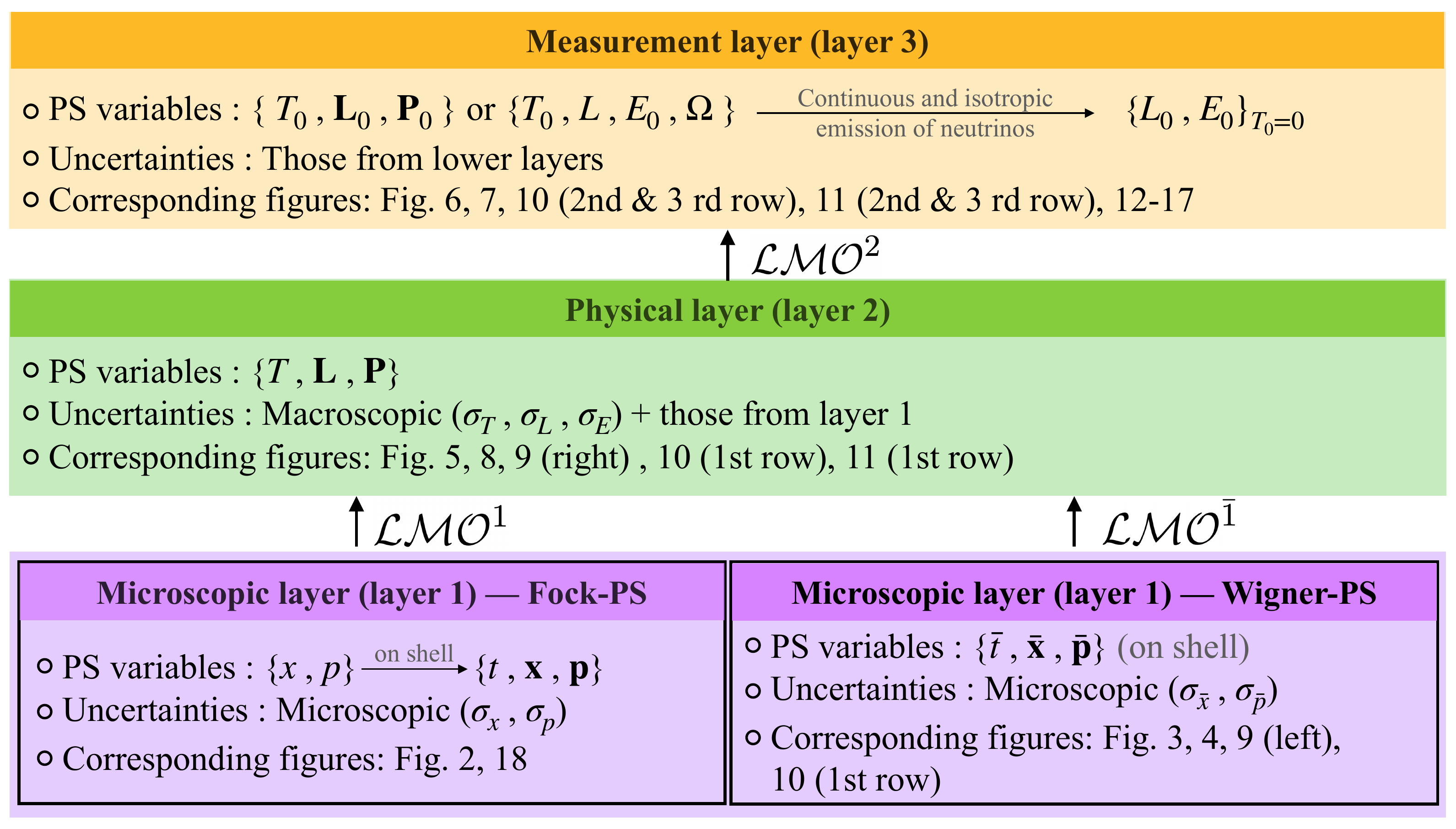}
		\caption{\label{fig:layers} Illustration of the layer structure, and the notation of each phase space variable deciding the kinematics of states: $t,\bar{t},T,T_0$ are the temporal variables; $\vx,\bar{\vx},\vL,\vL_0$ are the spatial variables; and $\vp,\bar{\vp},\vP,\vP_0$ are the momentum variables; $E_0$ is the energy and $\Omega$ represents the solid angle. The layers are linked by the layer-moving operator in Eq.~\eqref{LMO}, and the uncertainties are discussed in the text.}
	\end{figure}

	In Fig.~\ref{fig:layers}, we show the layer structure for calculating the expectation value of some observable, which is composed of three layers of PS, from microscopic to macroscopic, including the ``microscopic layer" (layer 1), the ``physical layer" (layer 2) and the ``measurement layer" (layer 3).	 
	As illustrated in the introduction, this structure combines QFT and the concept of having an open quantum system while including statistical effects for actual measurements. 
	Quantum effects, such as the superposition of states in the Hilbert space are described on layer 1, and it is impossible for both coordinate and momentum uncertainty to be zero due to the uncertainty principle.
	These uncertainties are parameterized as $\sigma_p$ and $\sigma_x$ which are in general independent of each other; the former is for uncertainties from external states on the mass-shell, which are described as WPs in  momentum space;  the latter is for uncertainties around the vertices in coordinate space, i.e.\ how non point-like the effective vertices are in a simplified effective diagram with only the external states and the neutrino propagator such as Fig.~\ref{fig:Feyn}.
	On the other hand, layer 2 describes our ignorance towards the system in the classical regime, where the probability is summed (integrated) over, instead of the amplitude. 
	The uncertainties on this layer include those transmitted from the first layer, and additional ones, which are usually, but not necessarily, macroscopic. 
	These additional (macroscopic) uncertainties contain the energy uncertainty $\sigma_E$, such as the energy resolution and the energy reconstruction model, on top of the coordinate uncertainty $\sigma_L$, e.g.\ the neutrino production profile.
	More examples contributing to $\sigma_p$, $\sigma_x$, $\sigma_E$ and $\sigma_L$ will be discussed in the following sections when its corresponding layer is introduced for calculating the FTP of neutrinos. 
	
	Consider the double slit experiment observed by taking a photo of the interference pattern as an analogy: the two slits represent uncertainties on the first layer, while the resolution of the camera taking the photo is on the second layer. The former creates freedom for a superposition state while the latter is a classical effect. 
	For the case of neutrino oscillation, although the uncertainties on layer 2 are macroscopic, we still observe quantum coherence, due to the smallness of neutrino mass splitting. 
	In the following, we will discuss the layer structure more formally, including aspects in Fig.~\ref{fig:layers}, such as the representation of the PS and the layer-moving-operators (LMOs), $\mathcal{LMO}$, connecting each layer.
	In particular, an important remark is that, as we will show later, the uncertainty parameters on each layer are the width of a weighting function in the $\mathcal{LMO}$, which carries information of the environment entangled to the system (e.g.\ the propagating neutrino).
	The uncertainty parameters therefore characterize neutrino decoherence in experiments.
	
	Each layer in our structure could have different representations for the PS, for instance, the first layer could be represented by either the Fock-PS or the Wigner-PS; for the representation of the second layer, we take the expectation values of the PS variables on the first layer assuming massless neutrinos (more on this in Sec.~\ref{Sec.2-3}), which will be called the relativistic-PS; as for the third layer, we simply represent the PS variables in terms of the expectation values of the relativistic-PS, which should coincide with the actual measurement values. 
	Otherwise, it will have a dependence on the mass of the neutrino, and the states will collapse to a certain neutrino mass state, giving us no oscillation. Hence, we call such representation the measurement-PS.   
	Each PS is composed of its own temporal, coordinate and momentum space, while the energy is implied by the momentum variables through the dispersion relation, and the notations are given in Fig.~\ref{fig:layers}. 
	The layer structure could generically be applied for the calculation of the expectation value for any measurement, but we only focus on the calculation of the FTP for neutrino oscillation in this work.

	As illustrated in Fig.~\ref{fig:layers}, the layers are connected by the LMOs, $\mathcal{LMO}^i$, moving a quantity $B_i(x_i,p_i)$, such as the FTP or the transition amplitude, from layer $i$ to layer $i+1$. This is done by integrating out the PS variables, $x_i,p_i$, of layer $i$, while considering additional uncertainties by including the weighting function $W^i$, such that
	\begin{equation}\label{LMO}
	\mathcal{LMO}^i B_i (x_i,p_i)=\int d^4 x_i \int d^3 p_i \, 
	\big[W^i(x_i,p_i; x_{i+1},p_{i+1})B_i (x_i,p_i) \big] 
	=B_{i+1}(x_{i+1}, p_{i+1}).
	\end{equation}
	In particular, corresponding to the open quantum system concept, $B_1$ would be the system of interest, and $W_1$ includes the environment entangled with it.
	Furthermore, corresponding to our notation of phase space variables in Fig.~\ref{fig:layers}, $x_1 = (t,\vx)$, $p_1 = \vp$ for the Fock-PS; $x_{1} = (\bar{t}, \bar{\vx})$, $p_{1} = \bar{\vp}$ for the Wigner-PS, representing the PS for the occupation number of quasi-probability distributions \cite{Wigner:1932eb}; $x_2 = (T,\vL)$, $p_2 = \vP$ for the relativistic-PS, and $x_3 = (T_0,\vL_0)$, $p_3 = \vP_0$ for the measurement-PS. Each of these PS will be explored one by one in the following subsections for the neutrino case. 
	Also, $B_i$ on the first, second and third layer represents the system of interest, the observable and the measured value, respectively. 
	Moreover, $W_i$ is a probability density function (PDF) defined as
	\begin{equation}
	\int d^4 x_i \int d^3 p_i \, 
	W^i(x_i,p_i; x_{i+1},p_{i+1}) = 1.
	\end{equation}
	In fact, the normalization of $W_i$ does not matter for now, as we will show later in Sec.~\ref{Sec.3-1} that the FTP will automatically be normalized.
	Nonetheless, we define the weighting function as a PDF simply for the convenience to observe the width, relating to the definition of ``width" described in Appendix \ref{Sec.AppA}. 
	Moreover, $x_{i+1}$ and $p_{i+1}$ are the next layer variables, usually defined as {(functions of)} the expectation values of $x_i$ and $p_i$. Hence, the width of $W_1(x,p;T,\vL,\vP)$ gives microscopic quantum uncertainties of $t$, $\vx$ and $\vp$, as $\sigma_t$, $\sigma_x$, and $\sigma_p$, respectively. Equivalently, that of $W_2(T,\vL,\vP;T_0,\vL_0,\vP_0)$ gives macroscopic {statistical} uncertainties for $T$, $\vL$ and $\vP$ as $\sigma_T$, $\sigma_L$ and $\sigma_P$,  respectively.
	In fact, if the weighting function can be written as $W^i(x_i-x_{i+1},p_i-p_{i+1})$, the layer moving operator is an act of convolution between $W_i$ and $B_i$, see Appendix \ref{Sec.AppA}.
	In addition, $\mathcal{LMO}^i$ refers to calculating the expectation value of the quantity $B_i$, and the layer structure is mathematically fibre bundles \cite{Sen}. In other words, looking from upper layers to lower layers, each PS point $(x_{i+1},p_{i+1})$ can be expanded into a whole PS composed of $(x_{i},p_{i})$ on the lower layer. Also, on each layer, operations mapping one state to the other could be made, depending on what we wish to observe. Viewing the LMOs as vertical operators,  such operators can be referred to as horizontal operators such that the state remains on the same layer.

	Additionally, if $B_i=e^{i x_i p_i} C_i(x_i,p_i)$, we have $\mathcal{LMO}^i e^{i x_i p_i} C_i=e^{i x_{i+1} p_{i+1}} C_{i+1}(x_{i+1}, p_{i+1})$ (see Appendix \ref{Sec.AppA} for details), indicating that the uncertainty principle between $x_i,p_i$ remains fulfilled on each layer. This is because $\mathcal{LMO}^i$ along with  $e^{ix_ip_i}$ means to first project everything onto the $x_i$ or the $p_i$ space, and then integrate over that space, while the projection process secures the uncertainty principle. In fact, this is exactly the case for the position-space representation of the wavefunction for some considered particle. In this case, $B_1(x,p) = e^{ixp} \tilde{\Delta}(p)$, where $\tilde{\Delta}(p)$ is the propagator in momentum space.  
	We will demonstrate this explicitly for the case of neutrinos in Sec.~\ref{Sec.2-1}.
	As a matter of fact, if $B_i=\exp({i\eta(x_i,p_i)})$, for some phase structure $\eta(x_i,p_i)$, the LMO meets the condition of giving rise to a phase wash-out (PWO) effect described in Appendix \ref{Sec.AppA}. 
	The PWO effect is an averaging effect over the phase structure caused by the non-trivial width of the (normalized) weighting function, resulting in an additional suppression term $\Phi$ as
	\begin{equation}
	B_i(x_{i+1},p_{i+1})\Phi(x_{i+1},p_{i+1})
	%=\Phi(x_{i+1},p_{i+1})e^{i\eta(x_{i+1},p_{i+1})}
	= \mathcal{LOM}^i B_i(x_{i},p_{i}),
	\end{equation}   
	where $|\Phi(x_{i+1},p_{i+1})|\leq 1, \, \forall \, (x_{i+1},p_{i+1})$. Only when the weighting function is symmetric with respect to the phase structure would $\Phi$ be a real function (see again  Appendix \ref{Sec.AppA}).

	Furthermore, when there is a substructure of $B$, i.e.\ $B_{i}=\sum_{\nu} B_{\nu i}$, then the summation rule is simply 
	\begin{equation} \label{LMO_sum}
	\mathcal{LMO}^i B_i =
	\sum_\nu \int d^4 x_i \int d^3 p_i  \, 
	W_i^{\nu}(x_i,p_i;x_{i+1},p_{i+1})B_{\nu i}(x_i,p_i).
	\end{equation} 
	
	Finally, the determination of the measurement expectation value of the FTP ($P_3$), is by doing the statistical averaging ($\mathcal{LMO}^2$) over the FTP on the physical layer ($P_2$); $P_2$, however, can either be calculated by squaring the transition amplitude ($A_2$) on layer 2, or directly by moving up ($\mathcal{LMO}^{\bar{1}}$) the quasi-probability distribution ($P_{\bar{1}}$) from the Wigner-PS.
	Here, $A_2$ is calculated by integrating over all the quantum-mechanical configurations of the environment ($\mathcal{LMO}^1$), moving the system of interest in the Fock-PS $(A_1)$ up to the physical layer; and $\mathcal{LMO}^{\bar{1}}$ performs an effective statistical averaging over an effective FTP, the quasi-probability distribution, for the quantum-mechanical superposition effect in the Wigner-PS \cite{Wigner:1932eb,Case}. That is,
	\begin{equation}  \label{LMOP}
	\begin{split}
	&P_3(T_0,\vL_0,\vP_0)=\mathcal{LMO}^2 P_2(T,\vL,\vP)
	=\mathcal{LMO}^2\{A^*_2(T,\vL,\vP) \, A_2(T,\vL,\vP)\} \\
	&=\mathcal{LMO}^2\{\mathcal{LMO}^1 A^*_1(x,\vp)\,
	\mathcal{LMO}^{1} A_{1}(x,\vp)\}\\
	&=\mathcal{LMO}^2\{\mathcal{LMO}^{\bar{1}} 
	P_{\bar{1}}(\bar{t},\bar{\vx},\bar{\vp})\}.
	\end{split}
	\end{equation}	
	In the following subsections, we will introduce each layer and its part in calculating the expectation value for the measurement of the FTP for neutrino oscillation in vacuum.	
	
	%2-1%%%%%%%%%%%%%%%%%%%%%%%%%%%%%%%%%%%%%%%%%%%%%%%%%%%%%%%%%%%%%
	\subsection{Microscopic Layer (Layer 1): QFT Transition Amplitude} \label{Sec.2-1}

	In this subsection, we calculate the transition amplitude on the Fock-PS with the QFT approach. Although QM can also describe neutrino coherence and decoherence on the Fock-PS, it could not answer a number of questions while the QFT approach can, see for example  \cite{Akhmedov:2010ms,Beuthe:2001rc}. 
	Moreover, for the purpose of investigating the quantum decoherence effect and its implication for fundamental physics, it is necessary to use the QFT framework, even for the scenario of vacuum propagation, since the weighting functions on the first layer could originate  from  uncertainties of the interactions  around the vertices. Moreover, in order to have the weighting functions which are determined by the states entangled to the neutrinos explicitly, the phase space of this layer will be $\vp${/$x$}, the momentum{/space-time coordinate} of the neutrino given by its entangled states. 
	As for the weighting functions, we treat the external particles as WPs (also referred to as the Jacob-Sachs model \cite{Jacob} in \cite{Beuthe:2001rc}) through Eq.~\eqref{WP}, resulting in a microscopic uncertainty, $\sigma_p$, represented in the momentum space. Hence, $\sigma_p$ includes information such as the life-time of the external particles \cite{Akhmedov:2012uu} for neutrinos produced by decaying particles, or the mean free path of processes before the production of neutrinos \cite{Kersten:2015kio}.  
	In addition, regardless of the uncertainties of the external particles, the microscopic {space and time} uncertainties of the interaction {around} the vertices are taken into account by another microscopic uncertainty, $\sigma_x$.
	This parameter depends on the internal states and the scattering/collision process. 
	In principle, one can also write the first layer Fock-PS directly in terms of the neutrinos, as in Appendix \ref{Sec.AppB}. However, in this case we can only obtain an effective weighting function in either the energy-momentum space or the space-time coordinate space.   
	
	\begin{figure}
		\centering
		\includegraphics[width=0.6 \textwidth]{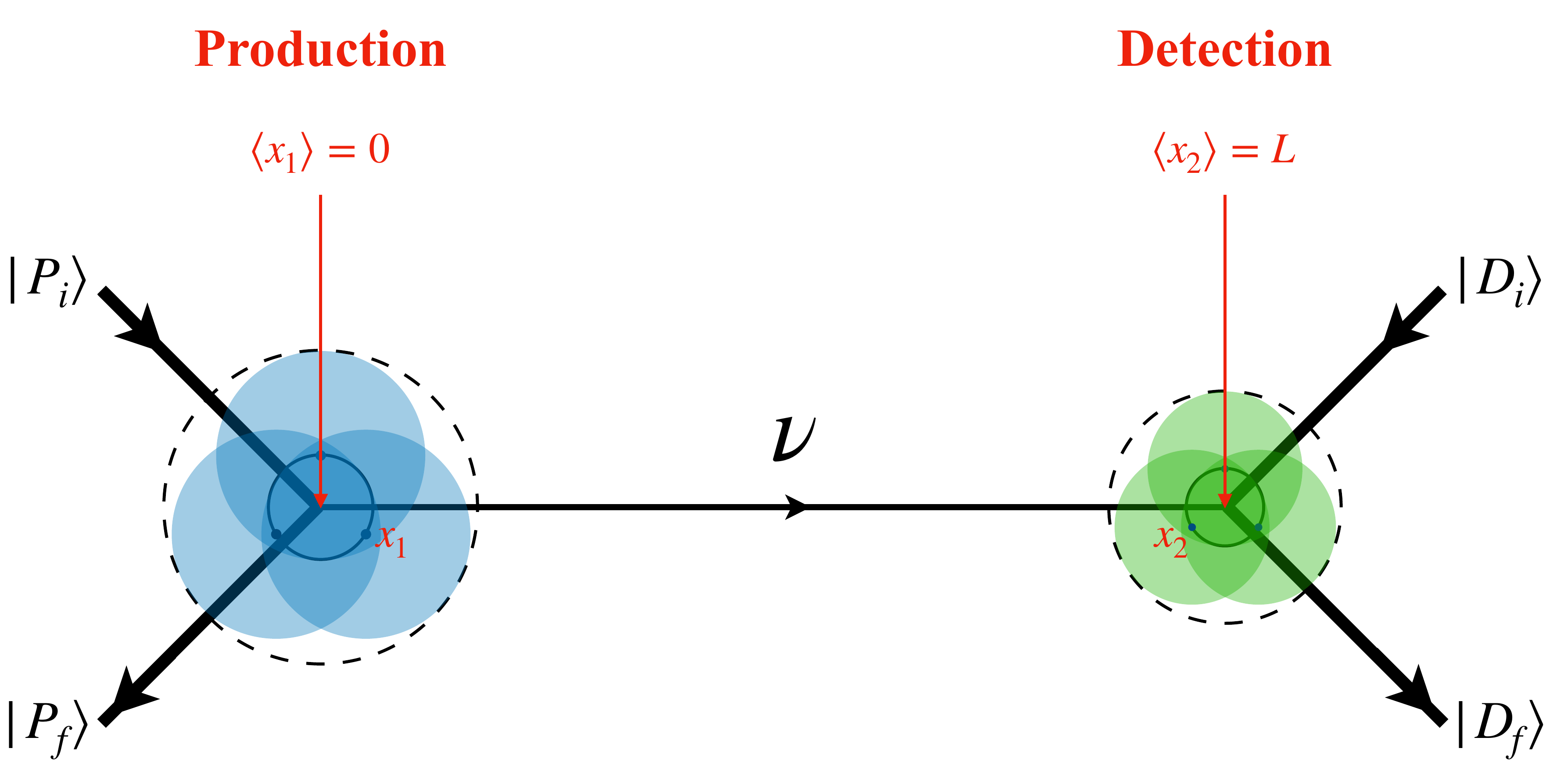}
		\caption{\label{fig:Feyn}A simplified Feynman diagram where the neutrino propagating a macroscopic distance is treated as a propagator of a full diagram, and the kinematics of the external particles are described with wavepackets. At the production/detection vertex site, the diameter of the shaded blue/green areas represent the uncertainties of external wavepackets projected onto the coordinate space, and their mean value is labeled as $x_1/x_2$. The inner circles with solid lines at both sites are the additional coordinate uncertainties from the blob vertices for the internal states regardless of the external particles. In other words, it represents the uncertainties of $x_1/x_2$ by $g_P(x_1)/g_D(x_2)$ in Eq.~\eqref{tot_amp}. Hence the total uncertainty on the coordinate space at this layer would be the diameter of the dashed-lined circles.}
	\end{figure}

	We calculate the transition amplitude for neutrinos in the first layer with the Fock-PS representation, by applying the $S$-matrix method following \cite{Beuthe:2001rc}, where the traveling neutrino is treated as an internal propagator in a diagram of a full process including the production and detection process as illustrated in Fig.\ \ref{fig:Feyn}. The kinematics of the neutrino's initial and final state are written in the form of WPs in the momentum space, which will eventually be combined as a weighting function with width $\sigma_p$. Hence, without loss of generality we can write 
	\begin{align} \label{WP}
	&|P_i\rangle=\int[dq]f_{Pi}(q,t)|q\rangle,  \qquad \quad |P_f\rangle=\int[dk]f_{Pf}(k,t)|k\rangle, \nonumber\\
	&|D_i\rangle=\int[dq']f_{Di}(q',t)|q'\rangle, \qquad   
	|D_f\rangle=\int[dk']f_{Df}(k',t)|k'\rangle,
	\end{align}
	for the initial/final state at the production site ($|P_{i/f}\rangle$) and the initial/final state at the detection site ($|D_{i/f}\rangle$), where $[dh]=d^3h/(2\pi)^3$, for each $h = \{q,k,q',k'\}$.  
	Additionally, the internal states (excluding the neutrino propagator) of the process are included by the distributions $g_P(x_1)$ and $g_D(x_2)$ which represent space-time uncertainties around the vertex at the production and detection site, respectively.
	Note that since these states are not restricted on the mass-shell, such uncertainties include four degrees of freedom, namely, a temporal uncertainty and three spatial ones, while the on-shell WPs only have three degrees of freedom.  
	However, such uncertainties are usually not explicitly referred to explicitly  in the literature, since in terms of WP separation, we will show that it can be combined with $\sigma_p$ as an effective value, and in terms of a localization term (such as in\cite{Giunti:2002xg}), it is microscopic compared the scale of the experiment. Nonetheless, since the external particles are better known and are, in principle, observable, one may still be able to extract the contribution of such uncertainties upon measurement.

	We can readily calculate the transition amplitude of a neutrino that is produced as flavor $\alpha$ while detected as flavor $\beta$ and propagates in mass eigenstates of mass $m_j$, and integrate over $x_1$ and $x_2$ for the sake of completion of our structure as  
	\begin{align}\label{tot_amp}
	%		\begin{split}
	&iA_{2,\ab} (T,\vL,\vP)
	\equiv i \sum_{j}U^*_{\alpha j}U_{\beta j} A_{2,j}(T,\vL,\vP)
	\nonumber \\
	&=i \sum_{j}U^*_{\alpha j}U_{\beta j}
	\int[dq]f_{Pi}(\bold{q})\int[dk] \, f^*_{Pf}(\bold{k})\int[dq'] \, f_{Di}(\bold{q}')\int[dk']f^*_{Df}(\bold{k}')  \nonumber\\ 
	& \times \int d^4 x_1 \, g_{P}(x_1) \int d^4 x_2 \, g_{D}(x_2)
	\int d^4 \, y_2 \, M_{Dj}(q',k')e^{-i(q'-k')(y_2-x_2)} \nonumber\\ %
	&\times \int \frac{d^4 p_\nu}{(2\pi)^4} \, \frac{\fsl{p_\nu}+m_j}{p_\nu^2-m_j^2+i\epsilon}e^{-ip_\nu(y_1-y_2)} 
	\int d^4y_1 \, M_{Pj}(q,k)e^{-i(q-k)(y_1-x_1)}. 
	\end{align}
	Here $M_P(q,k)$ and $M_D(q',k')$ are the plane-wave amplitudes determined by particles involved in the production and detection process, respectively. 
	With the goal of leaving only the neutrino momentum ($\vp=\bold{q}-\bold{k}=\bold{k}'-\bold{q}'$) and traveling distance and time ($x=x_2-x_1$) determined by the entangled states unintegrated, we derive the form of the layer structure as 
	\begin{equation} \label{Aj_full}
	A_{2,\ab}(T,\vL,\vP)= \int d^3 p \int d^4 x \, 
	F_j(\vp;\vP) G_x(x;X)
	\, A_{1,\ab}(x,\vp),
	\end{equation}	
	demonstrated explicitly in Appendix \ref{Sec.AppB}. 
	Here, $F(\vp;\vP)$ represents the effective PDF from the WPs of the external particles, and $G_x(x;X)$ is that from the vertices. 	
	Hence, the layer-moving-operator is
	\begin{equation}
	\mathcal{LMO}^1 = \int d^3p \int d^4 x  F_j(\vp;\vP) G_x(x;X),
	\end{equation}
	where $F(p;P)G_x(x;X)$ is the weighting function, and the width of these uncertainties, $\sigma_p$ and $\sigma_x$, are the observational parameters. From Fig.~\ref{fig:CompDist} in Appendix \ref{Sec.AppB} we see how these parameters are related to each of the original distributions in Eq.~\eqref{tot_amp}. 
	Here, the notation $G_x(x;X)$ means that $X$ is the expectation value of $x$ for the PDF $G_x(x)$, as well as other functions in this paper. 
	In general, the first layer transition amplitude takes the form of Eq.~\eqref{A_1}, where all configuration of the internal energy of the neutrino propagator has to be included. However, since the measurement is done macroscopically, we can consider the propagating neutrino to be on the mass-shell, hence the first layer transition amplitude for neutrino oscillation is
	\begin{equation}
	A_{1,\ab}(x,\vp)  
	=\sum_{j}U^*_{\alpha j}U_{\beta j} 
	e^{-itE_j(\vp)+i\vx \vp}. \label{Amp1}
	\end{equation}
	Note that although we impose here the on-shell approximation by hand, it will still be on-shell on the second layer even if we do not. This is due to the fact that the variables on the second layer are macroscopic while those on the first are microscopic. Hence, the dynamics becomes classical and we will obtain the energy-momentum dispersion relation. 
	Additionally, note that $e^{ipx}$ in $A_{1,\ab}$ ensures the uncertainty principle all the way up to the measurement layer, as illustrated previously. 
	
	Finally, we expand the neutrino energy around $\vp=\vPj$, the saddle point of the total weighting function on the momentum space, then keep terms up to the first order,  i.e.\ $E_j(\vp)\simeq E_j+\vj(\vp-\vPj)$, where $E_j\equiv E_j(\vPj)$ and $\vj \equiv \partial E_j(\vp)/ \partial \vp |_{\vp=\vPj}=\vPj/E_j$. Hence, corresponding to Eq.~\eqref{A_2j}, we obtain
	\begin{equation} \label{A_j_final}
	A_{2,j}(T,\vL,\vP)
	= e^{-iE_jT+i\vP_j\vL} \hat{\Phi}_j(\vL,\vP),
	\end{equation} 
	where $\vL$ and $\vP$ are the second layer PS variables, such that $\vP_j=\vP_j(\vP)$ (the explicit relation for the latter will be discussed in Sec.\ \ref{Sec.2-3}). 
	In principle, it is possible to do a full analysis of weighting functions from the neutrino decoherence phenomenon, and inspect the mechanism behind it. Nonetheless, this would require very high precision measurements and a wide spectrum in both coordinate and momentum space, and an analysis which is beyond the scope of this paper. Here, we investigate the weighting function through the width of the PDFs, $\sigma_x$/$\sigma_p$ for $G(\vx,\vL_j)$/$F_j(\vp,\vP_j)$ in Eq.~\eqref{AforWig} below. If the weighting functions are symmetric, then $\hat{\Phi}_j(\vL,\vP)$ is real, according to Appendix~\ref{Sec.AppA}. Otherwise, an additional parameter for the phase of $\hat{\Phi}_j(\vL,\vP)$ would be needed. 
	Additionally, note that $\sigma_x$ and $\sigma_p$ are not the total momentum and coordinate uncertainty of the system. 
	In fact, the total coordinate uncertainty is calculated in Appendix~\ref{Sec.AppB-2}, which turns out to be the convolution between the coordinate distributions at the production and detection site.
	The total coordinate distribution for each production/detection site is the convolution between $g_P(x_1)/g_D(x_2)$ and the momentum distributions of the external states projected onto the coordinate space ($\tilde{F}_{P/D}^{\rm tot}$). 
	In other words, the total coordinate uncertainty is $(g_P*\tilde{F}_P^{\rm tot})*(g_D*\tilde{F}_D^{\rm tot})$, where $*$ denotes convolution of two functions as noted in Table \ref{table1}. 
	This is illustrated at the vertices in Fig.~\ref{fig:Feyn}: while the external particles already cause some coordinate uncertainties by conjugating the momentum uncertainties onto the coordinate space (the individual blue and green circles), the internal process would give rise to spatial uncertainties on top of that (the inner solid circle lines) resulting in a total coordinate uncertainty (the outer dashed lines) larger than that of the individual uncertainties.

	In particular, for the case of Gaussian WPs, the formalism of $F_j(\vp,\vP)$ is derived widely in literature, e.g.\ \cite{Akhmedov:2010ms,Giunti:2002xg,Beuthe:2001rc}. 
	However, in the following we take a generic Gaussian form for the weighting function on the Fock-PS for simplicity,
	and also integrate out $x^0$ in Eq.~\eqref{Aj_full} for later convenience. 
	Therefore, from Appendix~\ref{Sec.AppB-2}, 
	\begin{equation} \label{AforWig}
	A_{2,j}(T,\vL,\vP)=e^{-iE_j T+i\vj \vP_j T}
	\int dp^3 \int dx^3 \, e^{i\vx\vp}
	G(\vx;\vL_j)F_j(\vp;\vP_j),
	\end{equation}
	where 
	\begin{equation} \label{Gauss}
	G(\vx;\vL_j) \propto  \exp \left[{\frac{-(\vx-\vL_j)^2}{4\,\sigma_x^2} }\right], \quad
	F_j(\vp;\vP_j) \propto \exp\left[{\frac{-(\vp-\Delta\,\vP_j)^2}{4\,\sigma_p^2}}\right],
	\end{equation}
	while $\vL_j=\vL-\vj T$ and $\Delta = 1 + 4 \sigma_x^2 \sigma_p^2$.

	The relation between $\sigma_x$ and $\sigma_p$ with the original input functions in Eq.~\eqref{tot_amp}, namely, $f_{Pi},f_{Pf},f_{Di},f_{Df},g_{P}$ and $g_{D}$ is calculated in the Appendix~\ref{Sec.AppB} and summarized in Fig.~\ref{fig:CompDist}. 
	At the end, $\sigma_x$ includes all the individual uncertainties, temporal and coordinate, production and detection in a convolutional type, hence, the larger one of which will dominate.
	On the other hand, $\sigma_p$ merges the momentum uncertainties of the production and detection in a product type referred to in Table~\ref{table1}; 
	however, at each site, the total momentum uncertainty  combines momentum uncertainties of individual external states in a convolutional way, therefore, the largest one among them would dominate. This relation coincides with the one from Ref.\ \cite{Giunti:2002xg} assuming Gaussian distributions.
	Note that the normalisation of the weighting functions is irrelevant here, since we will later show that the FTP will be automatically normalised on the measurement layer with the definition in Eq.~\eqref{def.P3}. 
	Hence, after inserting the Gaussian distributions, we easily find that
	\begin{equation} \label{Phi}
	\hat{\Phi}_j(\vL,\vP) \propto e^{i\vP_j\vL_j}
	\, \exp{\left[-\vPj^2\Delta\sigma_x^2-\frac{\vL_j^2\sigma_p^2}{\Delta}\right]}.
	\end{equation}        
	Thus, when $\sigma_x \ll 1/(2\sigma_p)$, i.e.\ when the width of the inner solid line in Fig.~\ref{fig:Feyn} is much smaller than that of the blue/green circle's width, $\sigma_x$ can be neglected, and vice versa.
	
	%2-2%%%%%%%%%%%%%%%%%%%%%%%%%%%%%%%%%%%%%%%%%%%%%%%%%%%%%%%%%%%%%
	\subsection{Microscopic Layer (Layer 1): Quasi-Transition Probability} \label{Sec.2-2}

	In this section we calculate the quasi-probability distribution (or Wigner function in {e.g.}\ \cite{Case}) of the FTP on the Wigner-PS.
	Such distributions bridge QM (or QFT in this case) to statistical probability distributions, such that the expectation value is calculated by direct PS integration.
	Moreover, in Section~\ref{Sec.3-2} we will illustrate that it is insightful and useful to look at state decoherence as a PWO effect from the Wigner-PS perspective.
	Although we do not obtain the quasi-probability distribution $P_{\bar{1}}$ from the Wigner transformation directly, we end up with the same formalism by doing a change of variables such that Eq.~\eqref{LMOP} is fulfilled. 
	This means we find $P_{\bar{1}}$ such that
	\begin{align}
	&P_{2,\ab}
	=\int d^3 \bar{x} \int d^3 \bar{p} \, P_{\bar{1}}(\bar{\vx},\bar{\vp})
	=A_{2,\ab} A^\dagger_{2,\ab}\nonumber\\
	&=\sumU \, e^{-i(E_j-E_k)T+i(\vj\vPj-\vk\vPk)T} \, A_{2,j}A^\dagger_{2,k},
	\label{P_wig}
	\end{align}
	where $A_{2,j/k}$ are given by Eq.~\eqref{AforWig}.
	Therefore, for
	\begin{equation}
	P_{\bar{1}}=\sumU e^{-i(E_j-E_k)T+i(\vj\vPj-\vk\vPk)T}P_{\bar{1},jk},
	\end{equation} 
	Eq.~\eqref{P_wig} implies that
	\begin{align} \label{P1bar_def}
	\int d^3 \bar{x} \int d^3 \bar{p} \, P_{\bar{1},jk}(\bar{\vx},\bar{\vp})
	=\int d^3 x \int d^3 p \, e^{i\vp \vx}\, G(\vx)F_j(\vp)
	\int d^3 x' \int d^3 p' \, e^{-i\vp'\vx'}\, G^*(\vx')F_k^*(\vp'),
	\end{align}	
	in which the right-hand side involves mixing of the two PS, $(\vx,\vp)$ and  $(\vx',\vp')$, while that on the left-hand side does not. 
	
	Therefore, $P_{\bar{1},jk}$ is the quasi-probability distribution representing the occupation number of having both the $j$th and the $k$th mass eigenstate simultaneously in the Wigner-PS. 
	The equation above is achieved when we replace  $(\vx,\vx')\rightarrow (\bar{\vx}=\frac{1}{2}(\vx + \vx'),\Delta \vx = \vx-\vx')$ and $(\vp,\vp')\rightarrow (\bar{\vp}=\frac{1}{2}(\vp + \vp'),\Delta \vp = \vp-\vp')$. Hence, Eq.~\eqref{P1bar_def} can be rewritten as  
	\begin{equation}
	P_{\bar{1},jk}(\bar{\vx},\bar{\vp})
	=\tilde{W}^G_{jk}(\bar{\vx},\bar{\vp})\tilde{W}^F_{jk}(\bar{\vx},\bar{\vp}),
	\end{equation}
	where $\tilde{W}^G_{jk}(\bar{\vx},\bar{\vp})$ and $\tilde{W}^F_{jk}(\bar{\vx},\bar{\vp})$ take the form of the Wigner quasi-probability distribution as follows:
	\begin{equation} \label{Wigner}
	\begin{split}
	&\tilde{W}^G_{jk}(\bar{\vx},\bar{\vp})=\int d^3(\Delta \vx)\, e^{i\Delta \vx \bar{\vp}} \,
	G(\bar{\vx}+\frac{1}{2}\Delta \vx;\vL_j)\,G^*(\bar{\vx}-\frac{1}{2}\Delta \vx;\vL_k), \\
	&\tilde{W}^F_{jk}(\bar{\vx},\bar{\vp})=\int d^3(\Delta \vp)\, e^{i\Delta \vp \bar{\vx}} \,
	F_j(\bar{\vp}+\frac{1}{2}\Delta \vp;\vP_j)\,F_k^*(\bar{\vp}-\frac{1}{2}\Delta \vp;\vP_k). \,
	\end{split}
	\end{equation}
	In this case $\mathcal{LMO}^{\bar{1}}=\int d^3\bar{x}\int d^3\bar{p}$ is simply the integration over the Wigner-PS, and the FTP on this layer is the quasi-probability distribution  $\tilde{W}^G_{jk}\tilde{W}^F_{jk}$, which includes both $W_1$ and $B_1$ in Eq.~\eqref{LMO}.
	In particular, if we assume all  weighting functions to be Gaussian on the  Fock-PS given in Eq.~\eqref{Gauss},
	the quasi-probability distributions become 
	\begin{equation} \label{Wigner_Gauss}
	\begin{split}
	&\tilde{W}^G_{jk}(\bar{\vx},\bar{\vp}) \propto \exp\left[i\bar{\vp}(\vL_j-\vL_k)\right]\exp\left[-\frac{(\bar{\vx}-\bar{\vL}_{jk})^2}{2 \sigma_x^2}  -2\,\bar{\vp}^2 \sigma_x^2\right], \\
	&\tilde{W}^F_{jk}(\bar{\vx},\bar{\vp}) \propto \exp \left[i\bar{\vx}\Delta(\vP_j-\vP_k)\right]\exp\left[-\frac{(\bar{\vp}-\Delta\bar{\vP}_{jk})^2}{2 \sigma_p^2} -2\,\bar{\vx}^2 \sigma_p^2\right] ,
	\end{split}
	\end{equation}
	where $\bar{\vL}_{jk}=\frac{\vL_j+\vL_k}{2}$ and $\bar{\vP}_{jk}=\frac{\vP_j+\vP_k}{2}$.
	When we move the FTP up to the physical layer by integrating over the Wigner-PS, there will be a PWO effect suppressing the plane wave term on the physical layer. 
	Moreover, the PWO effect is determined by the width of the weighting function relative to the wavelength of the phase structure, { which is $\vP_j-\vP_k$ for $\bar{\vx}$ and $\vL_j-\vL_k$ for $\bar{\vp}$ in this case. 
	}
	In fact, the wider the weighting function is relative to the wavelength of the phase structure, the more suppression will the PWO effect cause. 
	We can also view this as how many periods (e.g.\ $2\pi/(\vP_j-\vP_k)$ is one period in $\bar{\vx}$) there are determined by the phase structure within some width of the weighting function.
	We call the number of periods within some area on the PS  the phase density. Hence, the higher the phase density is, the more damping we will get from the PWO effect.	
	
	\begin{figure}
		\centering
		\includegraphics[width=0.7\textwidth]{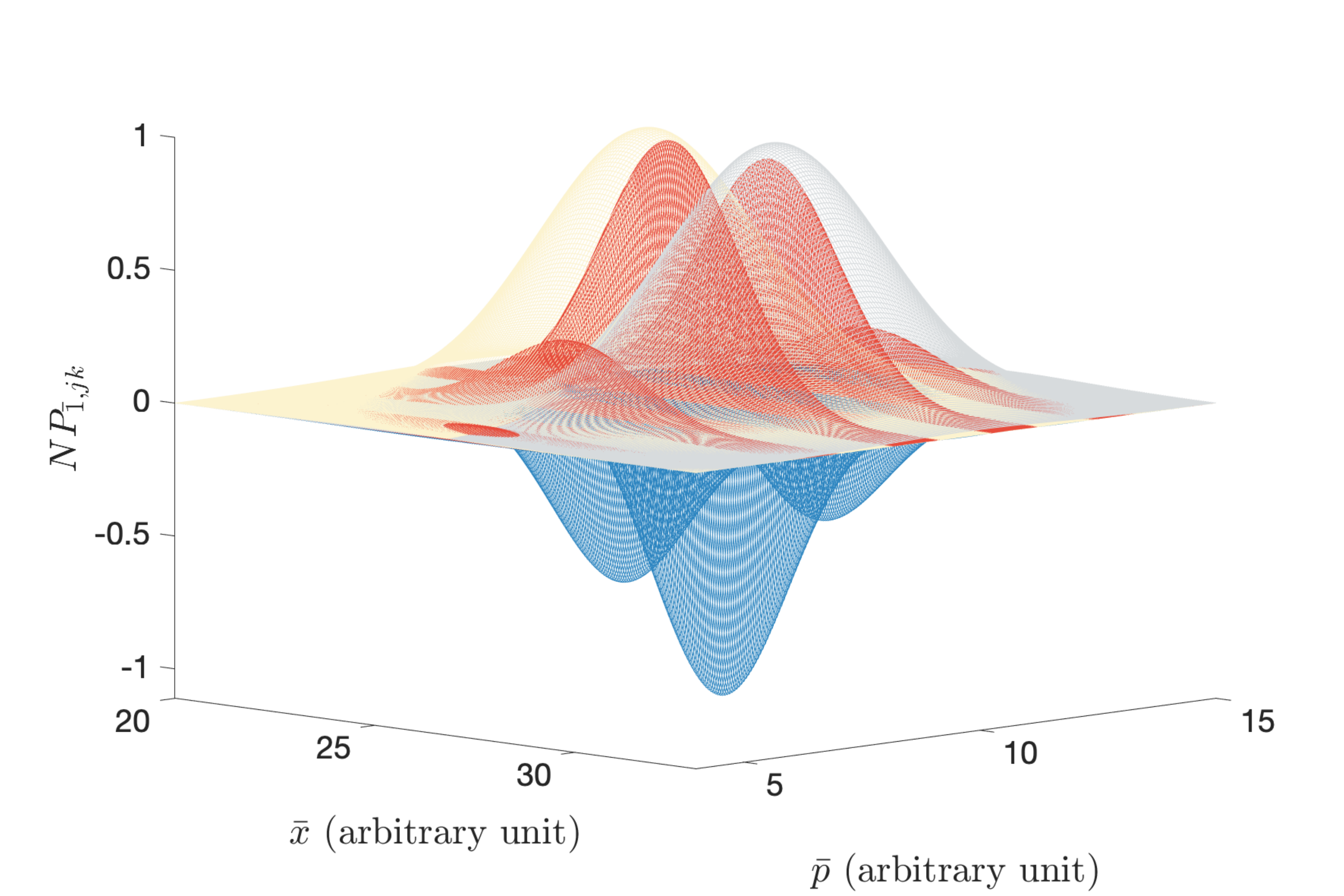}
		\caption{\label{fig:Wigner3D} 
			An illustration of the quasi-probability distribution in Eq.~(\ref{Wigner_Gauss}) on layer 1 in the Wigner-PS assuming Gaussian distributed weighting functions scaled by $N$ (as explained in the text, the normalization of the distributions is irrelevant, just the width and shape are).
			The yellow/grey areas represent $P_{\bar{1},jj}$/$P_{\bar{1},kk}$, and the red/blue area are for the positive/negative values of $P_{\bar{1},jk}$.   }
	\end{figure}

	\begin{figure}
		\centering
		\includegraphics[width=0.3333\textwidth]{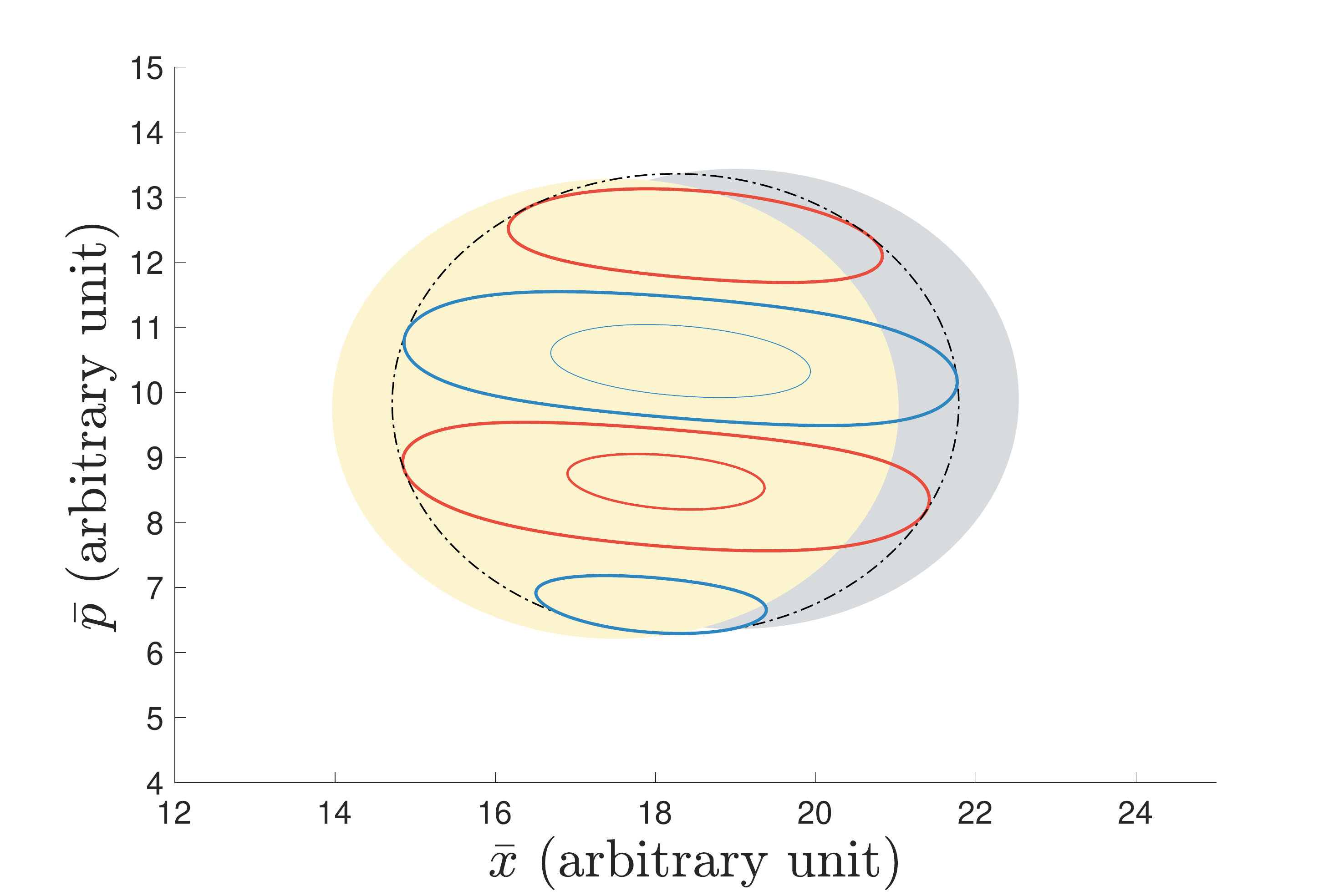}\hfill
		\includegraphics[width=0.3333\textwidth]{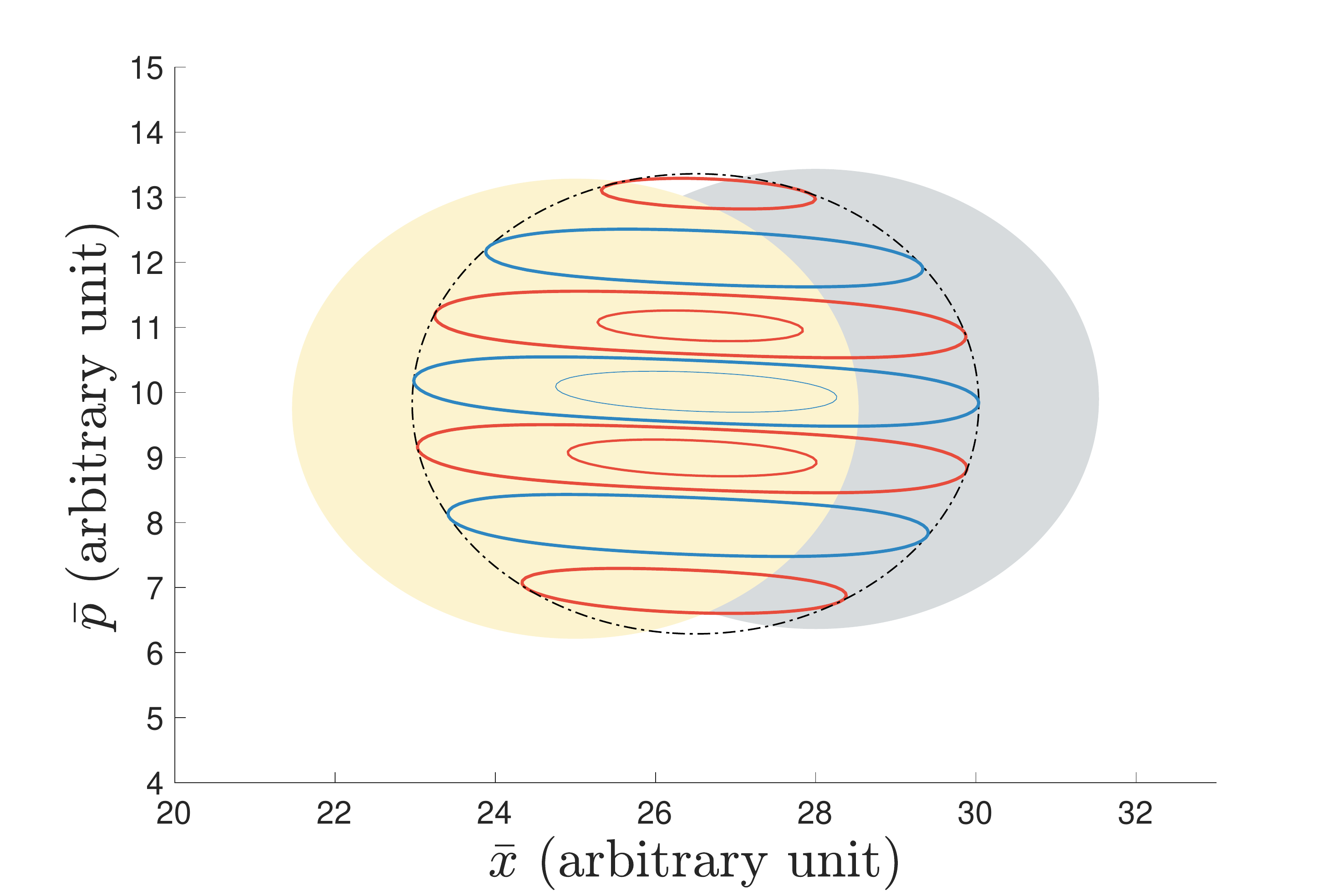}\hfill
		\includegraphics[width=0.3333\textwidth]{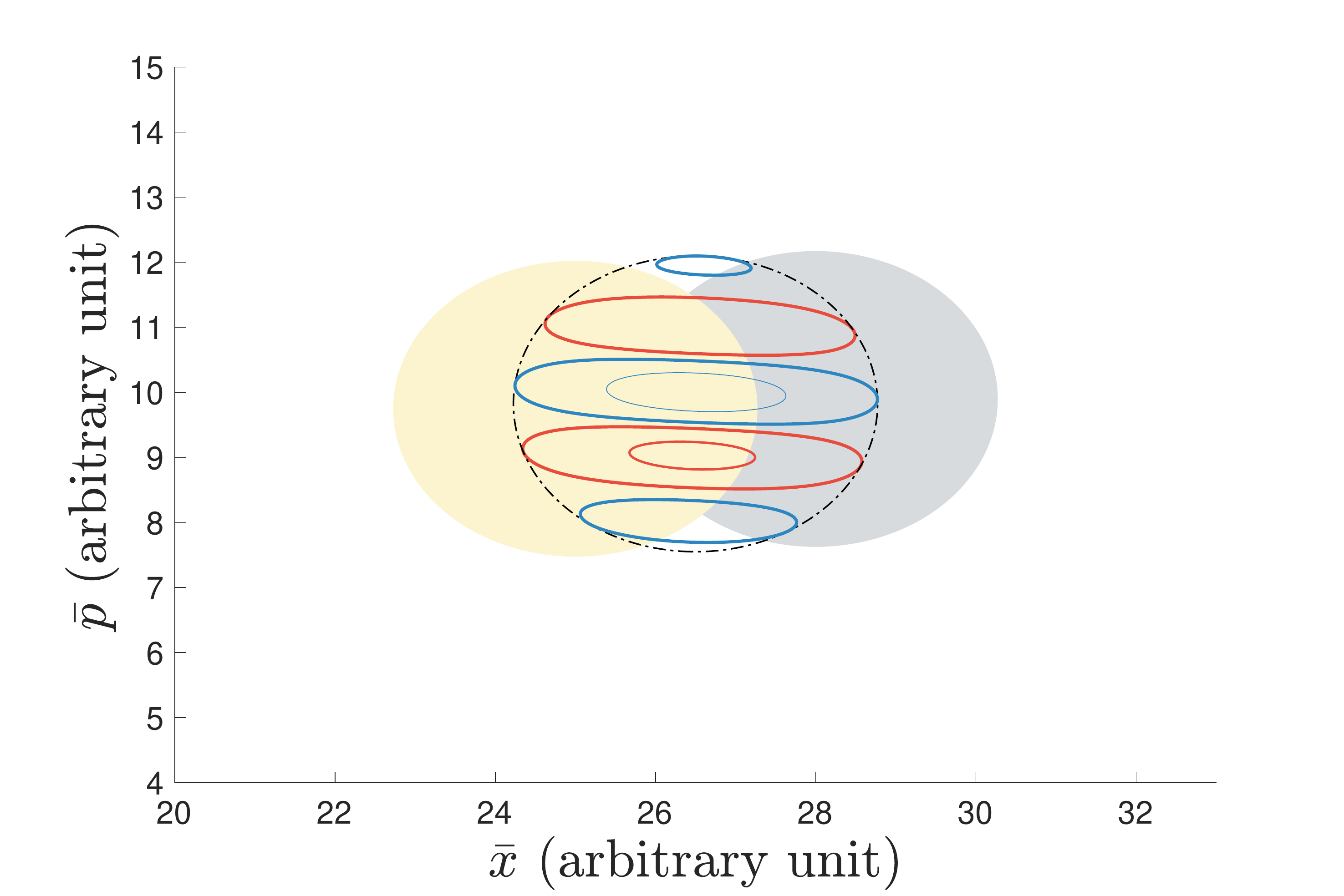}
		\caption{\label{fig:wigner} 
			An illustration of the quasi-probability distribution in Eq.~(\ref{Wigner_Gauss}) on layer 1 in the Wigner-PS assuming Gaussian distributed weighting functions.
			These plots show the projection of 3D plots like Fig.~\ref{fig:Wigner3D} onto 2D plots. 
			The edge of the shaded areas, the (outer) thick solid lines and the dashed lines are contour lines for the distributions within two standard deviations, while the (inner) thin lines are for one standard deviation. The shaded areas represent $P_{\bar{1},jj}$ and $P_{\bar{1},kk}$, the red/blue lines are for the positive/negative values of $P_{\bar{1},jk}$, and the black dashed line shows the weighting function for $P_{\bar{1},jk}$. The three plots differ by the traveling time ($T_{\rm plot}$) and width ($\sigma_{\rm plot}=\sigma_p=\sigma_x$) as: $T_{\rm left}<T_{\rm middle}=T_{\rm right}$ and $\sigma_{\rm left}=\sigma_{\rm middle}$, also $\sigma_{\rm right}=0.325 \,\sigma_{\rm left}$ or $1.376 \,\sigma_{\rm left}$.  }
	\end{figure}
	Fig.~\ref{fig:Wigner3D} and Fig.~\ref{fig:wigner}  are plotted as an example to illustrate the quasi-probability distribution on the Wigner-PS before integrating out the PS, which would lead to a PWO effect on the physical layer after the integration. 
	In particular, Fig.~\ref{fig:wigner} demonstrates the projection of 3D plots such as Fig.~\ref{fig:Wigner3D} onto a 2D contour plot{s}, where the areas within both the shaded and non-shaded circles represent contour lines of one standard deviation of a Gaussian distribution with Eq.~\eqref{Wigner_Gauss}. 
	For both plots, the more times we see positive and negative values (red and blue circles) alter, the higher the phase densities are within one 
	standard deviation range of the weighting function, and the larger the PWO effect will be.   
	In particular, relative to the left plot which is taken for some time $T_{\rm left}$ and some width $\sigma_{p}=\sigma_x=\sigma_{\rm left}$ for the weighting functions, the middle plot has the same width but a larger propagating time, {i.e.}\ $T_{\rm middle} > T_{\rm left} $. Furthermore, the right plot has width as either $\sigma_{p}=\sigma_x=\sigma_{\rm right}=0.325 \, \sigma_{\rm left}$ or $1.376 \, \sigma_{\rm left}$, and $T_{\rm middle} = T_{\rm right} $.
	By comparing the left and middle plot, we see that on this layer, as time evolves, although the FTP of the two mass eigenstates separate, the width of the overlapping FTP does not. However, the phase density does increase, hence, so does the PWO effect. On the other hand, since the width of the FTP is $\sigma_{\bar{x}}^2=\sigma_x^2/(1+4 \sigma_x^2 \sigma_p^2)$ for $\bar{\vx}$ and $\sigma_{\bar{p}}^2=\sigma_p^2/(1+4 \sigma_x^2 \sigma_p^2)$ for $\bar{\vp}$, the maximum value of $\sigma_{\bar{x}}^2 \sigma_{\bar{p}}^2$ is $1/4${. This indicates that the quasi-probability is localised and the uncertainty principle holds automatically. } 
	Hence, there are two solutions of $\sigma_{\rm right}$, for some value of $\sigma_{\bar{x}}=\sigma_{\bar{p}}< 1/2$.
	Also, the decrease in $\sigma_{\bar{x}}$ and $\sigma_{\bar{p}}$ would also reduce the phase density, relaxing the PWO effect.
	Note that on the Wigner-PS, when the quasi-probability distribution is localised to a point, it describes a classical monochromatic field \cite{Case}.

	%2-3%%%%%%%%%%%%%%%%%%%%%%%%%%%%%%%%%%%%%%%%%%%%%%%%%%%%%%%%%%%%%
	\subsection{Physical Layer (Layer 2): Transition Probability} \label{Sec.2-3}
	% Formalism 
	In this section, we move onto the physical layer, where experimental uncertainties are considered in addition. Since we consider isotropic emission of neutrinos for simplicity, the additional uncertainties introduced through the weighting function on this layer are the (macroscopic) energy uncertainty $\sigma_E$ and the (macroscopic) coordinate uncertainty $\sigma_L$. The former include the energy uncertainties, such as the energy resolution of the experiment ($\sigma_E \propto 1/\sqrt{E_0}$) and the energy reconstruction models (for example, see \cite{Martini:2012uc,DeRomeri:2016qwo}); whereas the latter include any uncertainties on the propagation distance. For instance: the core size and the distribution of multiple reactors for reactor neutrinos; the length of the decay pipe and the velocity of the parent particles before they decay into neutrinos for accelerator neutrinos; the uncertainty in the altitude of where the neutrinos are produced in the atmosphere for atmospheric neutrinos; or even more exotic effects that would introduce uncertainties to how long a neutrino propagates before it gets detected, such as space time fluctuations. 
	
	Before taking the additional uncertainties on layer 2 into consideration, we need to first move the FTP from layer 1 to this layer, which can either be transferred from the first layer by the Fock-PS via

	\begin{equation} 
	P_{2,jk}(T,\vL, \vP)=e^{-i(E_j-E_k)T+i (\vP_j-\vP_k) L} 
	\hat{\Phi}^*(\vL_k;P_k) \hat{\Phi}(\vL_j;P_j), 
	\label{P2_1}
	\end{equation}
	where the FTP is
	\begin{equation}
	P_{2,\ab}(T,\vL, \vP)=\sumU P_{2,jk}(T,\vL, \vP),
	\end{equation}
	or by the Wigner-PS via
	\begin{equation}
	P_{2,jk}(T,\vL, \vP)=\int d^3\bar{x} \int d^3\bar{p} \,
	\tilde{W}^G_{jk}(\bar{\vx},\bar{\vp}) \, \tilde{W}^F_{jk}(\bar{\vx},\bar{\vp}).
	\label{P2_1bar}
	\end{equation}
	Both approaches give us the same result as a useful consistency check.
	By taking the Gaussian distribution on the first layer in Eq.~\eqref{Gauss}, we arrive at 
	\begin{equation} \label{P2jk_Gauss}
	P_{2,jk}(T,\vL,\vP) \propto 
	e^{-i(E_j-E_k)T+i(\vP_j-\vP_k)\vL}
	\exp \left[-(\vP_j^2+\vP_k^2)\sigbx^2
	-(\vL_j^2+\vL_k^2)\sigbp^2\right],
	\end{equation}
	by either Eq.~\eqref{Phi} or Eq.~\eqref{Wigner_Gauss}, and it is also useful to rewrite the widths in terms of $\sigbx= \Delta\,\sigma_x$ and $\sigbp=\sigma_p/\Delta$, where $\Delta=1+4\sigma_x^2\sigma_p^2$.
	
	% approximation and comparison with the standard formula
	So far we have not specified how the second layer momentum variable $\vP$ in the relativistic-PS is related to the  expectation value of momentum  $\vP_j$ for each mass eigenstate.
	The relation is obtained below by comparing the massless case with the small mass case. 
	For a certain energy $E$, in the massless case, we have $E=|\vP|$; but for nonzero neutrino mass, some of the energy, $\delta E_j$, will be consumed by the mass, and this freedom is limited by the uncertainty of the energy on the first layer. In other words, this picture is equivalent to having some $\vP_j(E')$ instead of $E_j(\vp)$ while deriving Eq.~\eqref{A_j_final}, with $E$ being the mean of energies $E'$ of the neutrinos given by the external particles in the first layer. Therefore, we can write 
	\begin{equation} \label{approx_absP}
	|\vPj|\equiv E - \delta E_j . 
	\end{equation}
	If we expand $\vPj$ with respect to $m_j$ and keep only the leading order, we get 
	\begin{equation} \label{P_j}
	\vPj\simeq \vP-\vec{\xi}_j \frac{m_j^2}{2E} 
	=\vec{\xi}_p E -\vec{\xi}_j \frac{m_j^2}{2E},
	\end{equation}
	where $\vec{\xi}_p=\vP/E=\vP/|\vP|$, so that $|\vec{\xi}_p|^2=|\vec{\xi}_j|^2=1$. 
	This allows the identification of $\delta E_j$, by substituting Eq.\ (\ref{P_j}) to Eq.\ (\ref{approx_absP}) with $|P_j|=\sqrt{P_j^2}$. The result is
	\begin{equation}
	\delta E_j = E- |\vPj| \simeq \vec{\xi}_p\vec{\xi}_j \frac{m_j^2}{2E}
	\end{equation}
	to lowest order in $m_j$. The actual energy which the mass eigenstate carries is decided by the dispersion relation: 
	\begin{equation} \label{approx_E}
	E_j=\sqrt{|\vPj|^2+m_j^2} \simeq E - \delta E + \frac{m_j^2}{2E} 
	\simeq E + \frac{m_j^2}{2E}\left(1-\vec{\xi}_p\vec{\xi}_j \right). 
	\end{equation}
	Therefore, we can see when $\vec{\xi}_p\vec{\xi}_j=1$, i.e.\ the case where all mass eigenstates as well as the massless case are co-linear with each other, we have  equal energy of mass states; and when $\vec{\xi}_p\vec{\xi}_j=0$, we have equal momentum modulus instead. To have exact equal momentum, we need $\vec{\xi}_j=\vec{0}$. However, according to e.g.\ \cite{Akhmedov:2009rb,Giunti:2000kw}, neither of this should be the case due to Lorentz invariance. 
	Additionally, it is also useful to derive the group velocity 
	\begin{equation}\label{v_j}
	\vj = \frac{\vP_j}{E_j}\simeq \left(\vec{\xi}_p E -\vec{\xi}_j \frac{m_j^2}{2E}\right)
	\frac{1}{E} \left(1 + \frac{m_j^2}{2E^2}(1-\vec{\xi}_p\vec{\xi}_j)\right)^{-1}
	\simeq \vec{\xi}_p \left(1-\frac{m_j^2}{2E^2}\right).
	\end{equation}
	With the approximated relation from above, Eq.~\eqref{P2jk_Gauss} becomes
	\begin{equation} \label{P2jk_Gauss_approx}
	P_{2,jk}(T,L,E) = e^{i\psi_{jk}'(T,L,E)}D'_x(T,L,E)D'_p(E),
	\end{equation}
	where the phase structure, the momentum weighting function and the coordinate weighting function are 
	\begin{align}
	&\psi_{jk}'(T,L,E) = -\frac{\mjk}{2E}(T(1-\eta)
	+L \eta), \label{psi_T}\\
	&D_p'(E)\propto \exp\left[-2\sigbx^2\left(E\vec{\xi}_p-\frac{m_j^2 \vec{\xi}_j+m_k^2 \vec{\xi}_k}{2E}\right)^2
	-\frac{\sigbx^2}{2}\left(\frac{m_j^2 \vec{\xi}_j-m_k^2 \vec{\xi}_k}{2E}\right)^2\right],\label{Dp_T}\\
	& D_x'(T,L,E) \propto
	\exp\left[-\sigbp^2\left(\frac{L \mjk}{2\sqrt{2}E^2}\right)^2
	-2\sigbp^2\left(1-\frac{m_j^2+m_k^2}{2E^2}\right)\left(T- L\right)^2 \right], \label{Dx_T} 
	\end{align}
	respectively. Also, we have taken $\vec{\xi}_j=\vec{\xi}_k\equiv\vec{\xi} $ and $\eta = \vec{\xi}\, \vec{\xi}_p=\vec{\xi}\, \vec{\xi}_L$  as the alignment factor.
	Note that in Sec.~\ref{sec.3} we will show that terms such as the second term in the brackets of Eq.~\eqref{Dp_T} will be cancel out by normalization.
	Furthermore, for experiments with continuously emitted neutrinos within a sufficiently long period of time, we should integrate out $T$, since there is no temporal information. In this scenario,  $\psi_{jk}'$ would be replaced by $\psi_{jk}$ and $D_x'$ would be replaced by $D_x$ in Eq.~\eqref{P2jk_Gauss_approx}, where
	\begin{align}
	&\psi_{jk}(L,E) = -\frac{\mjk L}{2E}, \label{psi_L}\\
	& D_x(L,E) \propto \exp\left\{-\sigbp^2\left(\frac{L \mjk}{2\sqrt{2}E^2}\right)^2
	-\frac{1}{2\sigbp^2}
	\left[\frac{\mjk}{2E}(1-\eta)\right]^2 
	\right\}. \label{Dx_L} 
	\end{align}
	Note that $\psi_{jk}$ being independent of $\vec{\xi}_j$ and $\vec{\xi}_k$ implies that ``equal energy", ``equal momentum" or anything in between, will lead to the same phase structure if we have no temporal information. 
	In fact, when $\sigbx \sim 0$, we will arrive at the standard decoherence formula in \cite{Giunti:2002xg}, namely
	\begin{equation} \label{P2_standard}
	P_{2,jk}(L,E) \simeq \exp \left[i\frac{\mjk L}{2E}\right]
	\exp \left[-\left(\frac{L}{L_{kj}^{\rm coh}}\right)^2
	-(1-\eta)\left(\frac{\mjk}{2\sqrt{2}E\sigbp}\right)^2\right],
	\end{equation}
	where
	\begin{equation} 
	L_{kj}^{\rm coh}
	=\frac{2\sqrt{2} E^2}{|\Delta m_{kj}^2|\sigbp} 
	\end{equation}
	for freely propagating neutrinos.

	Up to now, we have not yet considered the weighting functions on the second layer, but only how the weighting functions on the first layer are transmitted onto the second layer. But before going into that, we simplify our discussion by changing the variables $\{T,\vL,\vP\}$ to $\{T,L=|\vL|,E=|\vP|,\Omega_L,\Omega_P\}$, where $\Omega_L/\Omega_P$ are the solid angles for $\vL/\vP$.
	Therefore, the layer-moving-operator from the physical layer to the measurement layer is
	\begin{equation}
	\mathcal{LMO}^2= \int dL \int dE \int d\Omega \int dT \,
	H_L(L;L_0)\, H_E(E;E_0)\, H_T(T;T_0)\, H_\Omega(\Omega;\Omega_0).		
	\end{equation}
	Then by assuming that neutrinos are isotropically emitted and that we have no temporal information, i.e.\ we consider only uncertainties of $E$ and $L$, $P_{2,jk}$ is moved to the third layer as
	\begin{equation} \label{P3}
	P_{3,jk}(L_0,E_0) \propto \int dL \int dE \, 
	H_L(L;L_0)\, H_E(E;E_0)P_{2,jk}(L,E).
	\end{equation} 
	For a counting experiment, the transformation from the second layer to the third layer 
	\begin{equation}
	P_{3,jk}(X_0)=\int dX \, P_{2,jk}(X)H_X(X-X_0),
	\end{equation}
	performs the convolution between the FTP on the second layer which includes uncertainties from the first layer, and the weighting function $H_X(X;X_0)$, which is the PDF of the true value $X$, for some measured value $X_0$. 
	% Uncertainty of L	
	For example, the measured rate at $L_0$ does not only have contributions from neutrinos actually propagating the distance $L_0$ but is a sum of all possible contributions within a time window given by the uncertainty of time, which is taken as infinity for the case of continuous emission of neutrinos for a sufficiently long period of time. 	
	In particular, with the commutative and associative properties of convolution, the coordinate uncertainty, $H_L(L;L_0)$, comes from the convolution of the spatial PDF of the production process and the detection process.  Therefore, the largest uncertainty would dominate, which is usually  the production PDF, i.e.\ the source profile or the PDF of neutrino production.

\begin{figure} 
		\centering
		\includegraphics[width=0.8\textwidth]{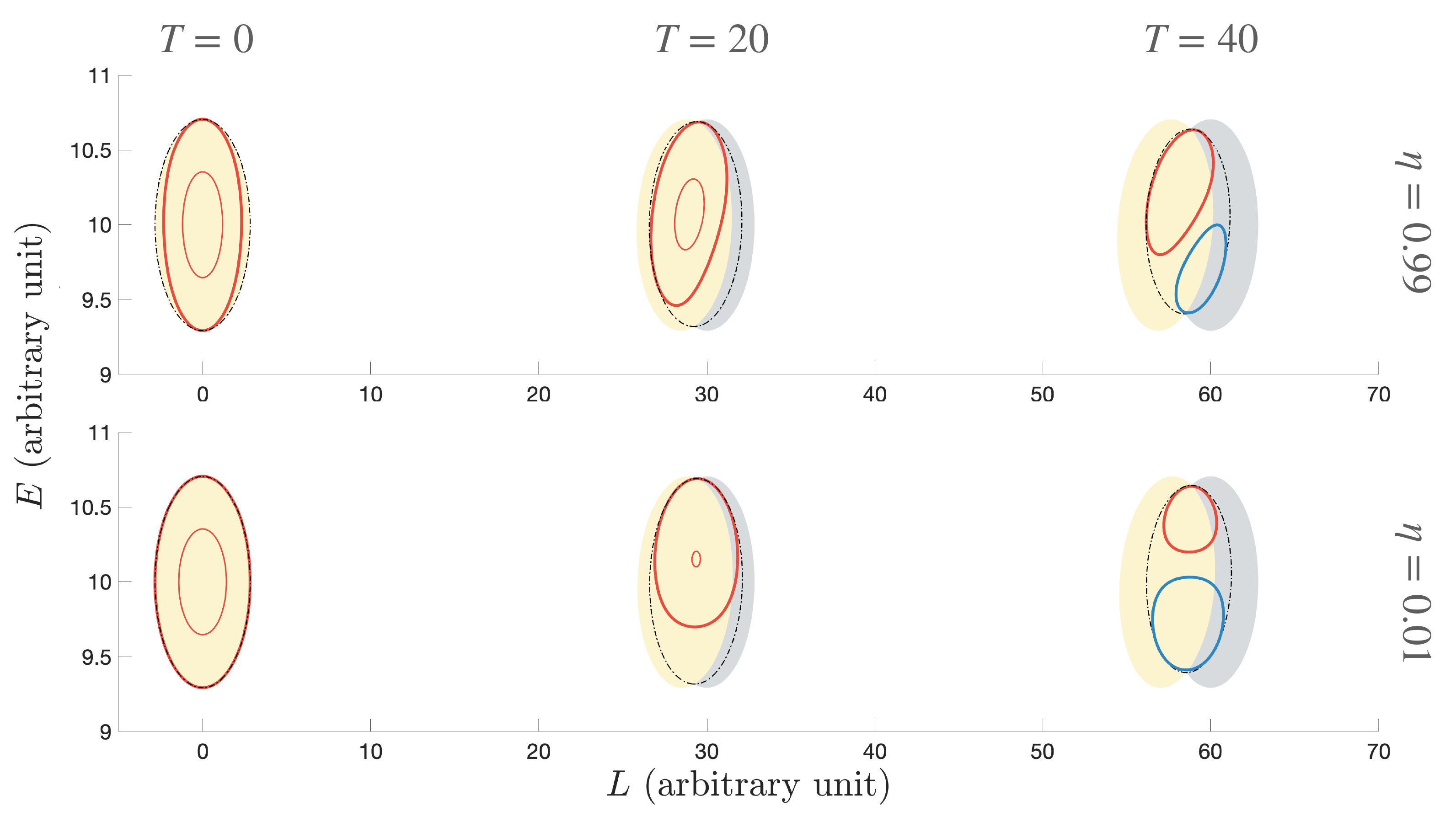}
		\caption{\label{fig:layer2}
			Same as Fig.~\ref{fig:wigner}, but demonstrating the FTP in Eq.~\eqref{for_fig4} on the physical layer, also assuming Gaussian distributed weighting functions. The differences between each FTP are labeled on the figure, showing the time evolution of the FTP and the effect of the alignment factor.}
	\end{figure}

	Since Eq.~\eqref{P3} also results in a PWO effect, Fig.~\ref{fig:layer2} is plotted in the same way as Fig.~\ref{fig:wigner} to see the phase density of the weighting function. The difference between these two plots is that on the  relativistic-PS, the separation of $P_{2,jj}(\vL,\vP)$ and $P_{2,kk}(\vL,\vP)$ affects the width of $P_{2,jk}(\vL,\vP)$ (black dotted circle), unlike the case of the Wigner-PS in Fig.~\ref{fig:wigner}. 
	In Fig.~\ref{fig:layer2} we illustrate the time-dependent case for $P_{2,jk}$ on the physical layer for an example, the phase structure is the same as Eq.~\eqref{psi_T}, as well as the spatial uncertainty with Eq.~\eqref{Dx_T}. However, since the energy range of interest is much larger than the central value of $D_p'$ which is suppressed by the neutrino mass, in Eq.~\eqref{Dp_T} we consider the second layer weighting function $H_E(E;E_0)$ for the energy uncertainty. Hence, the contour lines in Fig.~\ref{fig:layer2} are for one standard deviation of
	\begin{equation} \label{for_fig4}
	P_{2jk}(T,L,E;E_0)=\exp{\left[\frac{-(E-E_0)^2}{2\sigma_E^2}\right]}D'_x(T,L,E)\,
	e^{i\psi_{jk}'(T,L,E)}.
	\end{equation}       	
	We also show two cases of the alignment $\eta$ between $\xi_p$ and $\xi= \xi_j=\xi_k$ in Eq.~\eqref{P_j}, to see how it affects the time-dependent phase structure.  
	In addition, the heavier mass eigenstate $P_{2,jj}$ (the yellow shaded area) is a bit tilted at large $T$, due to the energy dependence of the group velocity in Eq.~\eqref{v_j}. Nonetheless, this tilting is negligible in reality, since the mass splitting is a lot smaller compared to the uncertainty of the energy, $\sigma_E$. 
	We will show that Fig.~\ref{fig:wigner} and Fig.~\ref{fig:layer2} correspond to state decoherence and phase decoherence, respectively, and illustrate the time-independent version of the plots in Sec.~\ref{sec.3}.   

	%2-4%%%%%%%%%%%%%%%%%%%%%%%%%%%%%%%%%%%%%%%%%%%%%%%%%%%%%%%%%%%%%
	\subsection{Measurement Layer (Layer 3): Transition Probability} \label{Sec.2-4}

	In this section, we finally reach the measurement layer where experimental data is collected. On top of the uncertainties arising from for the previous layers, i.e.\ the phase space uncertainties (PSUs), the final data also has to take into consideration the count uncertainties (CUs). While the PSUs are uncertainties of the PS variables (for instance, $\sigma_x$, $\sigma_p$, $\sigma_L$ and $\sigma_E$ discussed previously), the CUs are uncertainties of the neutrino flux, for example, the statistical uncertainties and the background uncertainties.    
	The PSUs and the CUs will be treated differently in our likelihood/$\chi^2$ analysis in Sec.\ \ref{sec.4}: the CUs will be treated as the conventional ``uncertainties" in the analysis, whereas the PSUs are included in the theoretical prediction. The total count rate for energy-distance binned data is
	\begin{equation}
	N_{\rm tot}=\int_{L_0\text{ bin}}  dL_0 \, \frac{1}{4 \pi L_0}
	\int_{E_0\text{ bin}} dE_0\, \Phi_0(E_0) D(E_0)  P_{3,\ab}(L_0,E_0), 
	\end{equation}
	where the CUs are included in $\Phi_0(E_0)$, which is the neutrino flux at $L_0 = 0$ taking into account the decay rate of the production process; $D(E_0)$ is the detection rate including the cross section involved in the detection  process. On the other hand, the PSUs are included in $P_{3,\ab}(L_0,E_0)$, the FTP on the measurement layer, 
	and will be discussed in the following paragraph.

\begin{figure} 
		\centering
		\includegraphics[width=0.765 \textwidth]{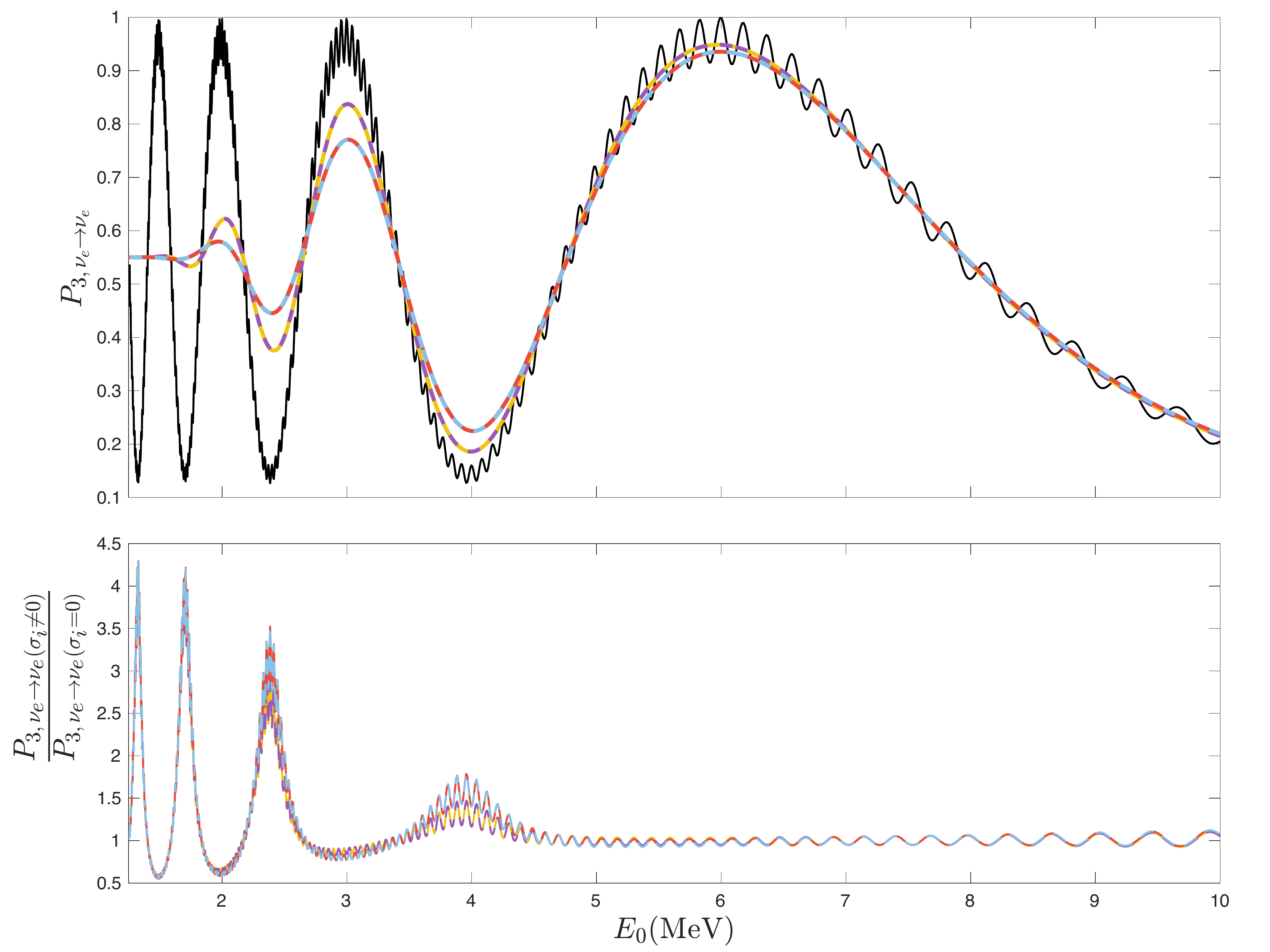}
		\caption{\label{fig:P3_spectrum}Three flavor FTP spectrum from electron neutrino to electron neutrino on the measurement layer at $L_0 = 200$ km (this distance is chosen such that decoherence effects are visible in this energy range for some reasonable uncertainties). The black line represents completely coherent FTP, and the coloured lines all have $\sigbp=0.4$ MeV and $\sigbx$ is negligible compared to $\sigma_L$. We have $\sigma_L=0$ m, $\sigma_E=0$ MeV for the yellow line, $\sigma_L=5$ m, $\sigma_E=0$ MeV for the purple line, $\sigma_L=0$ m, $\sigma_E=0.1 \sqrt{E_0}$ MeV for the red line, and $\sigma_L=5$ m, $\sigma_E=0.1 \sqrt{E_0}$ MeV for the blue line. Here $\sigma_L$ is chosen according to a typical reactor core size, $\sigma_E$ is a typical detector resolution and $\sigbp$ is taken at a value such that it is comparable with $\sigma_E$.}
		\label{L0_P3}
	\end{figure}

\begin{figure}
		\centering
		\includegraphics[width=0.48\textwidth]{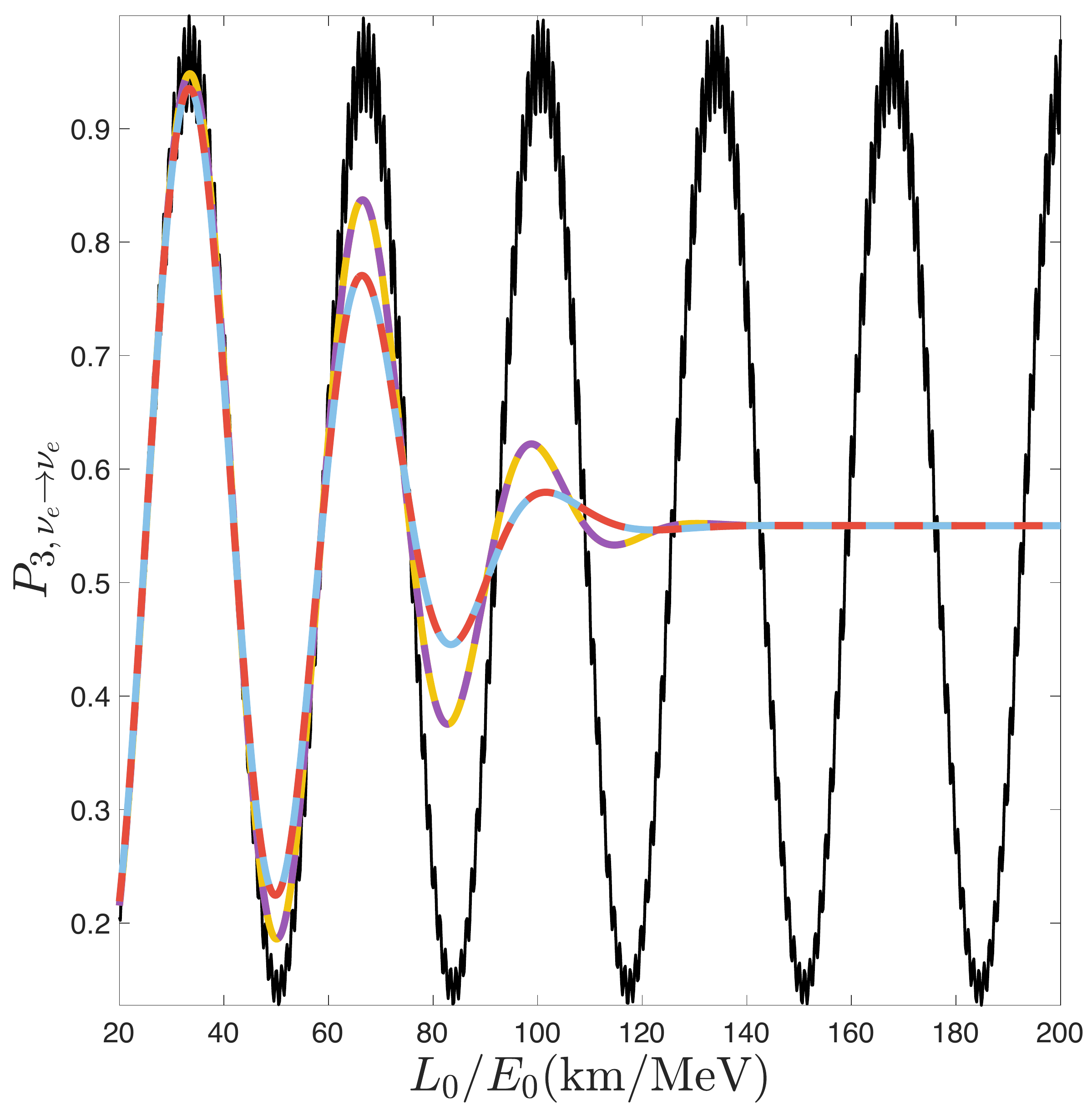}\hfill
		\includegraphics[width=0.48\textwidth]{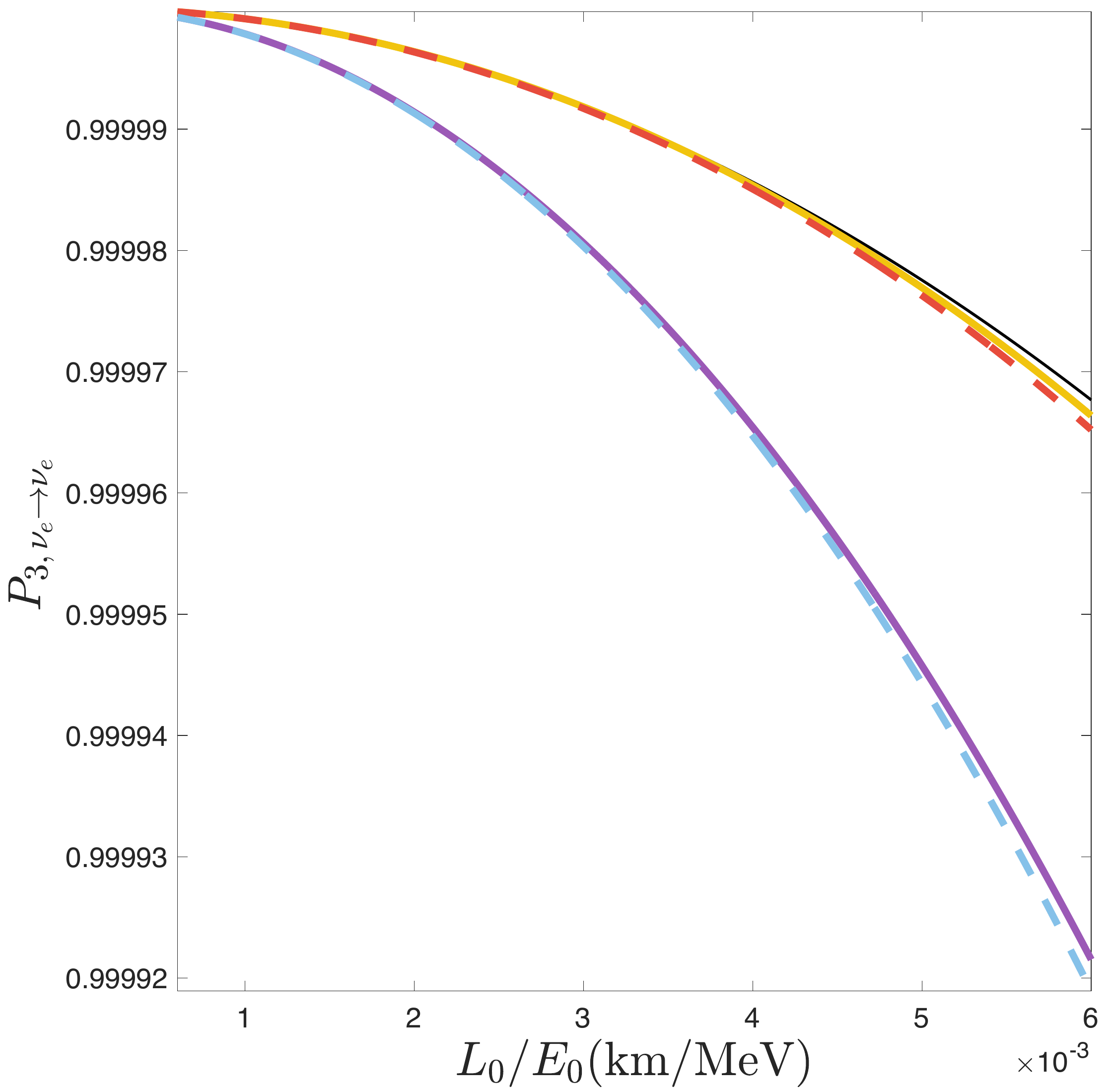}
		\caption{\label{fig:P3_L_E} FTP from electron neutrino to electron neutrino as a function of $L_0/E_0$ on the measurement layer for near (right) and far (left) detector. The lines are labels in the same way as Fig.~\ref{fig:P3_spectrum}. We can identify more sensitivity to $\sigma_L$ at near detectors while effects of  $\sigbp$ and $\sigma_E$ are  more pronounced at far detectors. 
		}
	\end{figure}

	Below, we demonstrate effects of PSUs on the FTP by integrating over the approximated time-ignorant FTP on the second layer, and take Gaussian PDFs for both $H_L(L;L_0)$ and $H_E(E;E_0)$.
	However, since only the integration over a Gaussian $H_L$, 
	\begin{align} \label{P3_HL}
	& P_{3,jk}^{ \,\rm semi} (L_0,E)=
	\int d L \, \exp\left[ \frac{-(L-L_0)^2}{4\sigma_L}\right]P_{2,jk}(L,E) \nonumber\\
	& =  \exp\left[i\frac{\mjk L_{0,jk}^{\rm eff}}{2E}\right] 
	\exp\left[-\frac{\left(L_{0,jk}^{\rm eff}\right)^2}{\left(L_{kj}^{\rm coh}\right)^2}
	-\left(\frac{\mjk \Delta_L}{2E}\right)^2
	-\left(\frac{\mjk}{2\sqrt{2}E}\right)^2
	\left[\left( \frac{1-\eta}{\sigbp}\right)^2+\sigbx^2\right]
	\right] \nonumber\\
	&\simeq
	\exp\left[i\frac{\mjk L_0}{2E}\right] 
	\exp\left[-\left(\frac{\mjk \sigma_L}{2E}\right)^2
	-\left(\frac{L_0}{L_{kj}^{\rm coh}}\right)^2\right] ,
	\quad \text{ ($\sigbx, 1/\sigbp \ll \sigma_L \ll L^{\rm coh}_{jk}, L_0$)},
	\end{align}
	can be calculated analytically,
	the integration over $E$ will be done numerically, and the FTP on the measurement layer, $P_3(L_0,E_0)$ is plotted in Fig.~\ref{fig:P3_spectrum} and Fig.~\ref{fig:P3_L_E}. In Eq.~\eqref{P3_HL}, the total spatial uncertainty (width of $H_L$) for a vacuum propagating neutrino is $\sigma_L^2=\sigma_S^2+\sigma_D^2$. Here $\sigma_S$ describes the production profile at the source and $\sigma_D$ represents the spatial resolution of the detector, hence, $\sigma_L$ is normally dominated by $\sigma_S$, as mentioned previously, and
	\begin{equation}
	\Delta_L^2 = \frac{\sigma_L^2 \left(L_{kj}^{\rm coh}\right)^2}{4\sigma_L^2+\left(L_{kj}^{\rm coh}\right)^2}, \qquad
	L_{0,jk}^{\rm eff}=\frac{L_0\left(L_{kj}^{\rm coh}\right)^2}{4\sigma_L^2+\left(L_{kj}^{\rm coh}\right)^2}.
	\end{equation}
	In the second line of Eq.~\eqref{P3_HL}, we can see that the two last terms in the exponent are locality terms, indicating that the more local uncertainty ($\sigbx$, $1/\sigbp$ and $\sigma_L$) we have, microscopic or macroscopic, the more the FTP will be smeared out, and since $\sigbx$ and $1/ \sigbp \ll \sigma_L $ the macroscopic one dominates as we can see in the third line. As for the first term in the second exponent, the coherence length is modified by $L_{kj}^{\rm coh} \rightarrow \sqrt{(L_{kj}^{\rm coh})^2+4\sigma_L^2}$, but since $L_{kj}^{\rm coh}$ is inversely proportional to the mass splitting of neutrinos, it is usually much larger than $\sigma_L$ for ground-based experiments. In this case, the microscopic uncertainty would dominate for this term. 
	However, this would not be the case for neutrinos produced in continuously emitting celestial objects for a long period of time, {see e.g.\ \cite{Farzan:2008eg}}. For instance, the size of the Sun's core would be much larger than the coherence length in the three neutrino paradigm. Nonetheless, the coherence length might be stretched out for neutrinos produced in an extreme environment, such as supernovae \cite{Esteban-Pretel:2007jwl,Akhmedov:2017mcc}, or by going to ultrahigh energy for long oscillation length  ($\propto 1/E_0$) \cite{Hooper:2004xr}.     
	Additionally, since the integration $\int dE H_E(E;E_0)P_{3,jk}^{ \,\rm semi} (L_0,E)$ satisfies the factorization condition in Appendix~\ref{Sec.AppC}, we can write 
	\begin{equation}
	P_{3,jk}(L_0,E_0) \simeq 
	\exp\left[-\left(\frac{\mjk \sigma_L}{2E_0}\right)^2
	-\left(\frac{\mjk \sigbp L_0}{2\sqrt{2}E_0^2}\right)^2\right]
	\int dE \, 
	%e^{i \mjk\frac{ L_0}{2E}} 
	\exp\left[i\frac{\mjk L_0}{2E}\right]
	\, H_E(E;E_0).
	\end{equation}
	Thus, the effects of $H_L$ barely depend on $L_0$, unless $\sigma_L$ accumulates w.r.t.\ $L_0$, since it would then only affect the locality term.

	The FTP of neutrinos detected at a longer distance is plotted in Fig.~\ref{fig:P3_spectrum} and the left plot in Fig.~\ref{fig:P3_L_E}, with oscillation parameters taken from NuFit 5.1 global fit results \cite{Esteban:2020cvm}.
	From these figures we can see that the FTP is barely sensitive to $\sigma_L$ compared to $\sigma_p$ and $\sigma_E$, and $\sigma_E$ would not only cause damping to the oscillation, but also a phase shift, which is more pronounced at low energies. On the other hand, the right plot of Fig.~\ref{fig:P3_L_E} is plotted for a very short traveling distance, and the sensitivity for $\sigma_L$ is much larger than that for $\sigma_p$ and $\sigma_E$. Although the effect is very small here compared to the long distance case, it benefits from larger statistics. Moreover, since the phase-density is higher for lower energies, due to a smaller oscillation length, the PWO effect is more significant and the coherence length is shortened, hence neutrino decoherence effects are more enhanced for lower energies, as we can see in Fig.~\ref{fig:P3_spectrum} and Eq.~\eqref{P3_HL}.

	%3%%%%%%%%%%%%%%%%%%%%%%%%%%%%%%%%%%%%%%%%%%%%%%%%%%%%%%%%%%%%%%
	\section{Neutrino Decoherence} \label{sec.3}
	%3-1%%%%%%%%%%%%%%%%%%%%%%%%%%%%%%%%%%%%%%%%%%%%%%%%%%%%%%%%%%%%%
	\subsection{Formalism} \label{Sec.3-1}
	In this section we show how neutrino decoherence can be further classified into two categories, depending on either coherence is lost by the separation of the mass eigenstates, or by statistical averaging.  
	Moreover, in the language of the layer structure, we will see that both categories of decoherence effects result from PWO effects, therefore, we formulate the effects of neutrino decoherence with a damping term ($\phi_{jk}$) and a phase shift term ($\beta_{jk}$), which can be written as
	\begin{equation} \label{Decoh_para}
	P_{3,jk}(X_3)= e^{i\left[ \psi_{jk}(X_3)-\beta_{jk}(X_3;\vec{\sigma})\right]}\phi_{jk}(X_3;\vec{\sigma}),
	\end{equation}  
	for
	\begin{equation}
	P_{3,\alpha \rightarrow \beta}(X_3)=\sumU P_{3,jk}(X_3),
	\end{equation} 
	where $\psi_{jk}$ is the phase structure on the second layer in Eq.~\eqref{psi_T}. 
	Also, $\vec{\sigma} = \{\vec{\sigma}_x, \vec{\sigma}_p,\vec{\sigma}_L,
	\vec{\sigma}_E,\vec{\sigma}_T\}$ represents the set of parameters that describe the weighting functions, such as the width and asymmetry parameters for the respective variables.
	In general $X_3=\{T_0,L_0,E_0,\Omega_{L0},\Omega_{P0}\}$ are the third layer's temporal variable, spatial variable, energy variable, spatial solid angle and momentum solid angle. As for isotropic neutrinos $X_3=\{T_0,L_0,E_0\}$, and on top of that, if the neutrinos are continuously emitted for a sufficiently long period of time, $X_3=\{L_0,E_0\}$. In this paper, we will only consider the two latter cases for simplicity. In particular, if all the weighting PDFs are Gaussian distributions, then the damping term can be parameterized as
	\begin{equation} \label{damp_Gauss}
	\phi_{jk}(X_3)= e^{-\left[\mjk \gamma(X_3;\vec{\sigma})\right]^2}.
	\end{equation} 	
	% (Operational) definition of P and justification of this definition
	Moreover, since $P_{3,jk}$ is on the measurement layer, it is necessary to properly define the operational FTP.
	Here, we define the FTP by each $P_{3,jk}(X_3)$, as the ratio between the total count with and without oscillation, namely
	\begin{equation} \label{def.P3}
	P_{3,jk}(X_3)=
	\frac{\int dX_2 \, H(X_2;X_3) \, \Gamma_{2,jk}(X_2;X_3)}
	{\sqrt{\int dX_2 \, H(X_2;X_3) \, \Gamma_{2,jj}(X_2;X_3)}
		\sqrt{\int dX_2 \, H(X_2;X_3) \, \Gamma_{2,kk}(X_2;X_3)}},
	\end{equation}
	where $\Gamma_{2,jk}(X_2;X_3) \propto P_{2,jk}(X_2;X_3)$ is the un-normalized FTP on the second layer.
	Here $X_2$ are the second layer variables analogous to $X_3$.
	We note that the so-defined $P_{3,jk}(X_3)$ is not affected by the scale of the weighting functions on each layer, but is only dependent on their widths and shapes.
	
	One might be concerned that the definition in Eq.~\eqref{def.P3} is not justified since we measure the FTP by the total $P_{3,\ab}$ instead of each $P_{3,jk}$. However, theoretically speaking, it is possible to measure $P_{3,jk}$ if we measure the denominator by measuring the exact neutrino mass, and the numerator from a well controlled oscillation experiment.
	Note that there would be no oscillation for a mass measuring experiment, for one would know exactly which mass eigenstate the neutrino propagates on. Therefore, a mass measuring experiment is only useful for determining the normalization of an oscillation experiment, whereas the numerator in Eq.~\eqref{def.P3} can be found in oscillation experiments for neutrino mixing where only two mass eigenstates (with eigenvalues $m_j$ and $m_k$) are allowed/sensitive.
	Nonetheless, with the smallness of the mass splitting, the shape of the weighting functions on each layer can be approximated as being independent of $j,k$. Therefore $\Gamma_{2,jj}(X_2;X_3) =\Gamma_{2,kk}(X_2;X_3) = \Gamma_{\alpha}^{\rm pro}(X_2;X_3)\sigma_{\beta}^{\rm det}(X_2;X_3)$, where $\Gamma_{\alpha}^{\rm pro}(X_2;X_3)$ is the production rate for flavor $\alpha$, and $\sigma_{\beta}^{\rm det}(X_2;X_3)$ is the detection cross section for flavor $\beta$. In this case, the difference between the weighting functions only comes from the different group velocities $\vj$, which is only relevant when $j\neq k$, hence, the FTP would become
	\begin{equation}
	P_{3,\alpha \rightarrow \beta}(X_3)= 
	\frac{\sumU\int dX_2 \, H(X_2;X_3) \, \Gamma_{2,jk}(X_2;X_3)}
	{\int dX_2 \Gamma_{\alpha}^{\rm pro}(X_2)\sigma_{\beta}^{\rm det}(X_2)H(X_2;X_3)}.
	\end{equation}
	This expression and its condition for the uncertainties to be independent of $j,k$ coincide with \cite{Akhmedov:2010ms}.
	Additionally, the definition of Eq.\,(\ref{def.P3}) automatically normalizes, 
	\begin{equation}
	\sum_{\alpha}P_{3,\alpha \rightarrow \beta}=\sum_{\beta}P_{3,\alpha \rightarrow  \beta}=1,
	\end{equation}
	without any further condition, since it implies $P_{3,jj}=1$ for any $j$, then
	\begin{equation}
	\sum_{\alpha}P_{3,\alpha \rightarrow \beta} 
	= \sum_{j,k} \sum_{\alpha}  U_{\alpha j}^*U_{\alpha k} U_{\beta j}U_{\beta k}^* \, P_{3jk} 
	= \sum_{j,k} \delta_{jk} U_{\beta j}U_{\beta k}^* \, P_{3,jk} 
	= \sum_{j} U_{\beta j}U_{\beta j}^* \, P_{3,jj}
	=1,
	\end{equation}
	and similarly for $\sum_{\beta}P_{3,\alpha \rightarrow  \beta}=1$. In fact, for $ 0 \leq  U_{\beta j}U_{\beta j}^* \leq 1$ and $ 0 \leq P_{3,jj}\leq1$  $\forall j$, if and only if $P_{3,jj}=1$ will the normalization condition be satisfied.

	Furthermore, the definition of Eq.~\eqref{def.P3}  also serves the purpose of analyzing the decoherence effect, by writing it as
	\begin{equation} \label{P3_def.decoh}
	P_{3,jk}(X_3)
	\equiv S_{3,jk}(X_3) \, \Phi_{3,jk}(X_3) \, e^{i\psi_{jk}(X_3)}, 
	\end{equation}
	where 
	\begin{equation} \label{def.SD}
	S_{3,jk}(X_3)=
	\frac{\int dX_2 H |\Gamma_{2,jk}|}
	{\sqrt{\int dX_2 \, H \, \Gamma_{2,jj}}
		\sqrt{\int dX_2 \, H \, \Gamma_{2,kk}}}, 	
	\end{equation}
	and 
	\begin{equation} \label{def.PD}
	\Phi_{3,jk}(X_3) =
	\frac{\int dX_2 \, H\, \Gamma_{2,jk}}{\int dX_2 H |\Gamma_{2,jk}|} \, e^{-i\psi_{jk}}.
	\end{equation}
	Here, we have written $H\equiv H(X_2;X_3)$, $\Gamma_{2jk} \equiv \Gamma_{2jk}(X_2;X_3)$, $\theta_{jk}\equiv \theta_{jk}(X_3)$ for simplicity. Both functions in Eq.~\eqref{def.SD} and Eq.~\eqref{def.PD} are the decoherence terms, which are both unitary for the fully coherent case, and with modulus $\leq 1$ in general. In particular, $S_{3,jk}$ represents the probability of the two mass states overlapping with each other, while $\Phi_{3,jk}$ gives us the PWO effect introduced in previous sections and in Appendix \ref{Sec.AppA}. Therefore, we call the former part  ``the state decoherence (SD) term", where $S_{3jk}(X_3) = 1$ indicates that two mass eigenstates are fully overlapping. The latter part will be called ``the phase decoherence (PD) term", where $\Phi_{3,jk}(X_3) = 1$ represents the case where there is no PWO effect on the physical layer. 
	
\begin{figure} 
		\centering
		\includegraphics[width=0.7\textwidth]{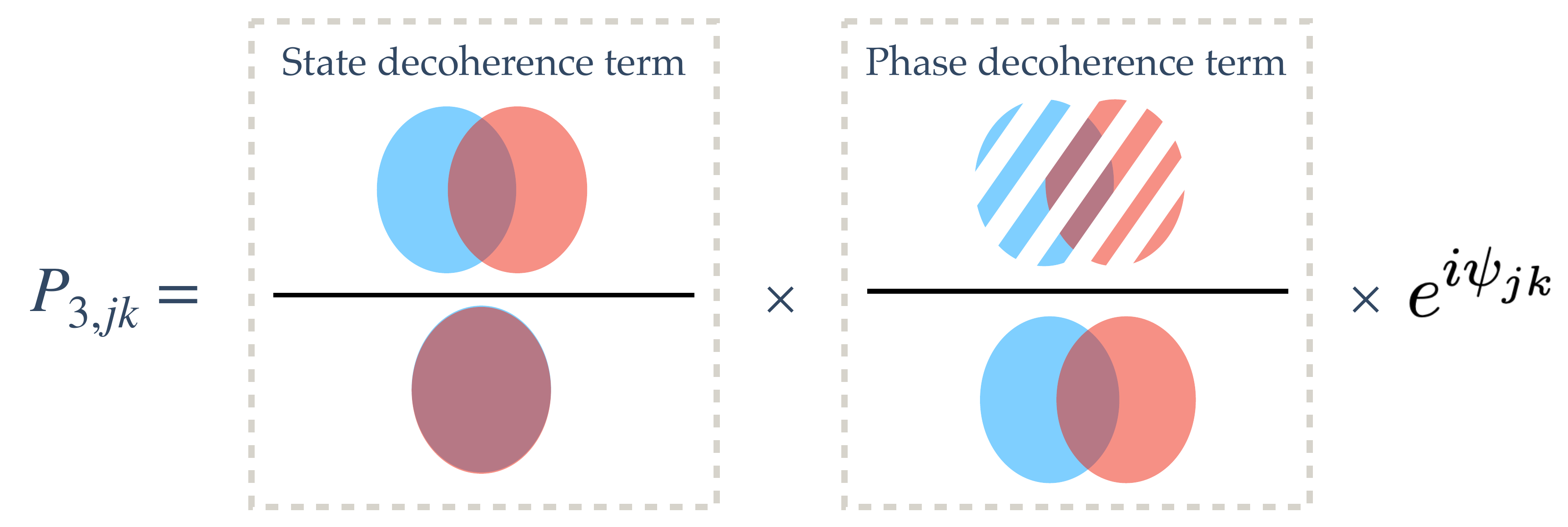}
		\caption{\label{fig:Decoh} Demonstration of how we separate the FTP defined in Eq.~\eqref{def.P3} into two terms: the state decoherence term and the phase decoherence term in Eq.~\eqref{def.SD} and Eq.~\eqref{def.PD} in terms of probability. The blue and red shaded circles represent two different mass eigenstates on the physical layer, and while the state decoherence term represents the separation of the two mass eigenstates, the phase decoherence term demonstrates a phase wash-out effect.}
	\end{figure}

	The trick to separate the decoherence term into SD and PD in Eq.~\eqref{P3_def.decoh} is demonstrated in Fig.~\ref{fig:Decoh}, where we insert $1=\int dX_2 H |\Gamma_{2,jk}| \, e^{-i\psi_{jk}}/\int dX_2 H |\Gamma_{2,jk}| \, e^{-i\psi_{jk}}$ into Eq.~\eqref{def.P3}. The blue and red shaded circles represent two different mass eigenstates on the physical layer, and the total FTP would be the purple area of the numerator in the phase decoherence term normalized by the purple area in the denominator of the state decoherence term. Finally, we see that SD is the probability of the two mass eigenstates being in superposition with each other on the relativistic-PS, while PD indicates our ignorance to the system on the relativistic-PS by averaging over all possibilities.    
	In particular, corresponding to Eq.~\eqref{P3_HL}, the term with the coherence length describing the WP separation represents SD, and the localization term shows PD. Therefore, we can already expect that the SD term will be dominated by the (microscopic) uncertainties on the first layer, while the PD term will be dominated by the (macroscopic) uncertainties on the second layer.
	
	%3-2%%%%%%%%%%%%%%%%%%%%%%%%%%%%%%%%%%%%%%%%%%%%%%%%%%%%%%%%%%%%%
	\subsection{State Decoherence} \label{Sec.3-2}
	% General description
	In this section, we will present how the SD term can be further analyzed and approximated. Eventually, we will show that although the SD represents state separation on the physical layer (2nd layer) it is equivalent to a PWO effect on the Wigner-PS (1st layer) under most conditions. Therefore, the dominating uncertainties for SD are those on the first layer, namely $\sigma_x$ and $\sigma_p$. 
	\begin{itemize}
		\item $\sigma_x$ (coordinate uncertainty on the Fock-PS $\ni$ layer 1): This uncertainty originates from the off-shell intrinsic uncertainties related to the finite space-time extension of the vertices, which depends on the neutrino interaction at the production and detection site. 
		In fact, the Fourier transformation of the weighting function $g_P$/$g_D$ results in an effective form factor as a function of neutrino momentum, which can be seen by integrating out $x_1$/$x_2$ in Eq.~\eqref{tot_amp}. Hence, depending on the interaction, $\sigma_x$ could be the size of the charge radius of the proton or neutron, which is at $\mathcal{O}(0.1-1)$ fm according to \cite{NuSTEC:2017hzk}, or at most the inter-atomic distance at $\mathcal{O}(0.1-1)$ nm, which is discussed in \cite{Kayser:2010pr}.
		In Sec.~\ref{Sec.2-1}, we included uncertainty sources as the unrelated spatial and temporal uncertainties (since they are not restricted to the mass-shell) for both the production and detection site in the calculation of the transition amplitude. 
		In Appendix \ref{Sec.AppB}, we have shown that $\sigma_x$ collects the sources of uncertainties in a convolution way, hence, it is dominated by the largest uncertainty source. 
		
		\item $\sigma_p$ (momentum uncertainty on the Fock-PS $\ni$ layer 1): This uncertainty arises from the WP description of the external states on a quantum mechanical level, which also appears when we calculate the transition amplitude. 
		For instance, $\sigma_p$ would be related to the mean free path of interactions before the external particles interact with the neutrinos or the life-time of the parent particles, see e.g.\ \cite{Giunti:2002xg,Akhmedov:2012uu}.
		Moreover, since the external particles are on the mass-shell, the energy uncertainties are correlated to the momentum uncertainties, and unlike $\sigma_x$, $\sigma_p$ collect the sources of uncertainties, including the production's and detection's momentum and energy uncertainties, in a product way, hence the smallest one among them would dominate. However, the total momentum uncertainty at each site is the convolution of the initial and final state WP, hence, the smallest one among them would dominate (see Appendix~\ref{Sec.AppB}).  
	\end{itemize}	
	Note that $\sigma_p$ and $\sigma_x$	 and independent of each other and can both be represented in either coordinate space or momentum space. The difference between these two uncertainties is whether they describe the uncertainties from the external states or the effective vertices (internal states).  
	Moreover, from Sec.~\ref{Sec.2-3} and Sec.~\ref{Sec.2-4}, we have seen that it is more effective to consider the width of the quasi-probability distribution: $\sigbx=\sigma_x/\Delta$ and $\sigbp=\sigma_p/\Delta$, where $\Delta=1+4\sigma_x^2\sigma_p^2$. 
	\begin{itemize}
		\item $\sigbx$ (coordinate uncertainty on the Wigner-PS $\ni$ layer 1):  By Eq.~\eqref{S_jk}, the damping term in Eq.~\eqref{damp_Gauss} for a time-independent Gaussian distributed PDF is
		\begin{equation} \label{gamma_x}
		\gamma_x = 
		\frac{\sigbx }{2 \sqrt{2}E_0} . 		
		\end{equation}
		However, as we will see in the next section, this structure is exactly the same as that for $\sigma_L$, which is usually macroscopic, and can therefore be neglected. 
		
		\item $\sigbp$ (momentum uncertainty on the Wigner-PS $\ni$ layer 1):  By Eq.~\eqref{Dx_L}, the damping term in Eq.~\eqref{damp_Gauss} for a time-independent Gaussian distributed PDF is
		\begin{equation}\label{gamma_p}
		\gamma_p = 
		\frac{\sigbp L_0}{2 \sqrt{2}E_0^{2}} \,, 
		\end{equation}
		and the time-dependent one can be taken care of by replacing $L_0$ with $T_0$.
		Therefore, the decoherence effect resulting from $\sigbp$ is should be searched for at long distance. Such effects have been widely studied, and explored, e.g.\ for reactor neutrinos in \cite{deGouvea:2020hfl,deGouvea:2021uvg}, which excludes $(2\sigbp)^{-1} <2.08 \times 10^{-4}$ nm at 90 \% CL. 
		In fact $(2\sigbp)^{-1}$ represents the total coordinate uncertainty combining the production and detection region. 
		Therefore, in the following, we sometimes take $\sigbp$ around 0.1 MeV as a ``reasonable" value, since it corresponds to $(2\sigbp)^{-1}\sim 10^{-3}$ nm, which is within the range of the proton/neutron charge radius and 
		the inter-atomic distance ($\mathcal{O}$(0.1–1) nm), see \cite{Kayser:2010pr} for an estimation for reactor neutrinos.  	
	\end{itemize}

	For the derivation of SD, we start by writing down the SD term following Eq.\ (\ref{def.SD}) as
	\begin{equation} \label{StatDecoh_def}
	S_{3,jk}(X_3) 
	=\frac{\int dX_2 H|\Gamma_{2,jk}|}
	{\sqrt{\int dX_2 \, H\, \Gamma_{2,jj}}
		\sqrt{\int dX_2 \, H\, \Gamma_{2,jj}}} 
	=\frac{\int dX_2 H  S_{2,jk}\Phi_{2,jk}}
	{\sqrt{\int dX_2 \, H \, \Phi_{2,jj}}\sqrt{\int dX_2 \, H \, \Phi_{2,jj}}},			
	\end{equation}
	by replacing $\Gamma_{2,jk}$ with
	\begin{equation}
	\Gamma_{2,jk}=|\Gamma_{2,jk}|e^{i\psi_{jk}}
	= S_{2,jk} \, \Phi_{2,jk} e^{i\psi_{jk}}
	\text{ and }
	\Phi_{2,jk}(X_2)=\int d^3\bar{x} \int d^3\bar{p} \, |\Gamma_{1,jk}(\vbx,\vbp;X_2)|,
	\end{equation}
	where $S_{2,jk}=S_{2,jk}(X_2)$, $\Phi_{2,jk}=\Phi_{2,jk}(X_2)$ and $\psi_{jk}=\psi_{jk}(X_2)$. Moreover, we have $\psi_{jj}(X_2)=0$ for any $j$  according to Eq.~\eqref{P2jk_Gauss_approx}, and $\Gamma_{\bar{1},jk}(\vbx,\vbp;X_2) \propto P_{\bar{1},jk}(\vbx,\vbp;X_2) $ is the un-normalized transition probability distribution on the Wigner-PS, such that $ \Gamma_{2,jk}(X_2)=   \int d^3\bar{x} \int d^3\bar{p} \, \Gamma_{\bar{1},jk}(\vbx,\vbp;X_2)$.
	Therefore, the remaining term becomes
	
	\begin{equation} \label{StateDecoh_Wigner}
	S_{2,jk}(X_2)
	=e^{-i\psi_{jk}}\frac{\int d^3\bar{x} \int d^3\bar{p} \, \Gamma_{1,jk}(\bar{\vx},\bar{\vp};X_2)}
	{\int d^3\bar{x} \int d^3\bar{p} \, |\Gamma_{1,jk}(\bar{\vx},\bar{\vp};X_2)|}
	\equiv e^{-i\psi_{jk}} \frac{\int d^3\bar{x} \int d^3\bar{p} \, \bar{D}_{jk}(\vbx,\vbp;X_2) e^{i\bar{\eta}_{jk}(X_2)}}
	{\int d^3\bar{x} \int d^3\bar{p} \, \bar{D}_{jk}(\vbx,\vbp;X_2)},
	\end{equation}
	indicating that $S_{2jj}(X_2)=1$, since $\psi_{jj}=\eta_{jj}=0$ $\forall j$.
	Hence, we only need to consider the $\bx$ and $\bp$ dependent terms for $\bar{D}_{jk}$. 
	In fact, Eq.~\eqref{StateDecoh_Wigner} has the same formalism as Eq.~\eqref{PWO} (i.e.\ it is a PWO effect) with the phase structure $\bar{\eta}_{jk}$ averaged over within the normalized $\bar{D}_{jk}$ region.  
	For instance, considering Gaussian distributed weighting functions (Eq.\eqref{Wigner_Gauss}), then 
	\begin{equation}
	\bar{D}_{jk}(\vbx,\vbp;T,\vL,E)=
	\exp \left[\frac{-\left(\vbx-\bar{\vL}_{jk}/\Delta\right)^2}{2\sigbx^2}\right]
	\exp \left[\frac{-\left(\vbp-\bar{\vP}_{jk}\right)^2}{2\sigbp^2}\right],
	\end{equation}
	and the phase structure is in general
	\begin{equation}\label{SD_phase_T}
	\bar{\eta}_{jk}(\vbx,\vbp;T,E)=-iT\vbp\,(\vj-\vk)+i\Delta\,\vbx\,(\vP_j-\vP_k),
	\end{equation}  
	where the relation between $T,\vL,E$ and $\bar{\vL}_{jk}, \bar{\vP}_{jk},\vPj, \vPk$ is given in Sec.~\ref{sec.3}.
	In the following, we again consider all distributions as isotropic, and simplify our discussion to a one-dimensional scenario.
	From Appendix \ref{Sec.AppC}, we find that for $\sigma_S$ and $\Delta_{X_2}$ being the width of $S_{2,jk}(X_2)$ and $Y_{2,jk}(X_2;X_3) \equiv H(X_2;X_3)\Phi_{2,jk}(X_2)$ respectively, the first approximation in Eq.~\eqref{SD_factor} can be made in the limit of $X_3 \gg \Delta_{X_2}$ and $\sigma_S \gg \Delta_{X_2}$. The picture of this factorisation condition is to neglect the uncertainty of $X_3$ on the second layer for the term that can be taken out of the integral, i.e.\ $X_2 \simeq X_3$, in $S_{2,jk}$ for some $\Delta_{X_2}$, but at the same time $\sigma_S$ can not be neglected, in order to have SD. Additionally, if the total uncertainty $\Delta_{X_2}$ is dominated by the physical layer uncertainty $H$, which is exactly the case for macroscopic measurements, then $Y_{2,jk}(X_2) \simeq Y_{2,jj}(X_2) \simeq Y_{2,kk}(X_2)$ and we arrive at the second approximation in the following equation:   
	\begin{equation} \label{SD_factor}
	S_{3,jk}(X_3) \simeq S_{2,jk}(X_2)\biggr|_{X_2=X_3} 
	\frac{\int dX_2 Y_{2,jk}}
	{\sqrt{\int dX_2 \, Y_{2,jj}}
		\sqrt{\int dX_2 \, Y_{2,kk}}}			
	\simeq S_{2,jk}(X_2)\biggr|_{X_2=X_3}.
	\end{equation}
	The evaluation of the comparison between the width sizes of $S_{2,jk}$, $\Phi_{2,jk}$ and $H$ can be done by taking Gaussian distributions for weighting function on each layer. 
	Precisely speaking, for both $S_{2,jk}$ (with width $\sigma_{S, X_2}$) and $\Phi_{2,jk}$ (with width $\sigma_{\Phi,X_2}$), we use Eq.~\eqref{Gauss}; as for $H(X_2)$ (with width $\sigma_{H,X_2}$), we simply write it as a Gaussian distribution around $X_3$. In this case, we have:
	
	\begin{equation}
	\Phi_{2,jk} (L,E,T)= \exp \left[-\frac{(L_j+L_k)^2\sigbp^2}{2}
	-\frac{(P_j+P_k)^2\sigbx^2}{2}\right], \label{phi_jk}
	\end{equation}
	and 
	\begin{equation}
	S_{2,jk} (L,E,T)= \exp \left[-\frac{(L_j-L_k)^2\sigbp^2}{2}
	-\frac{(P_j-P_k)^2\sigbx^2}{2}\right] \label{S_jk}.
	\end{equation}
	Therefore the width of $\Phi_{jk}$ w.r.t.\ $L$ and $T$ is $\sigma_{\Phi,L}=(2\sigbp)^{-1}$ and $\sigma_{\Phi,T} = [\sigbp(v_j+v_k)]^{-1} \simeq  E^2[\sigma_p(m_j^2+m_k^2)]^{-1}$, respectively. As for $S_{jk}$, the width w.r.t.\ $L$ and $T$ is $\sigma_{S,T} = [\sigbp(v_j-v_k)]^{-1} \simeq  E^2[\sigma_p(m_j^2-m_k^2)]^{-1}$ and $\sigma_{S,L} \rightarrow \infty$ for the time-dependent case, since $S_{2,jk}$ does not depend on $L$. As for the energy uncertainties, we have
	\begin{equation}
	\sigma_{\Phi, E}=\int dE 
	\exp \left[-\frac{1}{2}\left(2L-T\frac{m_j^2+m_k^2}{2 E^2}\right)^2\sigbp^2
	-\frac{1}{2}\left(2E+\frac{m_j^2+m_k^2}{2E}\right)^2\sigbx^2\right]
	\end{equation}
	and 
	\begin{equation} \label{sig_SE}
	\sigma_{S, E}= \int dE 
	\exp \left[-\frac{1}{2}\left(T\frac{m_j^2-m_k^2}{2 E^2}\right)^2\sigbp^2
	-\frac{1}{2}\left(\frac{m_j^2-m_k^2}{2E}\right)^2\sigbx^2\right] .
	\end{equation}
	With the $1/E$ dependence, the exponent in Eq.~\eqref{sig_SE} $\rightarrow 1$ as $E \rightarrow \infty$, so $\sigma_{S,E}$ diverges, and the width $\sigma_{\Phi,E}$ would also be stretched out. Therefore, $\sigma_{H,L} \ll \sigma_{\Phi,L},\sigma_{S,L}$ and $\sigma_{H,E} \ll \sigma_{\Phi,E},\sigma_{S,E}$. Hence, the macroscopic uncertainty $\sigma_{H,L},\sigma_{H,E}$ will dominate over the transferred microscopic ones of $\Phi_{2,jk}$ for $Y_{2,jk}(X_2;X_3)$ in Eq.~\eqref{SD_factor}. Thus, $\Delta_L \simeq \sigma_{H,L}$ and $\Delta_E\simeq \sigma_{H,E}$.
	Accordingly, the factorization condition for Eq.~\eqref{SD_factor} is fulfilled for the $L$ and $E$ part if the measurement values $L_0$ and $E_0$ are also much larger than $\sigma_{H,L}$ and $\sigma_{H,E}$, which is exactly the case for neutrino experiments which have the resolution to measure neutrino oscillation.
	
\begin{figure} [h]
		\centering
		\includegraphics[width=.9\textwidth]{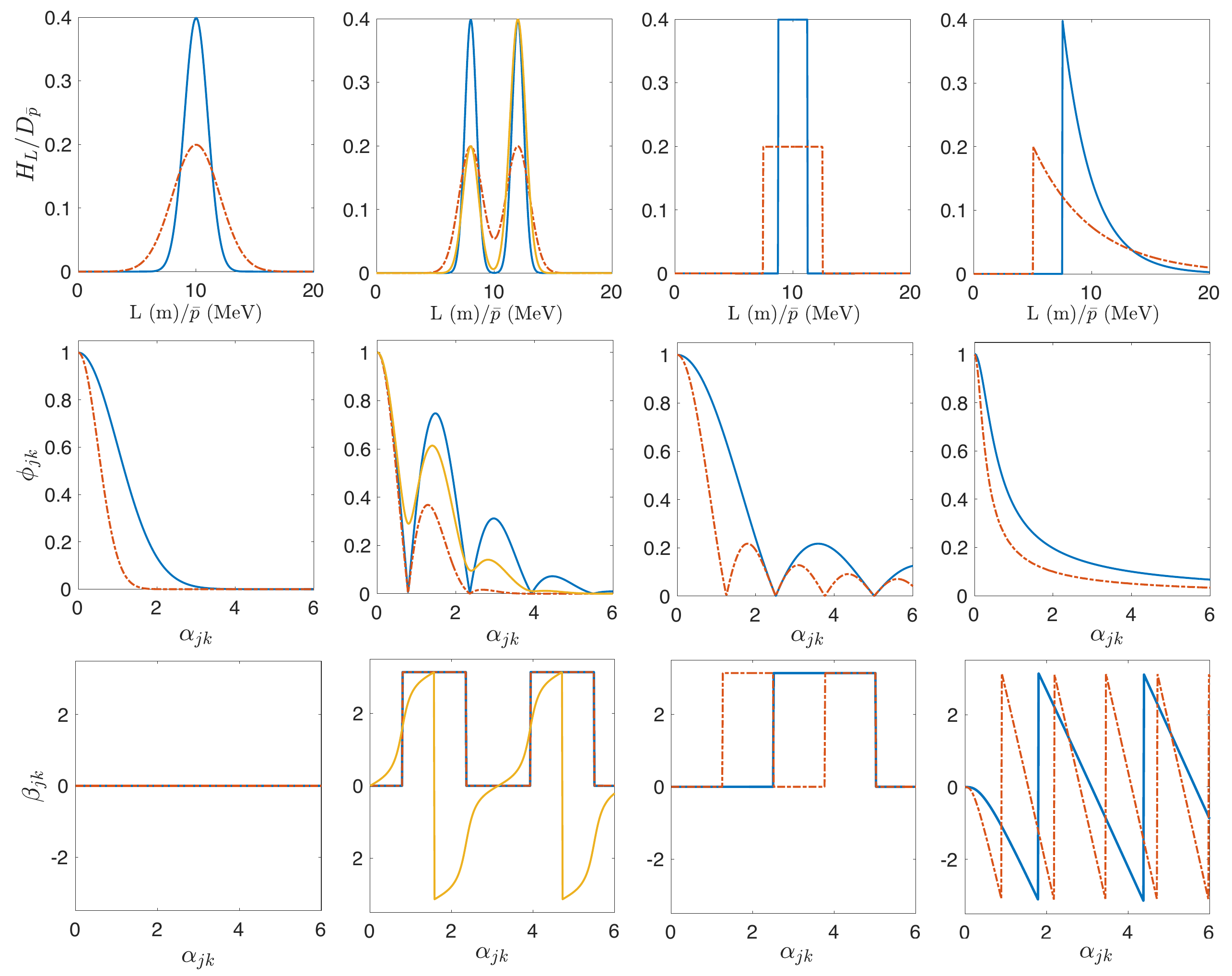}\hfill
		\caption{\label{fig:PWO_sigL}  Decoherence damping terms (second row) and phase shift terms (third row) for different shaped weighting functions (first row) of $H_L$-induced or $D_{\bp}$-induced decoherence. While both types of decoherence are described by Fourier transformation, the former transfers from $L$ space to the $\alpha_{jk}=\mjk/(2E_0)$ corresponding to Eq.~\eqref{PD_factor}; and the later from $\bp$ to $\alpha_{jk}=\mjk L_0/(2E_0^2)$ relates to Eq.~\eqref{SD_bp}. In particular, the same coloured lines represent weighting functions with the same widths.}
	\end{figure}

	As for the temporal part, we will discuss two scenarios: $\sigma_{H,T} \ll \sigma_{\Phi,T},\sigma_{S,T}$ and $\sigma_{H,T} \rightarrow \infty$. 
	If it  is neither of these two cases, one should integrate over $T$ in advance while taking $H_T$ into account.
	In the former case, the factorization condition for $T$ is also satisfied, and $H(X_2)$ dominates over $\Phi_{2,jk}(X_2)$ for all $X_2=\{L,E,T\}$, then $S_{3,jk}(X_3) \simeq S_{2,jk}(X_3)$. 
	Therefore, we can directly obtain the observational time-dependent SD weighting function $D'_{jk}(\bx,\bp;T_0,L_0,E_0)$ and phase structure $\eta'_{jk}$ by replacing $X_2$ with $X_3$. 
	Specifically, if all quantum uncertainties are Gaussian distributed,
	then $D'_{jk}=\bar{D}_{jk}(\bx,\bp;T_0,L_0,E_0)$ and $\eta'_{jk}=\bar{\eta}_{jk}(\bar{p},\bar{x};T_0,E_0)$. In Fig.~\ref{fig:wigner}, we illustrate such time-dependent PWO effect which increases as the WP of two states separate with time. In fact, this would be translated into WP separation on the physical layer in the sense of Eq.~\eqref{StatDecoh_def} and Fig.~\ref{fig:Decoh}. 
	\begin{figure} 
		\centering
		\includegraphics[width=0.93\textwidth]{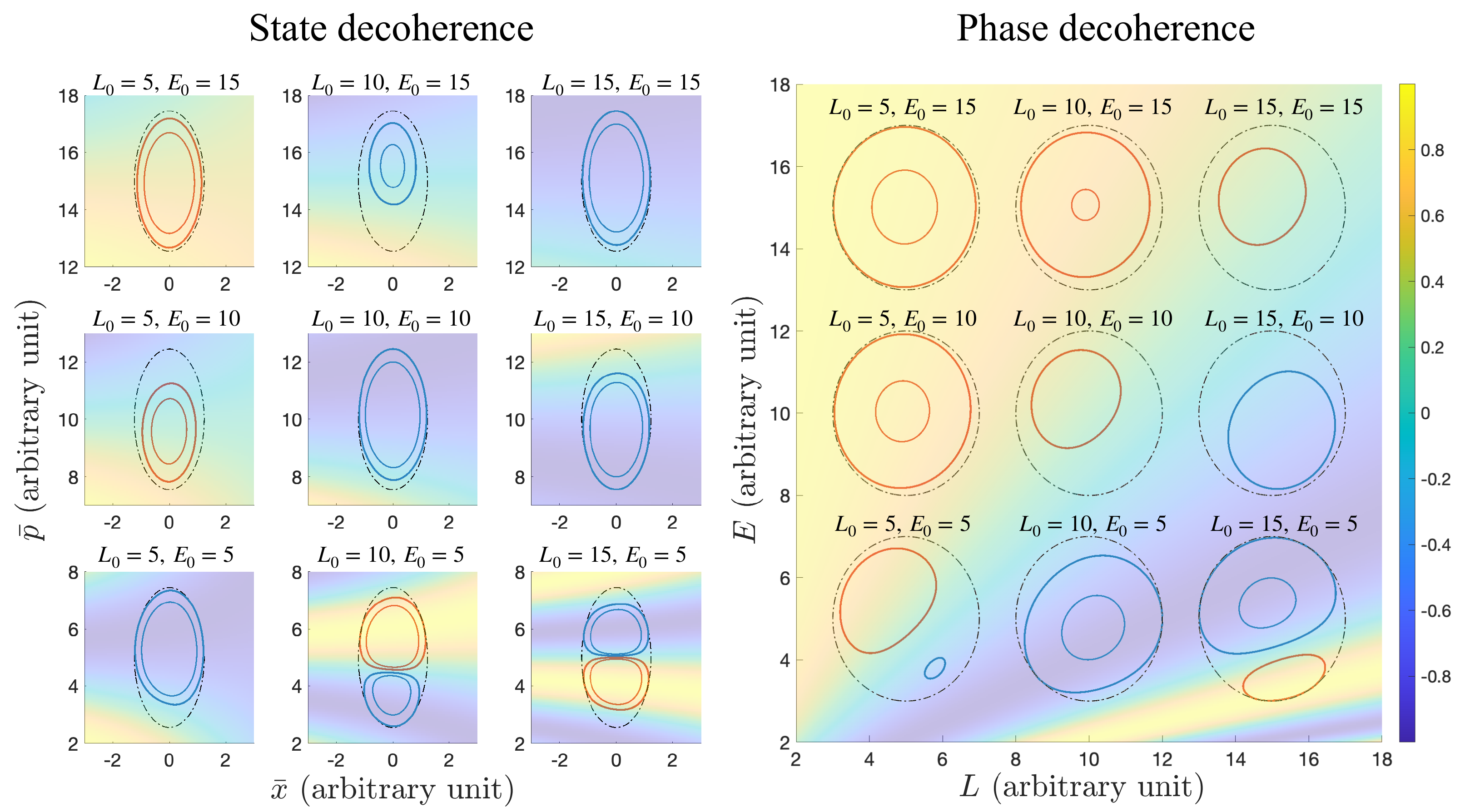}
		\caption{\label{fig:SD_PD} Demonstration of time-independent state decoherence (left plot) and phase decoherence (right plot) plotted in a similar fashion as Fig.~\ref{fig:wigner} and Fig.~\ref{fig:layer2}, both representing phase wash-out effects. The former is in the Wigner phase space on layer 1, while the latter is in the relativistic phase space on layer 2. 
			The coloured contour plot in the background is the oscillating phase structure plotted as $\cos(\eta_{jk})$ (left plot) and $\cos(\psi_{jk})$ (right plot), on the corresponding layer. The outer red (blue) circles are the level for two standard deviations of positive (negative) values of the time-independent $P_{\bar{1},jk}(X_1;X_3)$ (left plot) and $P_{2,jk}(X_2;X_3)$ (right plot) for weighting functions Eq.~\eqref{SD_weight} and Gaussian distributed $H_L/H_E$, respectively. The inner circle (if there is one), is the contour for one standard deviation. Also, the black dashed line is the contour for two standard deviations of the weighting functions.}
	\end{figure}
	On the other hand, if we have no temporal information during the detection process, i.e.\ $\sigma_{H,T} \rightarrow \infty$, then we should integrate out $X_2=T$ in Eq.~\eqref{StatDecoh_def} first, then look at the SD term in the same form of Eqs.~\eqref{StatDecoh_def}-\eqref{SD_factor} but with $X_3=\{L_0,E_0\}$, $X_2=\{L,E\}$ and $\Gamma_{2,jk}(X_2)$ replaced by $\int dT \Gamma_{2,jk}(L,E,T)$. In this case, after the approximation given in Eqs.~\eqref{approx_absP}-\eqref{approx_E} and including terms up to $\mathcal{O}(m^2)$, the Gaussian example in Eq.~\eqref{Gauss}, leads to the left plot in Fig.~\ref{fig:SD_PD}, 
	\begin{equation} \label{SD_phase_L}
	\eta_{jk}(\bx,\bp;L,E)|_{L=L_0,E=E_0}
	=i \frac{\mjk}{2E_0}\left[(\Delta\bar{x}-L_0)\frac{\bar{p}}{E_0}-\Delta\, \eta\, \bar{x}\right],	\end{equation}
	and the time-independent SD weighting function is
	\begin{equation} \label{SD_weight}
	D_{jk}(\bx,\bp;E_0)\simeq 
	\exp\left[ -2\bar{x}^2\sigbp^2
	-\frac{\left(\bar{p}-E_0+(m_j^2+m_k^2)/2E_0\right)^2}{2 \sigbp^2}
	-\left(\frac{\mjk\sqrt{\Delta}\,\sigbx}{2\sqrt{2}E^2}\right)^2\bp^2
	\right].
	\end{equation}
	In the left panel of Fig.~\ref{fig:SD_PD}, where the PWO effect for the above phase structure and weighting functions is shown, we take $\eta = 1$, $\sigbx = 0$ and amplify $\mjk$ to one order smaller than $E_0$ for illustration purpose.   
	Nonetheless, when $\mjk \ll E_0$, the integration over $\bx$ would only have negligible contribution to the $S_{3,jk}$, and the SD term becomes
	\begin{equation}\label{SD_bp}
	S_{3,jk}(L_0,E_0)\simeq
	\exp\left(i\frac{\mjk L_0}{2E_0}\right)
	\frac{\int d\bp \, D_{jk,\sigbp}(\bp;E_0)\exp\left(-i\frac{\mjk L_0}{2E_0^2}\bp\right)}{\int d\bp \, D_{jk,\sigbp}(\bp;E_0)},
	\end{equation}
	where $D_{jk,\sigbp}$ is a product of three Gaussian distributions. Moreover, for $E_0 \gg m_j^2/E_0$, the dominate one would be the distribution with width $\sigbp/2$ and centred at $E_0$, while the phase structure appears to be $\eta_{jk}\simeq-\mjk \, \bp\, L_0 /(2 E_0^2)$.
	This can be seen from the upper row in the left panel of Fig.~\ref{fig:SD_PD}, where the phase averaging mainly results from the integration over $\bp$.
	Hence, the SD is mainly decided by the Wigner distribution w.r.t.\ $\bp$, which we will call $D_{\bp}$-induced decoherence.
	In fact, the resulting SD term, $S_{3,jk}$, from such PWO effect not only agrees with the standard decoherence formula in Eq.~\eqref{P2_standard}, but also shows that such $L_0$ and $E_0$ dependence is a consequence of the phase structure.  
	Moreover, although we demonstrated the case where all weighting functions are Gaussian distributed, the phase structure also applies to arbitrary distributions if the saddle point approximation is adopted in Eq.~\eqref{Wigner}.
	Therefore, as the phase structure implies a Fourier transformation from $\bp$ to $\alpha_{jk}=\mjk L_0 /(2 E_0^2)$, we plot the damping term and the phase shift term from SD for some typical distributions in Fig.~\ref{fig:PWO_sigL}. 
	In particular, while the one Gaussian case represents a typical statistical distribution for a single process, the two-Gaussian case considers neutrinos produced simultaneously by two different processes with slightly different expectation values for momentum. The other two distributions are more suitable for describing an $H_L$-induced PD effect which will be introduced in the next subsection, where more discussions on this plot will be given along with the $H_E$-induced PD effect. 
	Nonetheless, since both decoherence effects are described by Fourier transformation, only with different space mappings, we demonstrate both effects in the same plot. In fact, this shows that different sources of decoherence effect are distinguishable by their $(L_0,E_0)$ dependence, which origins from the difference in their phase structures.

	To sum up, while the SD represents the separation of two mass eigenstates on the physical layer, which is quantified by the overlapping area between them (Fig.~\ref{fig:Decoh}), it is not the case if we move down to the Wigner-PS. From  Fig.~\ref{fig:wigner}, we can see that the width of $P_{\bar{1},jk}$ does not get smaller as the two mass eigenstates, $P_{\bar{1},jj}$ and $P_{\bar{1},kk}$ separate. Nevertheless, the phase density increases as the two mass eigenstates depart from each other, and the PWO effect is stronger, which would give the same result as calculating the amount of overlap of two mass eigenstates on the physical layer.  
	In addition, from the colorful background in Fig.~\ref{fig:SD_PD}, we see that the phase structure would vary with PS variables on the third layer, $T_0, L_0$ and $E_0$, while the width of the overlapping weighting function does not. For the time-independent case on the left plot, where $\sigma_{H,T}\rightarrow \infty$, it is straightforward to see that there is no dependence on $T_0$; yet, for the time-dependent case in Fig.~\ref{fig:wigner}, it turns out that there is no dependence on $L_0$. This is because in the difference $L_j-L_k$ the common factor $L$ cancels, which can be seen from either the phase structure in Eq.~\eqref{Wigner_Gauss}, or directly from Eq.~\eqref{S_jk}.     	
	
	%3-2%%%%%%%%%%%%%%%%%%%%%%%%%%%%%%%%%%%%%%%%%%%%%%%%%%%%%%%%%%%%%
	\subsection{Phase Decoherence} \label{Sec.3-3}
	As we will show in this section, the dominating uncertainties deciding the effect of PD are the macroscopic ones summarized below: 
	
	\begin{itemize}
		\item $\sigma_L$ (coordinate uncertainty on the relativistic-PS $\ni$ layer 2): This uncertainty mainly comes from the macroscopic spatial uncertainty of the full process, which is dominated by the uncertainty on the production profile of the neutrino source for the vacuum propagation case. 
		Unlike $\sigma_x$, $\sigma_L$ does not enter  the Feynman diagram (Fig.~\ref{fig:Feyn}) which calculates the transition amplitude. Instead, it accounts for uncertainty in the traveling distance of the transition probability on the second layer. 
		For instance, $\sigma_L$ could be dominated by the reactor core size ($\sim 3-5$ m) for reactor neutrinos and the distance mesons/muons travel before they decay into neutrinos in accelerator experiments. For Gaussian distributed PDF, the damping term in Eq.~\eqref{damp_Gauss} is to a good approximation
		\begin{equation}\label{gamma_L}
		\gamma_L \propto \sigma_L E_0^{-1},
		\end{equation}
		due to the smallness of mass splitting when we integrate out $E$ in Eq.~\eqref{P3_HL}. In other words, it is nearly independent of the traveling distance. Hence, it is favourable to search for such effects close to the neutrino source for the sake of higher statistics. 
		In Fig.~\ref{fig:PWO_sigL}, we show the effects of PD for different production profiles:
		The one Gaussian PDF can be used for neutrinos produced at rest such as reactor neutrinos and DAR neutrinos. 
		The two Gaussian PDFs are shown for the scenarios with multiple sources/detectors. 
		The box PDF should be adopted when constraints put forth by the experimental setup dominates. For instance, the pipe/rod volume of some accelerator/reactor producing neutrinos, which cuts the production profile to an idealized box shape. 
		Finally, the exponential decaying PDF is for neutrinos produced by decaying particles decaying at flight, for instance, accelerator neutrinos and atmospheric neutrinos.
		However, if the propagation process also contributes to $\sigma_L$, then $\sigma_L$ will accumulate over distance, such as through matter effects \cite{Coloma:2018idr}, or some  exotic effects \cite{Lisi:2000zt,Hooper:2004xr,Gomes:2020muc,Farzan:2008zv}. 
		In this case, we can write $\sigma_L^2 \propto L_0$, to agree with the dependence on the traveling distance in the damping term  calculated by the Lindblad equation, which the literature mentioned above adopts.

		\item $\sigma_E$ (energy uncertainty on the relativistic-PS $\ni$ layer 2): This uncertainty is mainly governed by the energy resolution and reconstruction model of the experiment. 
		Typically, for neutrinos detected with photomultiplier tubes, the energy resolution is given by $\sigma_{E}= \sigma^0_E \sqrt{E_0}$, where $\sigma^0_E$ is usually $ \mathcal{O}(0.1) \sqrt{\text{MeV}}$.  
		Additionally, there will also be different contributions due to the energy reconstruction model, for instance, the degree of quasi-elastic scattering in \cite{Martini:2012uc,DeRomeri:2016qwo}, leading to a tail in $H_E$.
		Moreover, due to the $1/E$ dependence in $\psi_{jk}$, which makes $\sigma_E$ asymmetric w.r.t.\ the phase structure, there will be a phase shift even for a Gaussian distributed PDF, as we have plotted numerically in Fig.~\ref{fig:PWO_sigE}. 
		Then from such numerical results, we can extract an $E_0$ and $L_0$ behavior.
		For example, we can see from Fig.~\ref{fig:P3_spectrum} that the effect of $\sigma_{E}$ also increases with $L_0$ in a comparable way as $\sigbp$.
		
		\begin{figure} [t]
			\centering
			\includegraphics[width=.9\textwidth]{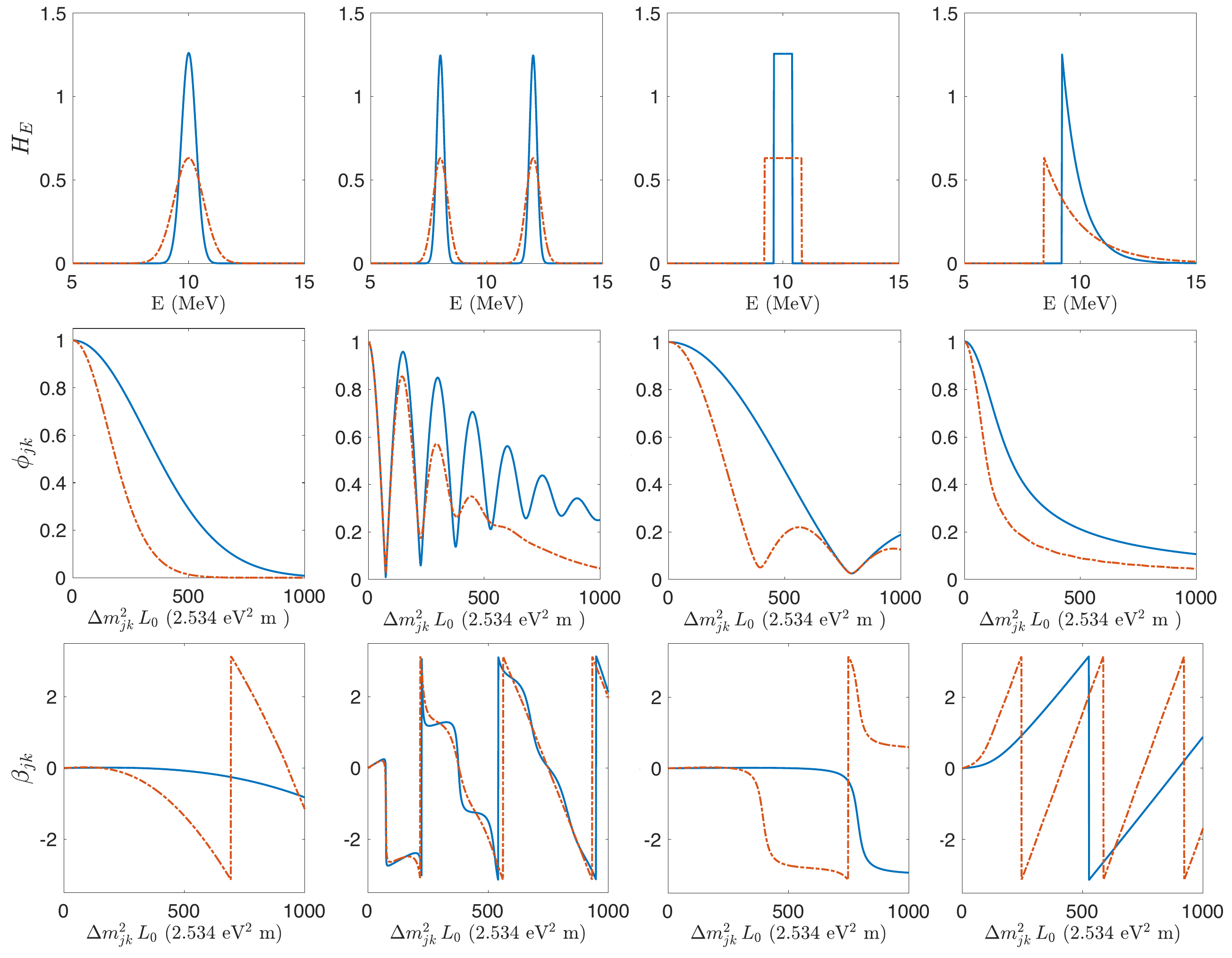}%\hfill
			\caption{\label{fig:PWO_sigE} Phase decoherence by the macroscopic energy uncertainty $\sigma_E$ on the physical layer as a phase wash-out effect. Including a damping term and a phase shift term in Eq.~\eqref{Decoh_para} for different shapes of $H_E$ with the same widths ($\sigma_E / \sqrt{E_0}= 0.1 \sqrt{\rm MeV}$ for the blue line, $\sigma_E /\sqrt{E_0}= 0.2 \sqrt{\rm MeV}$ for the red line), at energy $E_0 = 10$ MeV.}
		\end{figure}
	\end{itemize}

	Similar to state decoherence, phase decoherence also describes a PWO effect, which can be seen directly from its definition:
	\begin{equation} \label{PhaseDecoh_def}
	\Phi_{3,jk}(X_3) = e^{-i\psi_{jk}(X_3)}\,\int dX_2 Y'_{2,jk}(X_2;X_3)e^{\psi_{jk}(X_2)}.
	\end{equation}
	Here  
	\begin{equation}
	Y'_{2,jk}(X_2;X_3)=\frac{S_{2,jk}(X_2)\Phi_{2,jk}(X_2)H(X_2;X_3)}
	{\int d X_2 S_{2,jk}(X_2)\Phi_{2,jk}(X_2)H(X_2;X_3)}
	\end{equation}
	is a real and normalized PDF, which is dominated by the second layer macroscopic weighting function $H_{X_2}$ for $X_2=L,E$. When $\sigma_{H,T} \ll \sigma_{S,T},\sigma_{\Phi,T}$, $Y'_{2,jk}$ is completely dominated by the macroscopic weighting functions,
	in this case the phase structure will be 
	\begin{equation}
	\psi'_{jk}(T,\vL,E)=-i(E_j-E_k)T+i(\vP_j-\vP_k)\vL. 
	\end{equation}
	On the other hand, if we have no temporal information, we should again integrate over $T$ before looking at the decoherence effect. The phase structure up to $\mathcal{O}(m^2)$ for the time-dependent scenario is Eq.~\eqref{psi_T}, while the one for the time-independent one is Eq.~\eqref{psi_L}.

	From Eq.~\eqref{P3_HL} we can see that after $L$ is integrated out, the factorization condition in Appendix \ref{Sec.AppC} is satisfied, such that we can factorize Eq.~\eqref{PhaseDecoh_def} as
	\begin{equation} \label{PD_factor}
	\Phi_{3,jk}(L_0,E_0) \simeq e^{-i 2 \mjk \frac{ L_0}{E_0}}
	\int dL H_L(L;L_0) e^{i\mjk \frac{L}{E_0}}
	\int dE H_E(E;E_0) e^{i\mjk \frac{L_0}{E}}
	\end{equation} 
	for the time-independent case, where both $H_L(L;L_0)$ and $H_E(E;E_0)$ are  normalized PDFs. 
	Therefore, we can treat the (macroscopic) coordinate and energy uncertainties on the second layer as separate PWO effects, 
	naming the former $H_L$-induced PD, and the latter $H_E$-induced PD. In Fig.~\ref{fig:PWO_sigL} and Fig.~\ref{fig:PWO_sigE} these two PWO effects are demonstrated, respectively, where the weighting functions, $H_L$ and $H_E$ are taken as PDFs with the same width for the same colored lines. In other words, $\int d X_2 H_{X_2} = 1$ and $\max\{H_{X_2}(X_2)\}=1/(\sigma_{X_2} \sqrt{2 \pi})$, according to our definition of ``width" in Appendix \ref{Sec.AppA}.
	Therefore, the only parameter for the two figures presenting the PWO effect on the second layer is the width $\sigma_{X_2}$, for some $L_0$ or $E_0$, which we take $\sigma_L = 1$ m for the blue and yellow line, $\sigma_L = 2$ m for the red line in Fig.~\ref{fig:PWO_sigL}, at distance $L_0 = 10$ m; also, $\sigma_E / \sqrt{E_0}= 0.1 \sqrt{\rm MeV}$ for the blue line, $\sigma_E /\sqrt{E_0}= 0.2 \sqrt{\rm MeV}$ for the red line in Fig.~\ref{fig:PWO_sigE}, at energy $E_0 = 10$ MeV.  	
	In particular, analogously to the $D_{\bp}$-induced SD, the $H_L$-induced PD takes the form of  a Fourier transformation from $L$ to $\alpha_{jk}=\mjk/(2E_0)$, as plotted in Fig.~\ref{fig:PWO_sigL}. For instance, the Gaussian PDF transforms into a Gaussian distribution in the $\mjk/E_0$ space, the box PDF  transforms into a sinc function, and the PDF for  exponential decay (for neutrinos produced by decaying charged leptons) is transformed into a Lorentzian function. 
	As for the phase shift term, from Appendix \ref{Sec.AppC} we know that only the asymmetric functions (the yellow-lined two Gaussian PDF and the exponential decaying PDF) have non-zero and non-$\pi$ phase shift. 
	For the symmetric ones, while there is no phase shift for a single Gaussian PDF, the phase ranges would jump from 0 to $\pi$ (still no imaginary part) for the box PDF and the symmetric two-Gaussian PDF, due to negative values of the function from the  Fourier transformation.
	However, owing to the smallness of neutrino mass splitting, $\alpha_{jk}$ is typically small for both $D_{\bp}$-induced and $H_L$-induced decoherence effects. Hence, the range of interest would lie in a small range around $\alpha_{jk}\rightarrow 0$, where the phase is zero for asymmetric cases.
	On the other hand, regardless of how ``symmetric" the weighting PDF spectrum is, it is not symmetric w.r.t.\ $1/E$. Therefore, there is always a non-trivial $H_E$-induced phase shift, even for the Gaussian distributed PDF.   
	Moreover, it is also clear that the larger the width of the PDF is, the more significant the decoherence effect (both the damping term and the phase shift term) becomes. 
	In the end, if we find some phase structure dependence as the damping term and/or the phase shift term in Fig.~\ref{fig:PWO_sigL} and Fig.~\ref{fig:PWO_sigE}, we should be able to reconstruct the production profile and cross-check the energy reconstruction model.

	Comparing PD with SD, both come from PWO effect, but with different phase structures ($\eta'_{jk}/\psi'_{jk} $ for time-dependent SD/PD and $\eta_{jk}/\psi_{jk}$ for time-dependent SD/PD) and distributions ($D'_{jk}\, / \, H_TH_LH_E$ for time-dependent SD/PD and $D_{jk}\simeq D_{\bp} \, / \, H_L H_E$ for time-dependent SD/PD).
	In particular, for SD both the phase structure and the distribution varies with $L_0,E_0$, while in the PD case, only the mean of the distribution does. Such structure is illustrated in the right plot of Fig.~\ref{fig:SD_PD}, where the phase structure in the background only depends on the second layer PS variables, $L$ and $E$, but not on the third layer PS variables, $L_0$ and $E_0$. 
	This also implies that the second and third layer share the same coherent phase structure, but not with the first layer. The third layer PS only decides where the weighting functions are centred, and when it appears at a higher phase density area (lower energy or larger distance), there will be a stronger PWO effect.
	Illustrating the PWO effect for the time-independent case, Fig.~\ref{fig:SD_PD} is plotted in the same ways as Fig.~\ref{fig:wigner} and Fig.~\ref{fig:layer2}, while also showing the phase structures on the Wigner-PS and the relativistic-PS, respectively, in the colourful background. 
	The final observational effects on the measurement layer of SD, $H_L$-induced PD and $H_E$-induced PD are further plotted in Fig.~\ref{fig:PWO_sigL} and Fig.~\ref{fig:PWO_sigE}.

	The advantage of putting SD and PD effect in terms of PWO effect is that one can estimate the decoherence from $S_{3,jk}$ and $\Phi_{3,jk}$ numerically, and the only input would be the weighting functions. Therefore, the steps to estimate the SD/PD effect numerically are simply: 1) Determine the weighting functions on the Wigner-PS and relativistic-PS; 2) Multiply it with the corresponding phase structure derived in this section;  3) Integrate out the respective PS. Such integration are certain to converge, since the weighting functions are localised and distributed around the next level PS variables. For instance, even for the simplest phase structure -- the time-independent case on layer 2 -- the PD terms can only be evaluated numerically as we did for Fig.~\ref{fig:PWO_sigE}.
	Therefore, in principle, by analysing the waveform and spectrum of neutrino oscillation, we should be able to reconstruct the weighting functions once we identify its corresponding phase structure.

	%4%%%%%%%%%%%%%%%%%%%%%%%%%%%%%%%%%%%%%%%%%%%%%%%%%%%%%%%%%%%%%%
	\section{Phenomenology of Neutrino Decoherence} \label{sec.4}
	From the previous section, we saw that neutrino decoherence effects can be classified into SD and PD, which are both a consequence of the PWO effect. Therefore, both would result in damping terms and phase shift terms, where the latter is nontrivial only when the weighting function is not symmetric w.r.t.\ the phase structure. 
	In particular, SD is dominated by the (microscopic) uncertainties on the first layer ($\sigma_x$ and $\sigma_p$), while PD is governed by the (macroscopic) uncertainties ($\sigma_T$, $\sigma_L$ and $\sigma_E$) on the second layer. 
	Nonetheless, we can only observe $\sigbp$ for the first layer uncertainties since $\sigbx \ll \sigma_L$. 
	In addition, we only consider the time-independent case in this section, since current experiments shown in Fig.~\ref{fig:NFS_where} continuously emit neutrinos for a sufficient long period of time, therefore, $\sigma_T \rightarrow \infty$. 
	Additionally, we also parameterize the first layer uncertainties as $\sigbx$ and $\sigbp$, then only $\sigbp$ would be an valid observational parameter, while $\sigbx$ would be eaten by $\sigma_L$.   
	In this section we estimate how far we are experimentally from having a 90\% CL sensitivity for observing damping and/or phase shifting signatures in Eq.~\eqref{Decoh_para} from the $D_{\bp}$-induced SD, $H_L$-induced PD and $H_E$-induced PD. 
	
	For the damping signatures, we consider all weighting functions distributed as single Gaussians. 
	Hence the damping term is then parameterised by $\sigbp$, $\sigma_L$ and $\sigma_E$ in 
	\begin{equation}
	\phi_{jk}= \exp \left[
	-\left( \frac{\mjk \sigbp L_0}{2 \sqrt{2}E_0^{2}}\right)^2
	-\left( \frac{\mjk \sigma_L}{2 E_0}\right)^2
	-\left( \mjk \gamma_E(L_0,E_0;\sigma_E)\right)^2
	\right],
	\end{equation}  
	where the first two parts are taken from Eq.~\eqref{gamma_p} and Eq.~\eqref{gamma_L}, and the latter can only be found numerically as we showed in Fig.~\ref{fig:PWO_sigE}. 
	Furthermore, we estimate the sensitivity towards the three uncertainty parameters through a $\chi^2$-analysis for the traditional rate measuring method (RMM) to look for unexpected disappearance or appearance signals caused by SD and/or PD. 
	Moreover, since neutrino decoherence is more enhanced at low energy, from Fig.~\ref{fig:NFS_where} we find that for current experiments, reactor neutrinos should have the best sensitivity. 
	Hence, in the section below, as a benchmark experiment to estimate how far we are from detecting neutrino decoherence through the damping term, we choose the RENO experiment. This is because it has less uncertainty compared to the Double Chooz experiment and a less complicated structure for the distribution of reactors and detectors compared with the Daya Bay experiment, which would affect the damping signature through $\sigma_L$ in a non-trivial way, see Appendix \ref{Sec.AppD}. 
	 In addition, different dependence on $L_0$ and $E_0$ in the damping signatures in neutrino oscillation is discussed in e.g.\ \cite{Blennow:2005yk,JUNO:2021ydg}.
	
	The RMM, on the other hand, is a lot less sensitive to the phase shift terms, since a shift in the phase would barely vary the shape of the neutrino spectrum, as we will further illustrate in Fig.~\ref{fig:PA_sens}.
	Hence, for the phase shifting signatures, we introduce a method to measure the oscillation phase directly, which could be done by moving the detector around an expected oscillation minimum, namely the phase measuring method (PMM).
	Such method not only has the purpose of measuring asymmetries in the weighting functions, it is also a cleaner way to measure neutrino oscillation signatures as we will further discuss in Sec.~\ref{Sec.4.2}.      
	As for the theory input, only the $H_E$-induced PD will have a non-trivial contribution for Gaussian distributed weighting functions. We thus consider a two-Gaussian distributed $D_{\bp}$ to introduce an asymmetry for the quantum uncertainties, and ignore $H_L$-induced PD, since it is has only negligible contribution for ground-based experiments.
	At last, after introducing the PMM and evaluating theoretic inputs and parametrizations (an asymmetry parameter ``$a$" for the quantum uncertainties and the width $\sigma_E$ for the energy uncertainty), we estimate the statistically and systematic uncertainties required for having a 90\% CL sensitivity in the parameter space, considering a $\pi$DAR neutrino source.     
	
	%%%%%%%%%%%%%%%%%%%%%%%%%%%%%%%%%%%%%%%%%%%%%%%%%%%%%%%%%%%%%%%%%
	\subsection{Rate Measuring Method} \label{Sec.4.1}
	
	In order to observe the damping term $\phi_{jk}$ which, in general, results from any source of decoherence, we analyze the total count rate of neutrinos for ground-based neutrino experiments. 
	This is done by fitting our theory for SD through $\sigbp$, as well as PD through $\sigma_L$ and $\sigma_E$ to oscillation data, defining the $\chi^2$-function:
	\begin{equation}\label{NFS_Chi}
	\chi^2(\sigma_n) = \min\limits_{\vec{\alpha}} \sum_{i\,\rm{bins}}
	\frac{(R_i(\sigma_n,\vec{\alpha})-R_i^{\rm data})^2}{U_i}
	+\sum_{j} \left( \frac{b_j-b_j^0}{\sigma_j}\right)^2.
	\end{equation} 
	Here $R_i$ could either be the observed rate or the ratio of the detected rate of the near and far detectors; $b_j$ are the pull parameters, which include the oscillation parameters and the experimental uncertainties of the rate, and $U_i$ represents the statistical uncertainty of the bin. 
	
\begin{figure} [h]
		\centering
		\includegraphics[width=.59\textwidth]{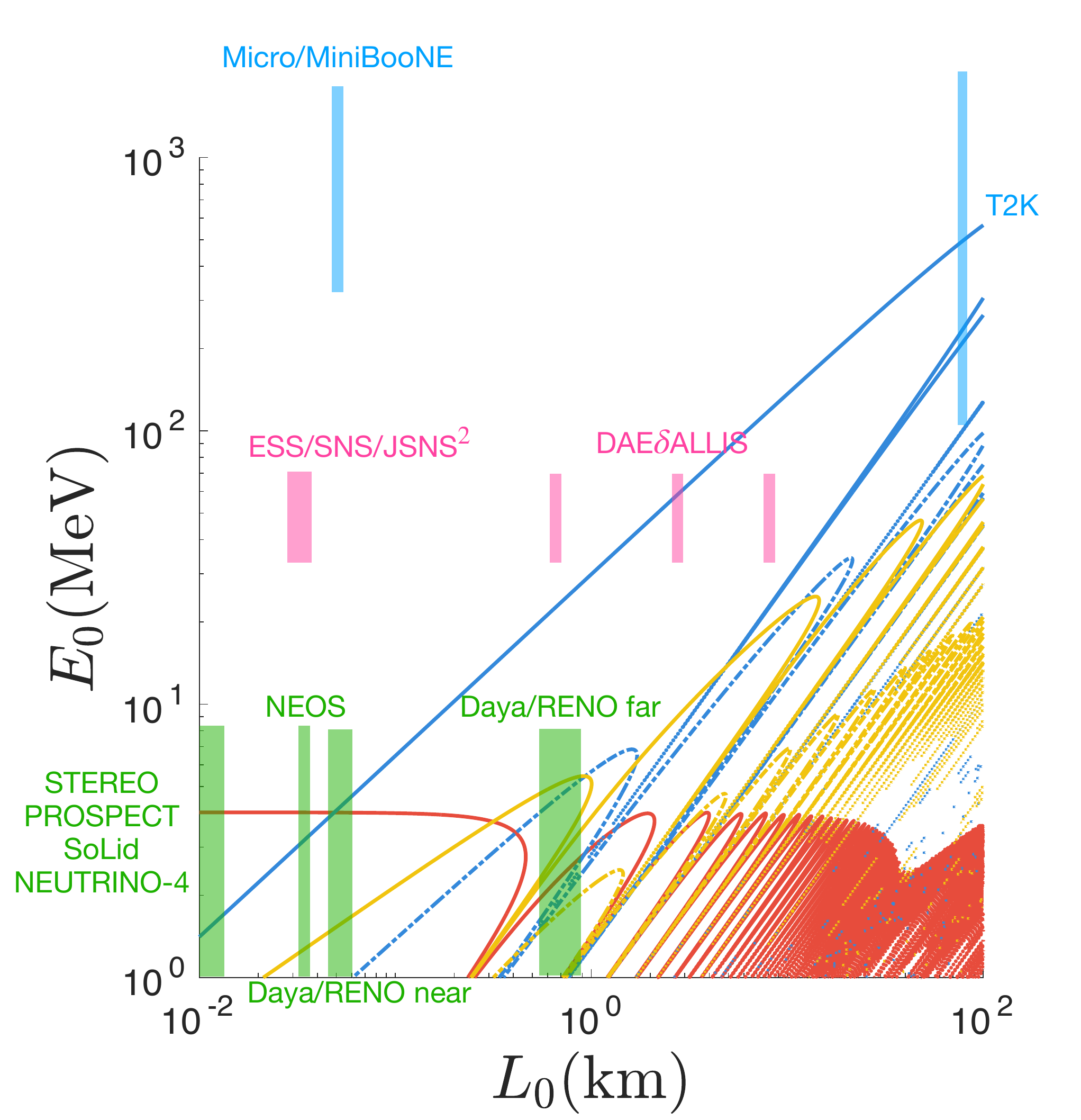}
		\caption{\label{fig:NFS_where}
			This figure is plotted to give us an idea on where to find neutrino decoherence effect among current ground-based experiments.
			The red, blue and yellow lines represent contours of $|P(\sigma_n \neq 0)-P(\sigma_n = 0)|=10^{-6}$ for the solid lines and $10^{-4}$ for the dashed lines, for $n=L,E$ and $\bar{p}$, respectively, where the sensitivity would be higher below the lines. We set $\sigma_L = 3$ m $\sigma_E = 0.1 \sqrt{E_0}$ MeV and $\sigbp = 0.1$ MeV as an example. Additionally, experiments (see \cite{ParticleDataGroup:2020ssz} for a review) with their corresponding baseline  and typical neutrino energies are labeled on the plot, for accelerator neutrinos (blue), decay-at-rest neutrinos (pink) and reactor neutrinos (green).}
	\end{figure}

\begin{figure} 
		\centering
		\includegraphics[width=.58\textwidth]{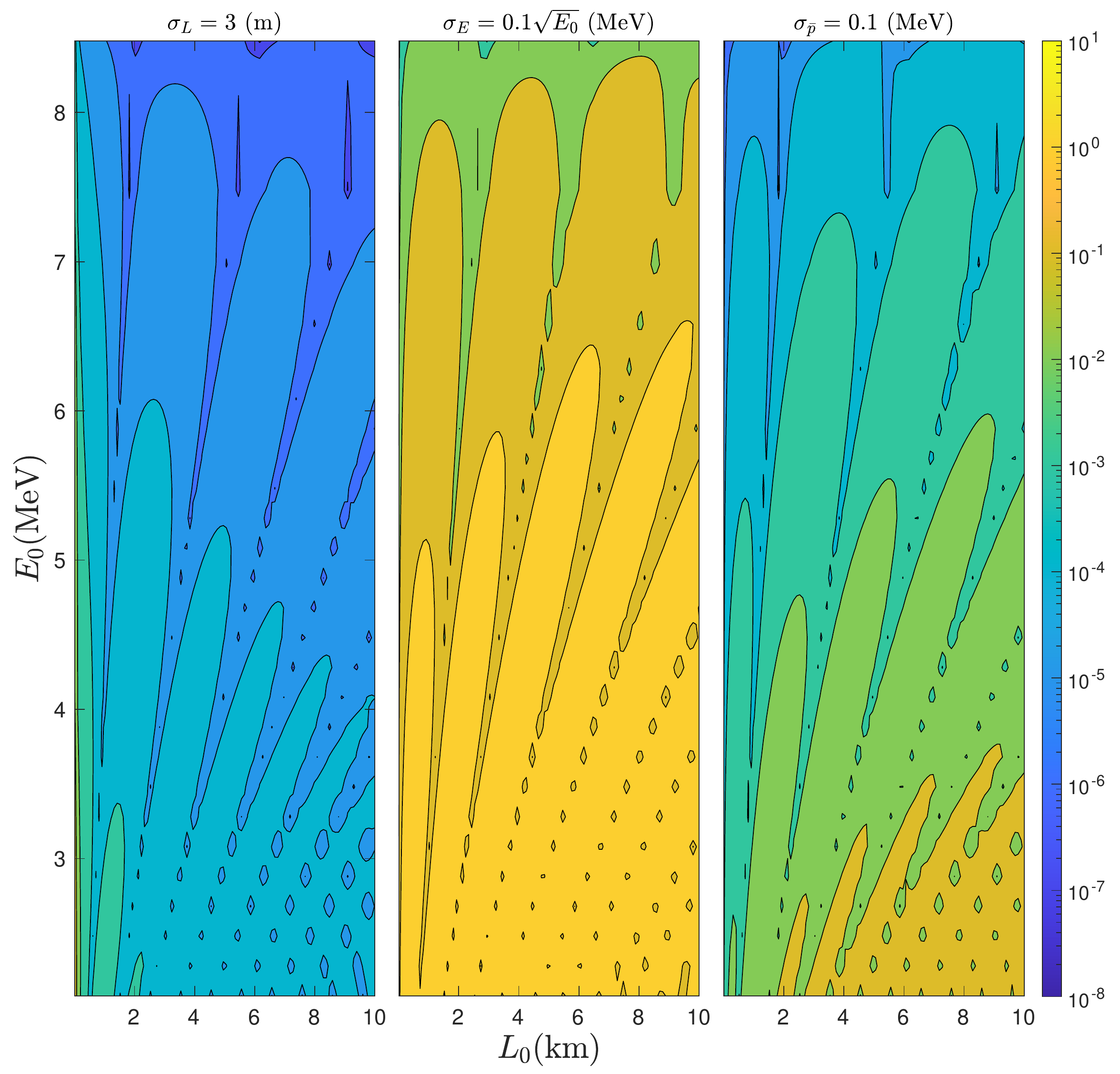}\hfill
		\caption{\label{fig:reactor_where} The contour plot of the flux difference, $|\Phi(\sigma_n \neq 0)-\Phi(\sigma_n = 0)|$ for reactor neutrinos, where the energy spectrum and the decrease with traveling distance for reactor neutrinos are taken into account. The formalism is given in Eq.~\eqref{flux}. }
	\end{figure}

	With the purpose of determining what experiments are more sensitive for each decoherence parameter, we plot Fig.~\ref{fig:NFS_where}, in which the red, blue and yellow lines represent contours of $|P(\sigma_n \neq 0)-P(\sigma_n = 0)|=10^{-6}$ for the solid lines and $10^{-4}$ for the dashed lines, for $n=L,E$ and $\bar{p}$ respectively, giving us a hint of what experiments to look at for a certain $\sigma_n$. For simplicity, $P$ represents the FTP on the third layer for $\bar{\nu}_e \rightarrow \bar{\nu}_e$ in this section. 
	We can see from the figure that for all decoherence effects, the influence would be larger at lower energies, because the oscillation structure is denser along the $\bp$ and $E$ axes, so the PWO effect is enhanced. Therefore, reactor neutrinos having the lowest energies for ground-based experiments would be the best candidate. Additionally, for vacuum oscillation, the decoherence effect by $\sigma_L$ is small and does not depend on $L_0$, hence, it is suitable for experiments near the source where the statistics are high, whereas $\sigbp$ and $\sigma_E$ are more pronounced at larger distance.   
	Moreover, a more realistic version of Fig.~\ref{fig:NFS_where} for reactor neutrinos is plotted in Fig.~\ref{fig:reactor_where} for the each $\sigma_n$, where the contour lines represent the rate difference between decoherent and coherent fluxes, $|\Phi(\sigma_n \neq 0)-\Phi(\sigma_n = 0)|$, considering the energy spectrum of neutrinos for RENO and also the diffusion over distance by
	\begin{equation} \label{flux}
	\Phi(\sigma_n;E_0,L_0) = \sqrt{N(E_0)}\,\,
	\left|P(\sigma_n \neq 0)-P(\sigma_n = 0)\right| \,
	\frac{L_{\rm bm}}{L_0},
	\end{equation}
	where $L_{\rm bm}$/$N(E_0)$ is the average distance/spectrum of the near detector for the RENO experiment. 
	
\begin{figure} 
		\includegraphics[width=.5\textwidth]{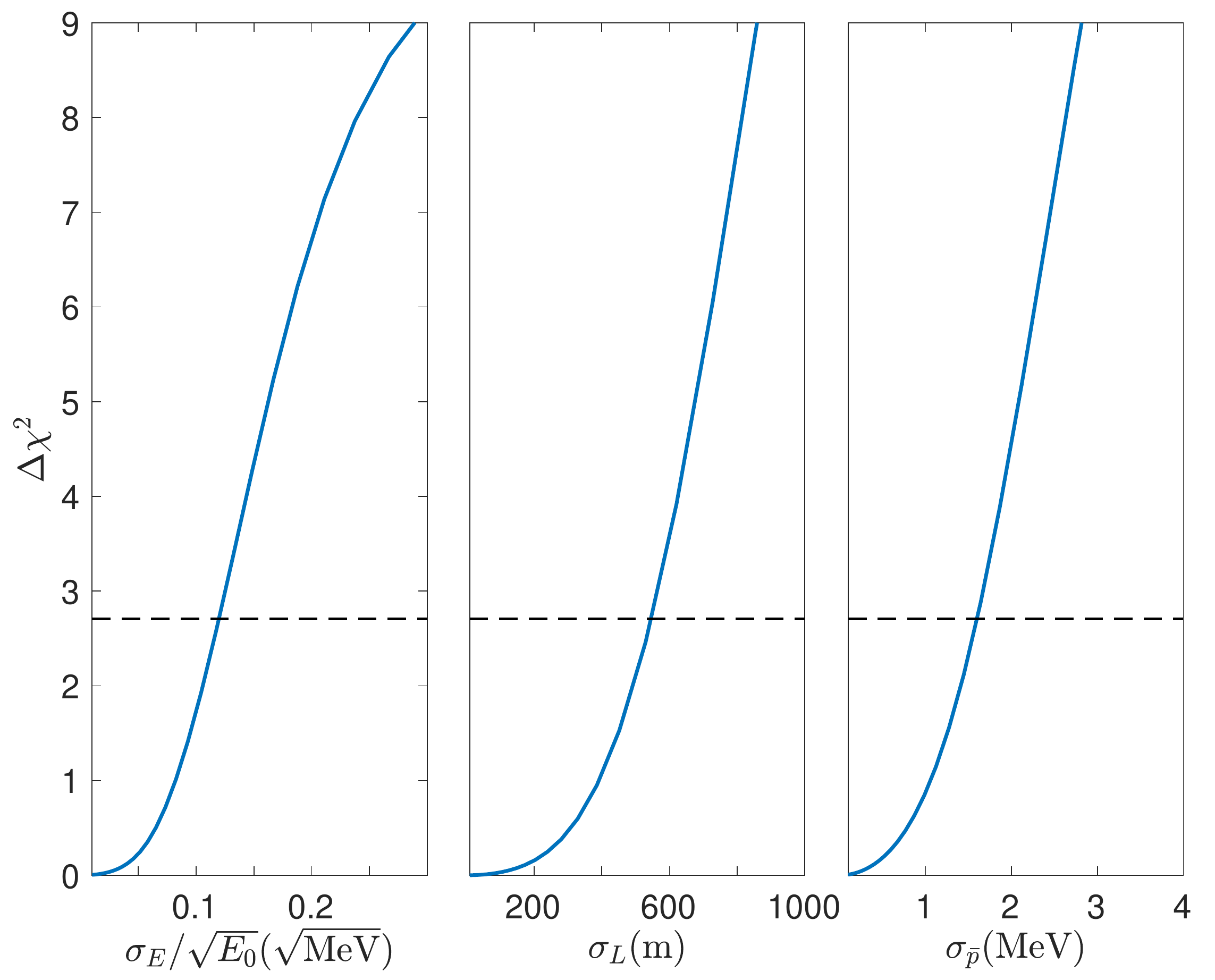}\hfill
		\includegraphics[width=.5\textwidth]{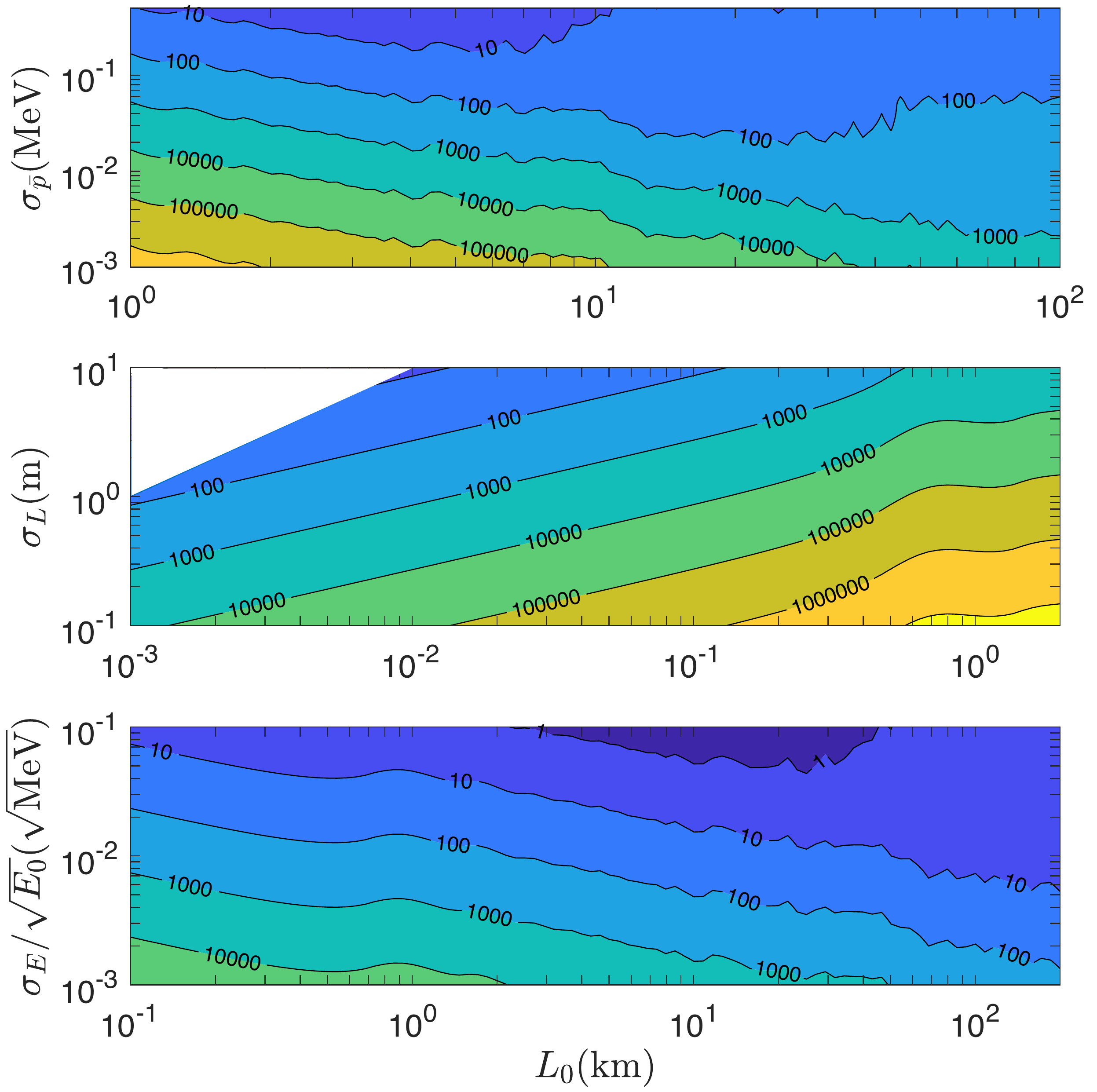}
		\caption{\label{fig:RENO} Taking RENO as a benchmark experiment for reactor neutrinos, the left three plots present the constraints on different decoherence parameters from a fit to RENO data.
			The right plot takes the obtained 90\% CL limit on $\sigma_n$ from the left plots. 
			It shows the contour lines for evaluating how-many-times statistics compared to the current RENO data are needed to achieve a 90\% CL sensitivity for some decoherence parameter and baseline, which is $\lambda$ in Eq.~\eqref{NFS_Chi_bm}. The while area in the middle figure is cut out since it would indicate that the detector is inside the reactor core.}
	\end{figure}

	We take the RENO experiment as a benchmark, and see how much more statistic is needed to have enough sensitivity for each of the three parameters. 
	The sensitivity we get from fitting the decoherence parameters with current RENO far-to-near ratio data in \cite{RENO:2018dro} by Eq.~\eqref{NFS_Chi} is given in three left plots of Fig.~\ref{fig:RENO}. Corresponding with the formalism in \cite{RENO:2018dro,RENO:2016,RENO:2010vlj} to Eq.~\eqref{NFS_Chi}, $R_i^{\rm data}=O_i^{\rm{F/N}}$, is the observed far-to-near ratio of IBD candidates in the $i$-th energy bin after background subtraction, taken from the supplemental material in \cite{RENO:2018dro}; and the theoretical input is  
	
	\begin{equation} \label{eq:T}
	R_i(\sigma_n,\vec{\alpha})=
	(1+\epsilon+f)\,
	\frac{1+b^{\rm{F}}}{1+b^{\rm{N}}} \, N^{\rm MC}_i(\tau)
	\frac{\sum_{l=1}^6P_i(L_{0,l}^{\rm{far}},\sin^2(2\theta_{13}),\Delta m_{ee}^2 ;\sigma_n)}{\sum_{l=1}^6P_i(L_{0,l}^{\rm{near}},\sin^2(2\theta_{13}),\Delta m_{ee}^2;\sigma_n)},
	\end{equation}
	where $f, \epsilon, \tau, b^F,b^N $ are the pull terms for the systematic uncertainties, namely the uncorrelated reactor-flux systematic uncertainty, the uncorrelated detection, the timing veto systematic uncertainty, and the background uncertainties for near and far detectors, respectively, given in \cite{RENO:2016}. 
	Also, $\tau$, the uncorrelated energy-scale systematic uncertainty, is inserted by scaling the energy, i.e.\ $E_0\rightarrow (1+\tau)E_0$, and $N^{\rm MC}_i$ is the near-to-far ratio without oscillation traced back from the Monte Carlo data given in \cite{RENO:2018dro}; 
	Furthermore, $L^{\rm near}_l$/$L^{\rm far}_l$ is the distance to the $l$-th near/far detector, and $\sin^2(2\theta_{13})$, $\Delta m^2_{ee}=\cos^{2} \theta_{12}\Delta m_{31}^2+\sin^{2} \theta_{12}\Delta m_{32}^2$ are the coherent oscillation parameters. 
	Finally, we plot the three left plots in Fig.~\ref{fig:RENO} by minimising the $\chi^2$-function with the Python package ``$i$minuit" over $\vec{\alpha} = (\sin^2(2\theta_{13}), \Delta m^2_{ee}, f, \epsilon, \tau, b^F,b^N )$ and $\vec{\beta}=(\vec{\alpha},\sigma_n)$ for  
	\begin{equation}
	\Delta \chi^2 (\sigma_n) = \min\limits_{\vec{\alpha}}\chi^2 (\sigma_n) - \min\limits_{\vec{\beta}}\chi^2.
	\end{equation}
	The oscillation parameters are effectively marginalized by first minimizing over all the parameter, $\vec{\beta}$, to get the best-fit values $s_0=0.087$, $m_0=2.66 \times 10^{-3}$ eV$^2$ with errors $\sigma_s=0.023$, $\sigma_m = 0.12 \times 10^{-3}$ eV$^2$ for $\sin^2(2\theta_{13})$ and $\Delta m_{ee}^2$, respectively. Then we add two more pull terms  to the $\chi^2$-function,
	\begin{equation} \label{oscPull}
	\chi^2 \rightarrow \chi^2 
	+ \left(\frac{\sin^2(2\theta_{13})-s_0}{\sigma_s}\right)^2
	+ \left(\frac{\Delta m_{ee}^2-m_0}{\sigma_m}\right)^2.
	\end{equation}    
	The fitting results are shown in the three left plots of Fig.~\ref{fig:RENO},
	from which we take the 90\% CL limit as our benchmark points. They read $\sigma_{E}^{\rm bm}=0.12 \sqrt{E_0}$ MeV, $\sigma_{L}^{\rm bm}=548$ m, and $\sigma_p^{\rm bm}=1.6$ MeV. In particular, $\sigbp^{\rm bm}=1.6$ MeV is consistent with the analysis for the RENO experiment in \cite{deGouvea:2020hfl,deGouvea:2021uvg}; 
	it is obvious that $\sigma_{L}^{\rm bm}$ is still far from being a realistic value, which should be a few meters; nonetheless, $\sigma_{E}^{\rm bm}$ seems to be close to the energy resolution which is $ \sigma_{E}/\sqrt{E_0}=0.08 \sqrt{E_0(\rm MeV)+0.3}$ \cite{RENO:2016}.
	
	Furthermore, in order to estimate how many times more statistics we need to reach the required sensitivity for a reasonable $\sigma_n$, we assume that the statistical uncertainty will be increased with some value $\sqrt{N_i}$ for each energy bin. 
	On top of that, if uncertainties of the pull parameters are small enough and the signal count is much larger than the background count, we may consider only having the statistical uncertainties left in $\chi^2$. In this case 
	
	\begin{equation} \label{NFS_Chi_bm}
	\begin{split}
	&\Delta \chi^2(\sigma_n) \simeq  \sum_{i\,\rm{bins}}
	N_i \,|P_i(\sigma_n)-P_i(\sigma_n=0)|^2 \\
	&=\sum_{i\,\rm{bins}}
	\lambda N_i^{\rm{RENO}} \,
	\frac{L_{\rm RENO}^2}{L_0^2}
	|P_i(\sigma_n^{\rm{bm}})-P_i(\sigma_n=0)|^2
	\frac{|P_i(\sigma_n)-P_i(\sigma_n=0)|^2}
	{|P_i(\sigma_n^{\rm{bm}})-P_i(\sigma_n=0)|^2}.
	\end{split} 
	\end{equation} 
	Therefore, $\lambda$ represents the enhancement of statistics needed to have a 90\% CL signal for some value $\sigma_n$.
	Moreover, for $\lambda = 1$, Eq.~\eqref{NFS_Chi_bm} means to sum over all the energy bins in Fig.~\ref{fig:reactor_where} for some $\sigma_n^{\rm bm}$.
	Hence, with the benchmark values we get from the three left plots of Fig.~\ref{fig:RENO}, and the flux difference from Fig.~\ref{fig:reactor_where} inserted into Eq.~\eqref{NFS_Chi_bm}, we arrive at the right plot in Fig.~\ref{fig:RENO}, where we see how much more statistics we need ($\lambda$ in Eq.~\eqref{NFS_Chi_bm}) compared to RENO to gain a 
	90\% CL sensitivity. 
	This increment ($\lambda$) could be achieved by lowering the energy threshold for neutrino detection, increasing the reactor power or simply waiting for more data to be collected over time.
	Unsurprisingly, we are more sensitive to $\sigma_E$ and $\sigbp$ at larger distances, despite that the statistics drops by $1/L_0^2$, while $\sigma_L$ prefers a shorter propagation distance.
	Hence, the ranges of $L_0$ are chosen accordingly. 
	Furthermore, since $L_0$ cannot not be smaller then $\sigma_L$, we leave the triangular area on the upper left blank, and the edge of that area means that the detection is exactly beside the source.

	\subsection{Phase Measuring Method}\label{Sec.4.2}

	Decoherence effects not only include damping signatures, but would also cause a shift to the coherent phase $\psi_{jk}$ on the measurement layer by $\beta_{jk}(\vec{\sigma})$ defined in Eq.~\eqref{Decoh_para}, when there is an asymmetry in the weighting function.
	In this section, we propose an illustration of a, in principle realistic, method which is extensively more sensitive to the phase shift terms compared to the damping terms in contrary to the RMM, by measuring where an oscillation extremum occurs on the third layer.  
	Specifically, for some neutrino energy $E_0$, the goal is to search for the deviation in distance (\footnote{We will shortly explain why the local minimum would be more suitable for this method, therefore, we write $L_{\rm min}$ as the extremum for now.}$L_{\rm min}$-$L_{\rm osc}$) cased by the phase shift. For the case of two neutrino mixing, we measure $L_{\rm min}=L_{\rm min}^{jk}$ in  
	\begin{equation}\label{phase_shift}
	\psi_{jk}(L_{\rm osc}^{jk},E_0)=2n\pi \rightarrow 
	\psi_{jk}(L_{\rm min}^{jk},E_0) + \beta_{jk}(L_{\rm min}^{jk},E_0;\vec{\sigma}_n) = 2n\pi, 
	\end{equation}
	and see how it differs from $L_{\rm osc}^{jk}=4\pi nE_0/\mjk$ due to an asymmetry in the decoherence effect.
	Moreover, we can search for where an extremum occurs, and the signal is thus concentrated at a single $L_{\rm min}$, instead of having an entire distribution.
	Henceforward, an advantage of such method is that since we search for an extremum, the non-oscillating part of the event rate would only influence the signal trivially. Therefore, as long as other factors, such as the production rate, detection rate, background, etc.\ do not have an extremum within the range in ($L_0$, $E_0$) of interest, it would barely contribute to the signal.
	
	At the first oscillation minimum in a two-neutrino oscillation case, we look for
	\begin{equation}\label{DeltaL}
	\Delta L_{\rm min}^{jk} (E_0;\vec{\sigma}_n)=  L_{\rm min}^{jk}(E_0;\vec{\sigma}_n) 
	-L_{\rm osc}^{jk}(E_0)
	\simeq -\frac{2 E_{0}}{\mjk}\beta_{jk}(E_0, \vec{\sigma}_n).
	\end{equation} 
	The approximation above is done by taking  $\beta(L_{\rm min}^{jk},E_0;\vec{\sigma}_n)\simeq \beta(L_{\rm osc}^{jk},E_0;\vec{\sigma}_n)$ as we have checked that higher orders in the expansion of the left-hand side around the right-hand side can be neglected.
	In fact, due to the large difference between the atmospheric and solar mass splitting, when we search around $L_{\rm osc}^{13}$, the total $L_{\rm min} $ is $ L_{\rm min}^{13}$ for the three neutrino mixing paradigm.  
	Therefore, complications arising from the interference between different mass splittings, such as that from the damping terms, can also be negligible.
	Nonetheless, we still consider a full three neutrino-mixing scenario in our simulation below, and find $L_{\rm min}$ numerically even for the fully coherent case. 
	Additionally, in the same fashion, it is also possible to find a certain $E_{\rm min}$ for some $L_0$ according to Eq.~\eqref{phase_shift}.
	However, it is usually not possible to look for effects by $\sigma_E$ through $\Delta E_{\rm min}$ or $\sigma_L$ through $\Delta L_{\rm min}$, since the former is usually much larger than the latter. Hence, while the uncertainties give rise to a phase shift, it is likely to lower the sensitivity of the corresponding variable even more.
	Also, since $\sigma_L$ is too small for ground-based neutrino sources even in terms of $\Delta E_{\rm min}$, we do not consider it in this section. 
	For the purpose of this paper, we focus on the phase shift caused by the decoherehnce effect. Nonetheless, the PMM simply measures the (effective) neutrino oscillation phase exclusively. 
	Hence, this method would also include measurements such as the neutrino mass splitting, the CP phase and the mass hierarchy or even the existence of an additional sterile neutrino. 
	Therefore, in principle, a global analysis of all relevant experiments including decoherence effects would need to be performed.
	Fortunately, as we will soon show, while $\Delta L_{\rm min}$ from the errors in the mass splitting (which would also indicate the mass hierarchy) scales with $E_0$, that from the asymmetry of the intrinsic quantum uncertainties would saturate to a constant value when $E_0$ is above $\sim 5-10$ MeV. Hence, there will be a distinctive dependence on $E_0$ between these contributions in $\Delta L_{\rm min}$. 
	Moreover, we will also show that the phase shift term from quantum decoherence effect is insensitive to traditional measurements of the neutrino spectrum, while other fundamental oscillation parameters are determined by these measurements with increasing precision.   
	Therefore, for simplicity and illustration propose, we fix the neutrino mass at values determined by global analyses \cite{Esteban:2020cvm}, assume a three flavor oscillation with normal mass ordering and take $\delta_{\rm CP}=\pi$ in this section.

\begin{figure} 
		\centering
		\includegraphics[width=0.9\textwidth]{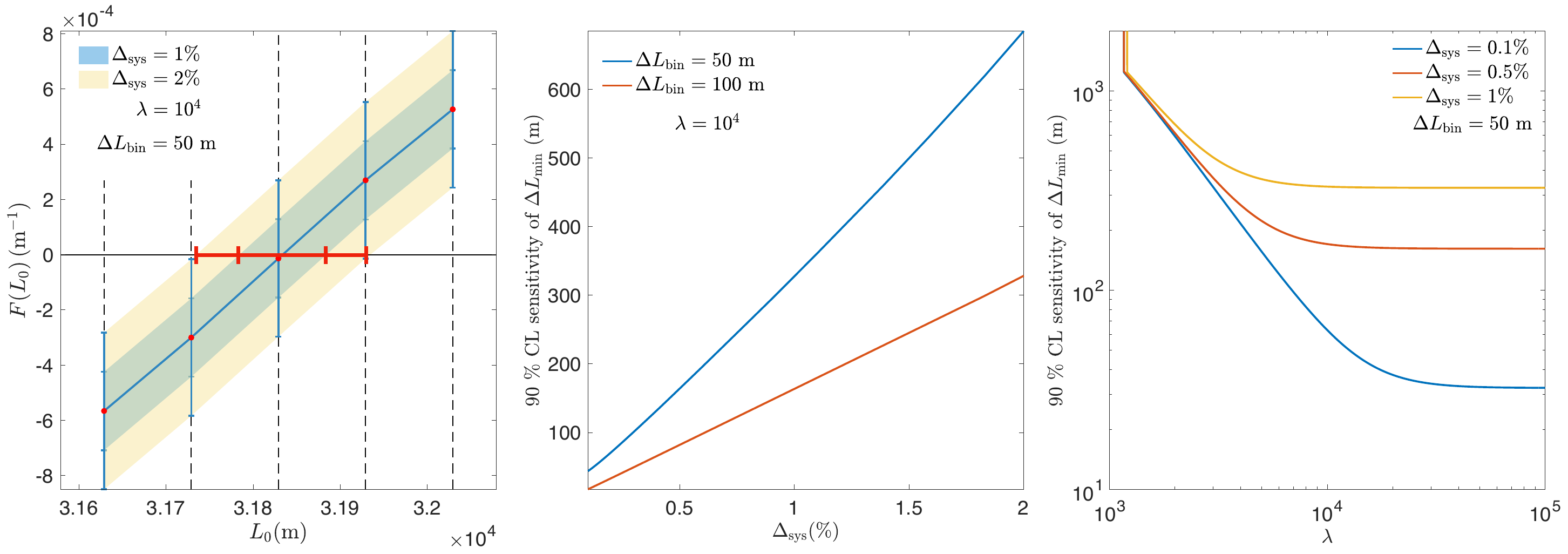}
		\caption{\label{fig:PA_demo}
			The left plot shows how we determine $L_{\rm min}$ for neutrinos with 30 MeV energy and its uncertainty (red error bars) from Eq.~\eqref{F_i} (blue error bars) considering two different systematic uncertainties labeled in the plot. Here we consider the JSNS setup by taking Eq.~\eqref{JSNS} for $N(E_0)$ in Eq.~\eqref{N_i}, and each bin is separated by the black dashed lines, hence the parameters which would influence the uncertainty of $L_{\rm min}$ are the systematic uncertainty ($\Delta_{\rm sys}$), the statistical uncertainties ($\lambda$: increment w.r.t.\ the JSNS set up) and the bin size ($\Delta L_{\rm bin}$). 
			The middle and right plot show how these parameters would influence the sensitivity for $\Delta L_{\rm min}$ through simulating the uncertainties of $L_{\rm min}$.
			In particular, the middle plot corresponds to Eq.~\eqref{bin_scale}, and the right plot shows when systematic/statistic uncertainties dominate over one another.
			Specific steps to determine $L_{\rm min}$ and more discussions on the plots are given in the text.
		}
	\end{figure}

	Operationally, the way to find $L_{\rm min}$ is to scan over $L_0$ (e.g.\ by moving the detector) around where we expect to observe the first local minimum for some neutrino energy $E_0$. 
	In particular, we consider counting neutrinos within some position bin $\Delta L_{\rm bin}$, i.e.
	\begin{equation} \label{N_i}
	N_i (E_0;\vec{\sigma}_n)= N(E_0)
	\int^{L_i + \Delta L_{\rm bin}/2}_{L_i - \Delta L_{\rm bin}/2} 
	dL_0 \, \frac{1}{4 \pi L_0^2} 
	\, P_{\nab}(L_0,E_0;\vec{\sigma}_n),
	\end{equation}  
	where $N(E_0)$ is the number of neutrinos produced times the detection rate, which is independent of the traveling distance $L_0$. Here, we have assumed that we do not lose or gain neutrinos during its propagation. Nonetheless, even if we do take such consideration into account, it would only affect the signal of $L_{\rm min}$ trivially, as long as it does not create a bump or dip for a certain $L_0$ within our range of interest. 
	Next, in order to find $L_{\rm min}$, we look at when the normalized (numerical) derivative of $N_i$,
	\begin{equation}\label{F_i}
	F_i(E_0;\vec{\sigma}_n) = 
	\frac{1}{\bar{N}_i}
	\frac{N_{i+1}(E_0;\vec{\sigma}_n)-N_i(E_0;\vec{\sigma}_n)}{L_{i+1}-L_i}, 
	%\frac{2}{N_{i+1}(E_0;\vec{\sigma}_n)+N_i(E_0;\vec{\sigma}_n)} 
	\end{equation}    
	is zero. We have $F_i$ plotted as the red dots in Fig.~\ref{fig:PA_demo}, and the vertical dashed lines represent the position bins, $L_i \pm \Delta L_{\rm bin}/2$.
	Here $\bar{N}_i=(N_{i+1}(E_0;\vec{\sigma}_n)+N_i(E_0;\vec{\sigma}_n))/2$ is the normalization factor which would eliminate the correlated uncertainties between the position bins, similar to the purpose of having near-far detectors.
	Note that the bin size is required to be $\ll L_{\rm osc}$, such that a ``local" minimum would be observed.   
	Therefore, $L_{\rm min}$ is where the blue line connecting all dots ($F(L_0)$) intersect with the black $F(L_0)=0$ line with uncertainty labeled as red horizontal bars in Fig.~\ref{fig:PA_demo}.
	As for the determination of such uncertainty, we adopt the following steps:
	\begin{enumerate}
		\item Propagate the uncertainty of the count numbers $N_i$ ($\Delta_{\rm sys}/\Delta_{\rm stat}$ for the systematic/statistic uncertianties) to the uncertainties of $F_i$ by the relation in Eq.~\eqref{F_i} as the blue error bars in the left plot of Fig.~\ref{fig:PA_demo}. In particular, only the uncorrelated uncertainties remain due to the normalization factor $\bar{N}$. 
		\item Connect (or fit) the error bars of $F_i$ (i.e.\ $F_i\pm \sqrt{\Delta_{\rm sys}^2+\Delta_{\rm stat}^2}$) and draw an uncertainty band as we have demonstrated in the left plot of Fig.~\ref{fig:PA_demo}. 
		\item The uncertainty of $L_{\rm min}$ is then the intersection between the uncertainty band and $F(L_0)=0$ labeled as red error bars in the left plot of Fig.~\ref{fig:PA_demo}. 
		This uncertainty is therefore determined by the relation between the signal ($F_i$) and its uncertainty, as we will discuss in the following. Note that it is possible for such intersection to be infinite when the error of the maximal $|F_i|$ surpasses its value.
		See the right plot in Fig.~\ref{fig:PA_demo} for instance, where the sensitivity of $\Delta L_{\rm min}$ goes to infinity when the statistics is too low.
	\end{enumerate} 
	Moreover, the uncertainties depend on the chosen bin size, demonstrated in the middle plot of Fig.~\ref{fig:PA_demo}, in fact, while $\Delta L_{\rm bin} \ll L_{\rm osc}$, increasing the position bin size would lower the uncertainty.
	The reason of this is two-fold: 1) the statistics for one bin would increase, reducing the statistical uncertainty, 2) $F_i$ being enhanced by the bin size reduces the uncertainty of $L_{\rm min}$. 
	As a matter of fact, for an oscillator such as $\sin(L/L_{\rm osc})$ (the role of the FTP), its derivative made discrete by bins (the role of $F_i$) is 
	\begin{equation}\label{bin_scale}
	\tilde{F}_i = \frac{1}{L_{\rm osc}}
	\int^{L_i+\Delta L_{\rm bin}/2}_{L_i-\Delta L_{\rm bin}/2}
	dL \cos\left(\frac{L}{L_{\rm osc}}\right)
	\simeq \frac{\Delta L_{\rm bin}}{L_{\rm osc}}.
	\end{equation}
	The approximation is valid when $(L_i\pm \Delta L_{\rm bin})/L_{\rm osc} \sim 2n \pi$ for some integer $n$, which is well justified since we only search around the oscillation minimum. 
	Here, we can see that for a fixed bin size, the signal is smaller at higher energies, hence the uncertainties would be larger. 
	For instance, for $\Delta L_{\rm bin}=\mathcal{O}(10)$ m and $\Delta_{\rm sys}=\mathcal{O}(1)\%$, the energy range with finite sensitivity lies within a few MeV. 
	Furthermore, although $\alpha_0$ in the expansion $N_i=\sum_{i}\alpha_iL_0^i$ is canceled out for the signal, it still contains uncorrelated uncertainties. 
	Thus, without $\alpha_0 \ll \alpha_i$, for some $i \neq 0$, there will be a significant increase in the uncertainty of $ L_{\rm min}$.
	In fact, this is what we have for observing neutrinos around the maximum oscillation value or if we look for disappearing neutrinos.
	Therefore, rare event measurements of appearance channels at minimum oscillation phase would be the better option for our method.

	By taking a 50 m (around the detector size of Hyper-K \cite{Abe:2011ts} and the DUNE far detector \cite{DUNE:2020ypp}) bin size, only reactor neutrinos and DAR neutrino are in the  energy range which leads to an appripriate oscillation length. In particular, the monochromatic neutrinos from $\pi$DAR are most suitable for the PMM with just one measurement, due to the following reasons:
	\begin{itemize}
		\item The monochromatic neutrinos are produced sharply around $30$ MeV, which is suitable for a 50 m bin size as we have demonstrated in Fig.~\ref{fig:PA_demo}.
		\item It provides a detectable appearance channel by producing $\nu_\mu$ which could oscillate into $\nu_e$. On the other hand, reactors only produce $\bar{\nu}_e$, hence its appearance channels are not detectable since $\bar{\nu}_\mu$ will be below the Cerenkov threshold in the sub MeV range.
		\item Since we consider a fixed $E_0$, the monochromatic feature automatically satisfies the condition without wasting any neutrinos spread out in the spectrum. Hence, statistics-wise, on top of the bright spallation source, it would be better than having $\mu$DAR neutrinos if we only consider measurements of $L_{\rm min}$ at a single $E_0$.      
		\item The systematic uncertainty would also be strongly reduced for $\pi$DAR neutrinos. First of all, the timing structure of DAR experiments \cite{Ajimura:2017fld,Baxter:2019mcx,Alonso:2010fs} would enable identification between $\pi$DAR neutrinos and $\mu$DAR neutrinos.
		In fact, $\bar{\nu}_\mu \rightarrow \bar{\nu}_e$ from $\mu$DAR would suffer from an intrinsic uncertainty since the $\bar{\nu}_\mu$ and $\bar{\nu}_e$ production are indistinguishable \cite{Ajimura:2017fld,Baxter:2019mcx,Alonso:2010fs}. Secondly, the energy reconstruction would be highly accurate for monochromatic neutrinos. Discussions on this topic can be found in \cite{Harnik:2019iwv}.
	\end{itemize}
	Similar to what we did for RMM, we consider a benchmark experiment, and ask how far we are to having enough sensitivity for some decoherence parameters. 
	In particular, we take numbers from the existing JSNS experiment \cite{Ajimura:2017fld}, i.e. $1.114 \times 10^{23}$ proton-on-targets for 1 MeV power within 3 year, from which $64\%$ would contribute to a $\pi$DAR process, producing monochromatic $\nu_\mu$ which would oscillate into $\nu_e$ and be detected by a 17-ton gadolinium loaded liquid scintillator.
	Hence, for Eq.~\eqref{N_i}, we obtain
	\begin{equation}\label{JSNS}
	\frac{N(E_0)}{4 \pi L_0^2} \simeq 0.43
	%S_{\mu}(E_0)C_{\rm IBD}(E_0) \,
	\frac{1}{\rm{m}^2} 
	\left( \frac{P}{1 \,\rm MeV}\right)
	\left( \frac{T}{3 \,\rm yr}\right)
	\left( \frac{\rm M_{D}}{17 \,\rm ton}\right)
	\left( \frac{31829\, \rm m}{L_0}\right)^2.
	\end{equation}
	Here, we adopted the cross-section for quasielastic scattering of $\nu_e$ on proton from \cite{Strumia:2003zx} as $7.5 \times 10^{-41}$ cm$^{-2}$ at 30 MeV, and assume that the detector is moved to the oscillation minimum (the actual JSNS detector is placed 24 m from the source). 
	In Fig.~\ref{fig:PA_demo} we adjust the equation above by moving $L_0$ around its first oscillation minimum, then increase it $\lambda$ times.
	In addition, similar to other DAR channels \cite{Ajimura:2017fld,Ajimura:2020qni}, the systematic uncertainties should be dominated by intrinsic uncertainties, i.e.\ the $\bar{\nu}_e$ produced by $\mu$DAR, which take up approximately $3\%$ of total amount of neutrinos produced at 30 MeV. Furthermore, one could also identify whether a neutrino comes from $\pi$DAR from the timing structure, for instance, in the JSNS setup, $\bar{\nu}_e$ from $\mu$DAR takes up only $<10\%$ of the early time bin which is dominated by $\nu_e$ from $\pi$DAR \cite{Ajimura:2017fld}. 
	Hence, we take various systematic uncertainties in the range of 0.1-2\% in Fig.~\ref{fig:PA_demo}.
	From the middle and right plot of Fig.~\ref{fig:PA_demo}, we find the sensitivity for some $\Delta L_{\rm min}$ by first estimating the uncertainty of $L_{\rm osc}$ (i.e.\ when $\Delta L_{\rm min}$=0) for some systematic and statistical uncertainty (from $\Delta_{\rm sys}$ and $\lambda$), then further identify what values of $\Delta L_{\rm min}$ would be rejected by such data at 90\% CL.
	Theoretical estimates for decoherence effects which lead to such $\Delta L_{\rm min}$ will be shown in the following paragraph.

	\begin{figure}
		\centering
		\includegraphics[width=1\textwidth]{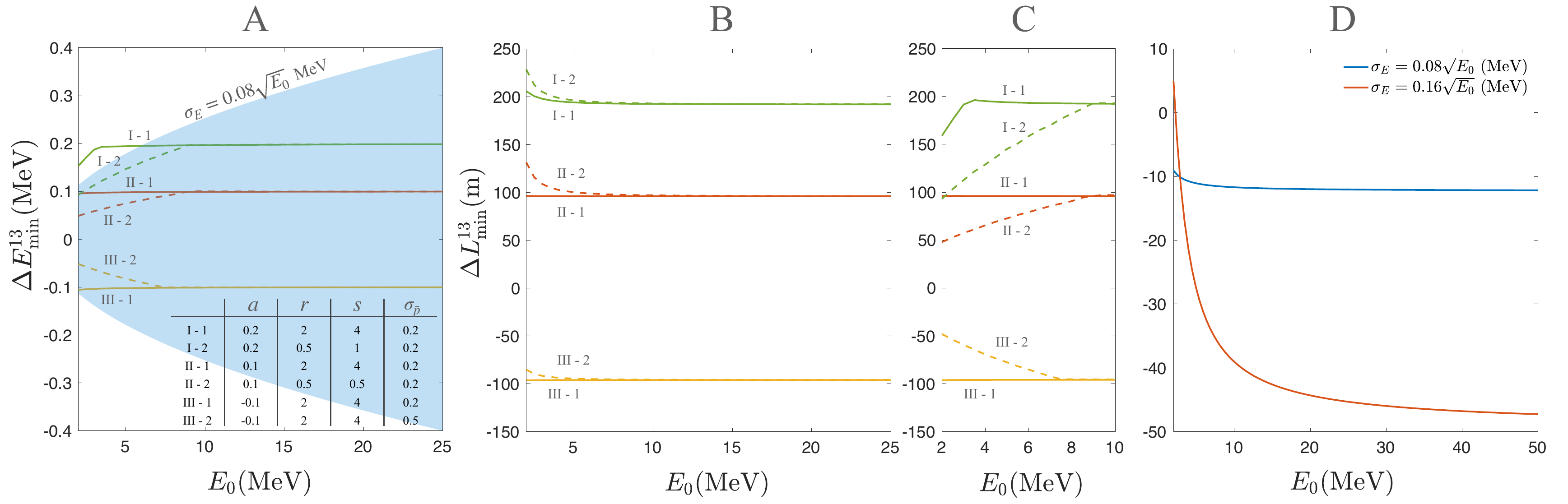}
		\caption{\label{fig:PA_theory}
			Plot A (B, C) shows the variation of $E_{\rm min}^{13}$ ($L_{\rm min}^{13}$) for a number of quantum uncertainty parameters for Eq.~\eqref{WtwoG}. The labels for each line (as well as its colour and style) on all three plots (and also in Fig.~\ref{fig:PA_sens}) correspond to the table on Plot A, while Plot A and C consider a constrain by $\sigma_E = 0.08 \sqrt{E_0}$ MeV (blue area in Plot A) in addition.
			Plot D shows the variation of $L_{\rm min}^{13}$ caused by $\sigma_E$ from a Gaussian distributed energy resolution.}
	\end{figure}
	
\begin{figure}
		\centering
		\includegraphics[width=.85\textwidth]{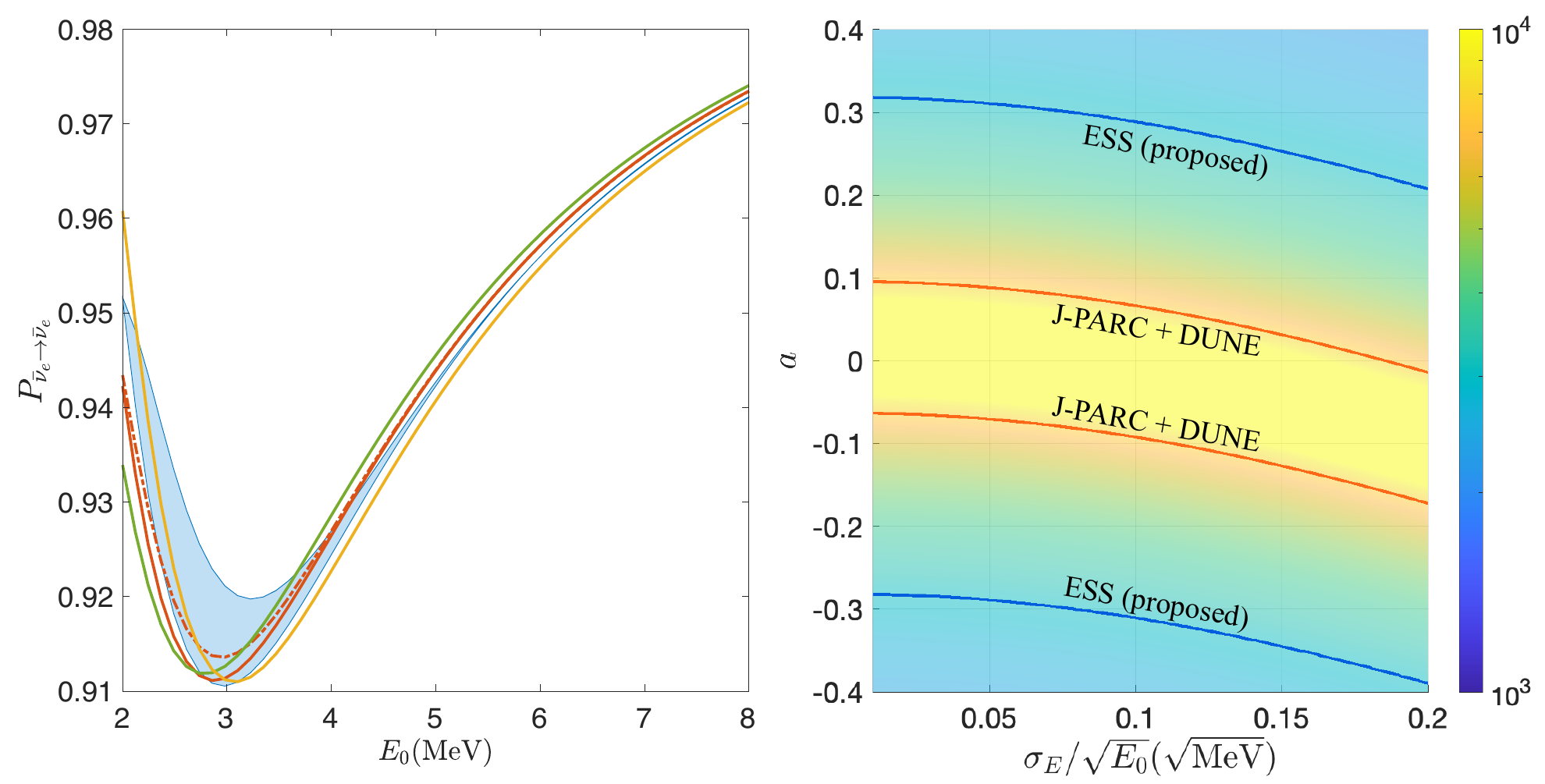}
		\caption{\label{fig:PA_sens}
		The left plot shows how the phase shift term is not sensitive to the rate measuring method compared to the damping term: The upper edge of the blue band represents the transition probability for $\sigbp$ at its upper limit given in \cite{deGouvea:2021uvg} for a Gaussian distributed $D_{\bp}$, and the lower edge is the fully coherent case. The colour code of the lines corresponds to the same decoherence parameters given in Fig.~\ref{fig:PA_theory}, which are within the parameter space in the right plot. 
		The right plot shows values of $\lambda$ in Eq.~\eqref{JSNS} required to achieve a 90\% CL sensitivity for the decoherence parameter space by the colour bar. The red line labeled ``J-PARC+DUNE" gives the required $\lambda$ by assuming a J-PARC-like source combined with a DUNE-like detector; and similarly, the blue line labeled ``ESS (proposed)" considers one year of data taking of the ESS source and the water Cherenkov detector proposed in \cite{ESSnuSB:2021azq}. Both cases are assumed to have the baseline $L_0$ at the first oscillation minimum at 30 MeV (31829 m).}
	\end{figure}

	The phase shift from decoherence effect for ground-based neutrinos would mainly come from the asymmetry of quantum uncertainties decided by the weighting function $D_{\bp}$ and the classical (statistical) energy uncertainty with weighting function $H_E$. In particular, we consider $H_E$ being dominated by the energy resolution (i.e.\ $H_E$ is Gaussian distributed) and $D_{\bp}$ as a two-Gaussian distribution generically formalised as 
	\begin{equation}\label{WtwoG}
	D_{\bar{p}}(\bar{p}\equiv |\bar{\vp}|;E_0) = \frac{1}{2\sqrt{\pi}\sigbp'(1+rs)}\left\{ 
	\exp\left(\frac{-(\bar{p}-E_0+dE_0)^2}{4\sigbp'^2}\right)
	+r\exp\left(\frac{-(\bar{p}-E_0-dE_0)^2}{4(s\sigbp')^2}\right)
	\right\},
	\end{equation} 
	where the width is $\sigbp=(1+rs)\sigbp'$, according to the definition in Appendix~\ref{Sec.AppA}.
	This formalism represents scenarios such as neutrino produced or detected with two types of interactions simultaneously, with different probabilities and widths ($r,s$) and have slightly different expectation values for $E_0$ ($E_0 \pm dE_0$, in particular).    
	Moreover, with the phase structure given in Eq.~\eqref{SD_bp}, the decoherence term is simply the Fourier transformation of $D(\bar{p};E_0)$ from $\bar{p}$ to $\alpha_{jk}=\Delta m_{jk}^2L_0/(2E_0^2)$, and the phase shift is 
	\begin{equation}
	\beta_{jk}(E_0,\sigbp)=\tan^{-1}\left( 
	\frac{1-r\,s\,e^{-(s^2-1)\alpha_{\bar{p},jk}^2\sigbp'^2}}{1+r\,s\,e^{-(s^2-1)\alpha_{\bar{p},jk}^2\sigbp'^2}}
	\tan(\alpha_{\bar{p},jk}\, dE_0)
	\right)
	\xrightarrow{\alpha_{\bar{p},jk}\ll1}
	\, \frac{1-rs}{1+rs}
	\frac{\Delta m_{jk}^2L_0}{2E_0^2}
	\, dE_0.
	\end{equation}
	Furthermore, the fact that we search around the first minimum ($\psi=2\pi$) and $\beta_{jk} \ll L_{\rm min}^{\rm osc}$ implies that
	$\Delta m_{jk}^2L_0/(2E_0)\simeq 2\pi$, hence 
	\begin{equation}\label{DL_sigp}
	\Delta L_{\rm min}^{jk} \simeq 
	\frac{2\pi}{2.53 \, \Delta m_{jk}^2} \, a,
	\end{equation}
	at high energies, where
	\begin{equation}
	a = \frac{1-rs}{1-rs} \, dE_0.
	\end{equation}
	This can be seen in Fig.~\ref{fig:PA_theory}, where lines having the same $a$
	merge to one constant value at higher energies which is independent of both $E_0$ and $\sigbp$. Such property is not generic for all sources of decoherence effect, in fact, only $\alpha_{\bar{p},jk}$ from the phase structure $\eta_{jk}=i\alpha_{\bar{p},jk}\bar{p}$, cancels out the energy dependence with $L_{\rm osc}^{jk}$ in Eq.~\eqref{DeltaL} exactly. 
	For instance, in Plot D of Fig.~\ref{fig:PA_sens}, $\Delta L_{\rm min}^{13}$ increases with energy only because $\sigma_E$ does as well. In fact, if $\sigma_E$ is not energy dependent, it would approach zero at large $E_0$. 
	Moreover, from Eq.~\eqref{DL_sigp} we can see that when $s=1$, i.e.\ the two bumps have the same width, $\sigbp$ would have no role in the phase shift.
	In addition, the variance of $H_E$ ($\Delta_E$, weighting function on the second layer with width $\sigma_E$) must be larger or equal to that of $D_{\bar{p}}$ ($\Delta_{\bar{p}}$). 
	In fact, when they are equal to one another, the energy would be measured to a quantum level. 
	Hence if one keeps on lowering $\Delta_E$, $\Delta_{\bar{p}}$ would be forced to lower accordingly and the uncertainty of $\bar{x}$ would increase in order to fulfil the uncertainty principle.
	In this case, if we consider $\sigma_E = 0.08 \sqrt{E_0}$, and scale $D_{\bar{p}}$ by scaling $\sigbp$ and $dE_0$ simultaneously to fit the constrain $\Delta_{\bar{p}}=\Delta_E$, we find a change from Plot B to Plot A and C in Fig.~\ref{fig:PA_theory}. Plot B, on the other hand, assumes that $\sigma_E$ is large enough (in this case, $\sigma_E \geq 0.17 \sqrt{E_0}$) such the the quantum uncertainties are un-squeezed. 
	Furthermore, we can see that the dashed lines are more influenced by the constraint from $\Delta_E$ than the solid lines, since they either have a larger $dE_0$ or $\sigbp$, both indicating a larger $\Delta_{\bar{p}}$. 
	From plot A, we can see that if $\sigma_E$ is small, it would squeeze $D_{\bar{p}}$ and lower the phase shift; on the other hand, if $\sigma_E$ is large, then the blue area covers all the lines and there will not be enough sensitivity.
	Therefore, while there is still a little space out of the sensitivity line, $\Delta E_{\rm min}$ it is also not a suitable approach to measure a $D_{\bp}$-induced phase shift.
	
	Finally, we estimate the sensitivity for the benchmarks in  Fig.~\ref{fig:PA_theory} for the PMM in the right plot of Fig.~\ref{fig:PA_sens}.
	Furthermore, in the left plot, we demonstrate how the RMM is not as sensitive to the phase shift term compared to the damping term. 
	The blue band is the range of $D_{\bp}$-induced state decoherence which is not constrained by the combined analysis of reactor experiments from \cite{deGouvea:2021uvg}, i.e.\ the upper edge of the band represents $W_{\bp}$ as a Gaussian with width $\sigbp=0.47$ MeV.
	On the other hand, while having $\mathcal{O}(100)$ m of $\Delta L_{\rm min}$ for the PMM, the colored lines (with the same parameter as those in Fig.~\ref{fig:PA_theory}) do not vary the FTP to an extend that is close to the limit set by the combined analysis (not to mention for just one single experiment).
	Moreover, while the RMM significantly depends on how the neutrino spectrum would be without oscillation, the phase shift, which slightly shifts the FTP, does not change the shape of the spectrum as the damping term does, hence, it can be easily compensated by non-oscillation related models.  
	On the contrary, for the PMM, the signal is amplified by the oscillation length and is nearly independent of non-oscillation related models.
	The main disadvantage is the lack of statistics since we aim at searching for appearing flavors at the oscillation minimum. 
	Nonetheless, from the right plot in Fig.~\ref{fig:PA_sens}, we see that with the increment mainly by the detector size, the statistics would be enough for a 90\% CL sensitivity for a range of decoherence asymmetry parameters of the quantum and classical uncertainties.  
	Specifically, compared to the 17 T detector mass and a cross section of $7.5 \times 10^{-41}$ cm$^2$ of JSNS, the DUNE detector would have an increased detector mass of 40 kT, and the  liquid argon material of the detector also enhances the cross section to $2.5\times 10^{-40}$ cm$^2$ at 30 MeV \cite{COHERENT:2020iec}, hence $\lambda\simeq 7.8\times 10^3$ in this case (red lines).
	As for the ESS setup proposed in \cite{ESSnuSB:2021azq}, while using a water Cherenkov detector implies a lower cross section ($3\times 10^{-42}$ cm$^2$ at 30 MeV \cite{Kolbe:2002gk}), the detector mass would be increased to 538 kT, and the spallation source is also brighter by having $2.7\times 10^{23}$ POT per year.

	%6%%%%%%%%%%%%%%%%%%%%%%%%%%%%%%%%%%%%%%%%%%%%%%%%%%%%%%%%%%%%%%
	\section{Conclusion} \label{sec.5}
	Owing to the increasing precision of neutrino oscillation experiments, neutrino decoherence effects may become approachable in future experiments, opening a new window to probe new physics.
	In this work we introduce the ``layer structure" (illustrated in Fig.~\ref{fig:layers}), which includes the concept of an open quantum system and classical statics while having QFT as the fundamental theory.  
	This structure is particularly useful for understanding mechanisms behind decoherence signatures in neutrino oscillation experiments. 
	For instance, quantum uncertainties such as coordinate uncertainties around the vertices and the lifetime of particles entangled with the system  are parameterised as $\sigbx$ and $\sigbp$. 
	These two uncertainty parameters are the width of weighting functions w.r.t.\ the coordinate and momentum variables in the Wigner phase space (on layer 1), respectively. 
	On the other hand, classical uncertainties caused by a lack of knowledge would also contribute to decoherence signatures, such as the production profile of neutrinos, energy resolution and errors of the energy reconstruction model. 
	The former cause dominates the uncertainty parameter $\sigma_L$ while the latter two are included in $\sigma_E$.
	These two parameters are the width of weighting functions w.r.t.\ the coordinate and energy variables in the relativistic phase space (on layer 2), respectively. 
	We have shown that decoherence effects from all these uncertainty parameters come from phase wash-out effects, which are determined by a phase structure and some distribution.  	
	For each uncertainty parameter, there is a certain phase structure and some localized distribution with width as the corresponding parameter, resulting in a phase wash-out effect suppressing and/or causing a phase shift in the oscillation signature.	
	The phase structure also characterises dependence on the traveling distance ($L_0$) and energy spectrum ($E_0$) for each uncertainty parameter, hence, enables us to identify the mechanisms behind decoherence signatures by analysing these parameters in the neutrino detection profile and/or spectrum.
	The phase structures are given in Eq.~\eqref{SD_phase_T} and Eq.~\eqref{psi_T} (Eq.~\eqref{SD_phase_L} and Eq.~\eqref{psi_L}) for uncertainties on the Wigner phase space and the relativistic phase space, respectively, for the time dependent (independent) case.
	Furthermore, we have classified neutrino decoherence in terms of its mechanism as state decoherence and phase decoherence. The former represents the separation of superposition (mass) states, and is dominated by quantum uncertainties; while the latter indicate averaging effect due to the information loss, and is mainly decided by macroscopic classical uncertainties.

	In particular, we calculate the case of Gaussian distributed weighting functions and estimate how much more statistics we need for certain $\sigbp$, $\sigma_L$ and $\sigma_E$, to be sensitive to them at  90\% CL in Fig.~\ref{fig:RENO}, by taking the RENO experiment as a benchmark. 
	We find from Fig.~\ref{fig:NFS_where} that  when the current far detector is located e.g.\ 14 (1.4457, 0.01) km away from the source\footnote{The actual distance of RENO's far detector is 1.4457 km.}, we need approximately 26 (2, 82) times more statistics to reach sensitivity to meaningful values $\sigbp = 0.1$ MeV ($\sigma_E = 0.08 \sqrt{E_0} \rm{\sqrt{MeV}}$, $\sigma_L = 3$ m).
	Furthermore, we propose a novel method, the phase measuring method, to measure the asymmetry of weighting functions by searching an oscillation minimum. Particularly, we estimate the sensitivity of this method for a two-Gaussian distributed, $D_{\bp}$-induced, quantum mechanical uncertainty as well as the statistical uncertainty from the energy resolution in Fig.~\ref{fig:PA_sens}. While the energy resolution ranges typically from $1-10 \%/\sqrt{E_0}(\sqrt{\rm MeV})$ for neutrino detectors, the asymmetry parameter $a$, could be caused by the quantum effect of having a superposition of different processes. For instance, having simultaneously quasi-elastic scatterings and inelastic scatterings for neutrinos scattering on nucleons, or by nuclear effects such as the Fermi motion \cite{Bodek:1980ar}. In fact, quantitative estimation of the asymmetry parameter would need further investigation. 
	To sum up, while the four uncertainty parameters $\sigma_x$, $\sigma_p$, $\sigma_L$ and $\sigma_E$ in our structure can be determined by some theoretical mechanisms, such as the wave packet size of the external particles, the type of collisions, matter effect, exotic effects like space-time fluctuation, etc; it could also be potentially measured experimentally through rate or phase measuring methods.
	Our considerations presented here provide the theoretical background for such analyses and can be applied to any experiment. Experimental improvements are necessary, for instance via better energy resolution or larger event numbers, or by other detection techniques made possible by e.g.\  developments in coherent elastic neutrino-nucleus scattering.

	Decoherence effect in neutrino oscillation also has many potential beyond ground-based experiments. For instance, atmospheric neutrinos might  be a promising possibility to see decoherence effects, especially the one mediated by $\sigma_L$.
	The production profile of $\sigma_L$ for atmospheric neutrinos would be of $\mathcal{O}(10)$ km and asymmetric. 
	Aside from the production profile, matter effects may also be included.
	Moreover, since the atmospheric neutrinos have a broad energy spectrum and can be detected at different zenith angles, we should be able to do a wide range tomography on the $(L_0,E_0)$ space to analysis decoherence effects. 
	Furthermore, there could be contributions to the uncertainties that are not directly measurable by other approaches, such as off-shell mediators contributing to $\sigma_x$. Therefore, a better understanding of the measurable uncertainties by other approaches would increase the sensitivity of probing new physics through neutrino decoherence.
	In addition to aiming at searching for mechanisms causing neutrino decoherence, one can also make use of the decoherence effect to investigate  other aspects, for instance, the search of sterile neutrinos or the measurement of CP violation, by designing and/or engineering these parameters, such as controlling $\sigma_{L}$ by the distribution of neutrino sources/detector, and $\sigbx, \sigbp$ by manipulating squeezed states used in quantum optics.

	%%%%%%%%%%%%%%%%%%%%%%%%%%%%%%%%%%%%%%%%%%%%%%%%%%%%%%%%%%%%%%%%	
	\section*{Acknowledgments}
	We would like to thank Evgeny Akhmedov and Janina Hakenmüller for useful discussions. TC acknowledges support by the IMPRS-PTFS.	
	
	\section*{Appendix}
	\appendix
	
	%A-1%%%%%%%%%%%%%%%%%%%%%%%%%%%%%%%%%%%%%%%%%%%%%%%%%%%%%%%%%%%%%

	\section{Fourier Transformation and Convolution Properties} \label{Sec.AppA}
	We review some useful properties of Fourier transformation (FT) and convolution in this section. These properties are particularly useful in our structure, for the layer moving operators involve integrations of complex functions, which can be parameterized as a plane wave term $\exp(i\eta(x,p))$ and a normalized-real probability density function (PDF) term. In particular, we will show how the layer variables are connected with each other via FT properties, how two sources of uncertainties are combined to an effective one with convolutional properties, and demonstrate the phase washout effect. Below, we will outline the properties in bold front followed by a more detailed demonstration.
	\begin{itemize}
		\item 
		\textbf{Property 1: The FT of an even function is real, while that of an odd function is purely imaginary.}  
		For an even function $f(x)$,
		\begin{equation}
		\begin{split}
		&\int^{\infty}_{-\infty} dx \, e^{-ipx} f(x)			=\frac{1}{2}\int^{\infty}_{-\infty} dx \, e^{-ipx} \left\{ f(x)+f(-x)\right\} \\
		&=\frac{1}{2}\int^{\infty}_{-\infty}dx \, e^{-ipx} f(x) - \frac{1}{2}\int^{\infty}_{-\infty}dx \,  e^{ipx} f(x)
		=\int^{\infty}_{-\infty} dx \, f(x)\cos{(px)} \in \mathbb{R}	,
		\end{split}
		\end{equation}
		and similarly for odd functions being imaginary after Fourier transformation.
		
		\item
		\textbf{Property 2: For any probability density function (PDF), $W(x;L)$, the FT can be written as}
		\begin{equation} \label{FT_PWO}
		P(L,p)=\int^{\infty}_{-\infty} dx \,e^{-ipx} W(x;L) = e^{-ipL} \tilde{W}(p) 
		\equiv e^{-i(pL-\beta(p))} |\tilde{W}(p)| ,
		\end{equation}
		\textbf{where the damping term $|\tilde{W}(p)| \leq \tilde{W}(0)=1 $ and the phase shift term $\beta(p)$ is non-zero and non-$\pi$ only when $W(x;L)$ is symmetric w.r.t.\ $L$.} 
		If the shape of $W(x)$ is symmetric, $W(x+L)$ would be even for $L=\int dx \, x \, W(x)$. Then, after shifting $x \rightarrow x+L$, we obtain Eq.~\eqref{FT_PWO} with $\tilde{W}(p)\in \mathbb{R}$ and no phase shift, i.e.\ $\beta=0$. 
		On the other hand, if the shape of $W(x)$ is not symmetric, we can always write it in terms of a an even function and an odd function, i.e.\ $W(x+L)=W_{\rm even}(x+L)+W_{\rm odd}(x+L)$. Hence on top of the even part, which is treated in the same manner as the symmetric case, the odd part would give rise to an imaginary part in $\tilde{W}(p)$, or in terms of rotation coordinate, a phase shift $\beta(p) \neq 0$ and $\pi$ relative to the oscillation phase, $pL$, on the next layer.   
		Moreover, $W(x)$ being a PDF indicates that  $\int dx \, W(x)=1$, and  $W(x)\geq 0$, therefore, $|\tilde{W}(p)|\leq \int dx\,  W(x)=\tilde{W}(0)=1$.
		A list of examples is given in Table~\ref{fig:PWO_sigL}, showing  how the asymmetry of the PDF induces a non-zero and non-$\pi$ phase shift. 
		Additionally, it is also clear from the plots that, in most cases, the larger the width (labeled as $\sigma_n$, for $n=\{p,L,E\}$) is for the PDF, the smaller will the width of $\tilde{W}(p)$ be. 
		Henceforward, since $|\tilde{W}(p)| \leq \tilde{W}(0)=1 $,  the larger $\sigma_n$ is, the smaller $|\tilde{W}(p)|$ will be, for some $p\neq 0$, and the more suppressed $P(L,p)$ will be.
		Another way to look at this effect is that a wider width of the PDF indicates that there is a wider range for $e^{ipx}$ to be averaged out upon the integration over $x$, namely, the PWO effect. 
		
		\item 
		\textbf{Property 3: The PWO effect is the generic case of Eq.~\eqref{FT_PWO}, {for a complex function $\Gamma(x;L)\equiv |\Gamma(x;L)|e^{i\eta(x)}$, and Eq.~\eqref{FT_PWO} is simply when $\eta(x)$ is linear in $x$.}} The PWO effect is written as
		\begin{equation} \label{PWO}
		\frac{\int dx \, \Gamma(x;L)}{\int dx \, |\Gamma(x;L)|}
		= e^{i (\eta(x)|_{x=L}-\beta)} \Phi(L),
		\end{equation}
		where $L$ is the central value of $\Gamma(x)$, such that $\Gamma_{\rm even}(x+L)$ is even. In fact, according to the layer structure presented in the main text, $L$ would also be the next level PS variable in our structure corresponding to $x$. Hence, similar to property 2, $|\Phi| \leq 1 $ and $\beta$ is non-zero only when $\Gamma(x;L)$ is symmetric w.r.t.\ $L$. In fact, this is why we call $\Phi$ the damping term and $\beta$ the phase shift term in this paper. Moreover, the wider $\Gamma(x)$ is relative to the wavelength for the phase structure $\eta(x)$, the smaller will $\Phi$ become.    
		
		\item
		\textbf{Property 4: For two distributions $f(x)$ and $g(x)$ with width $\sigma_f$ and $\sigma_g$, respectively, the width of  $(f*g)(L)$, $\sigma_{f*g}$ is larger than either $\sigma_f$ or $\sigma_g$, where ``$*$" represents the convolution of two distributions.} Whenever two function are related with the form 
		\begin{equation}
		\int dx \, f(x)g(x-L) \equiv (f*g)(L),
		\end{equation}   
		there is a convolution between these two functions. This usually occurs when there are multiple sources of uncertainties taken into consideration, for instance, the total uncertainties of the PS variables from both the initial state and the final state (Eq.~\eqref{F_P}), the production site and the detection site (Eq.~\eqref{G}), or the external process and the internal process (Eq.~\eqref{B17}). 
		The width of an arbitrary localized function $f(x)$ is defined here as
		\begin{equation}
		\sigma_f = \frac{1}{2\sqrt{ \pi}}\int dx \, |f'(x)|,
		\end{equation}  
		where $f'(x)$ is the normalized function of $f(x)$, and $f'(x)=f(x)/max\{|f(x)|\}$, such that its global maximum is unitary. 
		Also, $1/2 \sqrt{\pi}$ is inserted such that width of a Gaussian distributed function would have the width at one standard deviation and the other distributions are then defined accordingly. 
		On the other hand, the width of the product of two function, $\sigma_{fg}$, will be smaller than the individual widths of the functions $\sigma_f$ and $\sigma_g$, since
		\begin{align}
		&4 \pi \sigma_{fg}=\int dx \, |f'(x)g'(x)|
		\leq \int dx \, |f'(x)||g'(x)| \nonumber \\ 
		&\leq \int dx \, |f'(x)|= 2 \sqrt{\pi} \sigma_f
		\text{ and }
		\int dx \, |g'(x)|=2 \sqrt{\pi} \sigma_g,
		\end{align} 	
		for $|f'(x)|\leq 1$ and $|g'(x)|\leq 1$. 
		Moreover, by the convolution theorem, 
		\begin{equation}
		f*g=\mathcal{FT}^{-1} [\mathcal{FT}(f)\, \mathcal{FT}(g)],
		\end{equation}      
		we can see that comparing to the trivial case where $g$ is a delta function, and we have $f=\mathcal{FT}^{-1} [\mathcal{FT}(f)]$, the width of $\mathcal{FT}(f)\, \mathcal{FT}(g)$ would decrease when the width of $g$ is no longer zero, and hence $\sigma_{f*g}$ would increase. 
		For example, if $f(x)$ and $g(x)$ are Gaussian distributions, then $\sigma_{f*g}^2=\sigma_f^2+\sigma_g^2$.
		\item{\textbf{Property 5: Convolution of a complex function, $h(x) =f(x)\,e^{ipx}$, and a real function, $g(x)$, is
				\begin{equation} \label{conv_complex}
				\left( h*g\right)(y)= \int e^{ip'y} \, \tilde{f}(p'-p) \,\tilde{g}(p')
				\equiv e^{ipY(y)} I_1(y) I_2(p),
				\end{equation}
				where $\tilde{f}=\mathcal{FT}[f]$, $\tilde{g}=\mathcal{FT}[g]$ and the width of $I(y) \in \mathbb{R} $ is the same as that of  $(f*g)(y)$.}}
		
		By the convolution theorem,
		\begin{align}
		\mathcal{FT}[h*g]= \mathcal{FT}[h] \mathcal{FT}[g]=
		\int dx f(x) e^{-i(p'-p)x} \int dx g(x) e^{-ip'x}
		=\tilde{f}(p'-p) \tilde{g}(p'). \label{conv_complex2}
		\end{align} 
		Then by doing an inverse Fourier transformation from $p'$ to $y$ on Eq.~\eqref{conv_complex2}, we arrive at Eq.~\eqref{conv_complex}. Furthermore, the width of the product of two functions, $\tilde{f}(p'-p)\tilde{g}(p')$ is independent of the parallel shift from $p$, i.e.\ the width of $\tilde{f}(p'-p)\tilde{g}(p')$ is the same as $\tilde{f}(p')\tilde{g}(p')$, which is the case where the convolution is between  $f$ and $g$. For example, if $f(x)$ and $g(x)$ are Gaussian distributed, i.e.
		\begin{equation}
		f(x)=\exp \left[ \frac{-(x-\mu_f)^2}{4 \sigma_f}\right], \quad
		g(x)=\exp \left[ \frac{-(x-\mu_g)^2}{4 \sigma_g}\right] ,
		\end{equation}   
		then $\tilde{f}(p')= \exp\left(-ip'\mu_f- p'^2 \sigma_f^2\right)$, $\tilde{g}(p')= \exp\left(-ip'\mu_g - p'^2 \sigma_g^2\right)$, and 
		\begin{equation} \label{conv_fg}
		\tilde{f}(p'-p) \tilde{g}(p') =
		e^{-ip'(\mu_f+\mu_g)}
		\exp \left[-\left(\sigma_f^2+\sigma_g^2\right)\left(p'-\frac{p}{\Delta}\right)^2
		- \sigma_f^2 p^2\left(1-\frac{1}{\Delta}\right)\right],
		\end{equation}
		where $\Delta = (\sigma_f^2+\sigma_g^2)/\sigma_f^2$. 
		We can see that the width of Eq.~\eqref{conv_fg} w.r.t.\ $p'$ is independent of $p$.
		Therefore, according to Eq.~\eqref{conv_complex}, the convolution of $h$ and $g$ is then to do a Fourier transformation from $p'$ to $y-(\mu_f+\mu_g)$,  giving us
		\begin{equation}
		(h*g)(y)= 
		e^{i\frac{p}{\Delta}(y-\mu_f-\mu_g)}
		\exp \left[-\frac{(y-\mu_f-\mu_g)^2}{4(\sigma_f^2+\sigma_g^2)}
		- \sigma_f^2 p^2\left(1-\frac{1}{\Delta}\right)\right].
		\end{equation}
		Hence, the width w.r.t.\ $y$ is $\sigma_f^2 +\sigma_g^2$, which is the same the width of $(f*g)(y)$ shown in Property 4.
		Moreover, when $\sigma_f = 1/\sigma_p$ and $\sigma_g = 2  \sigma_x$, it follows $\Delta=1+4 \sigma_x^2 \sigma_p^2$, which agrees with Eq.~\eqref{Phi}. 
		
		\item \textbf{Property 6: In Table \ref{table1} \& \ref{table2} we classify how the width would evolve after Fourier transformation, product of functions, convolution of real functions and convolution of complex functions as in ``property 5".}  By the properties above, we summarize the width evolution of the first three types in the Table~\ref{table1}, specifying its relation with the original function(s) and give the example of assuming all original functions are Gaussian distributed.

		As for the last type, two functions are generated under such combination, namely, $H(y)$ and $I(p)$ in
		\begin{equation}
		\bigg| \int dx f(x)e^{ipx} g(x-y) \bigg|= H(y) I(p). 
		\end{equation}
		By property 5, we see that the width of $H(y)$ and $I(p)$ has properties shown in Table~\ref{table2}.

		\begin{table} [t!]
			\begin{center}
				\begin{tabular}{ |c|c|c|c| } 
					\hline
					Type (notation for $\sigma_h$)& Function relation & Width relation & Gaussian case\\ 
					\hline
					FT type ($\tilde{\sigma}_f$)& $h = \mathcal{FT}(f)$ & NC & 
					$\sigma_h=\frac{1}{2\sigma_f}$\\ 
					%$\sigma_h=1/(2\sigma_f)$\\ 
					\hline
					Product type ($\sigma_{fg}$)& $h=f\times g$ & PC, $\sigma_h < \{ \sigma_f,\sigma_g \}$ &
					$\frac{1}{\sigma_h^2}=\frac{1}{\sigma_f^2}+\frac{1}{\sigma_g^2}$\\ 
					%$\sigma_h^2=(\sigma_f^2\sigma_g^2)/(\sigma_f^2+\sigma_f^2)$\\ 
					\hline
					Convolution type I ($\sigma_{f*g}$)& $h=f*g$  & PC, $\sigma_h > \{ \sigma_f,\sigma_g \}$ 
					& $\sigma_h^2=\sigma_f^2+\sigma_g^2$\\
					\hline
				\end{tabular}
			\end{center} 
			\caption{\label{table1}Properties of the width of the function $h$ in terms of the original real function(s) $f$ (and $g$). Here NC/PC means that $\sigma_h$ is negatively/positively correlated to $\sigma_f$ (and $\sigma_g$).}
		\end{table}
		\begin{table} [t!]
			\begin{center}
				\begin{tabular}{ |c|c|c|}
					\hline
					Function (width notation), Type& Width relation & Gaussian case \\
					\hline
					$H(y)$ $(\sigma_H)$, Convolution type I & PC	 , $\sigma_H > \{ \sigma_f,\sigma_g \}$ 
					&$\sigma_H = \sigma_{f*g}$\\
					\hline
					$I(p)$ $(\sigma_I)$, Convolution type II & NC		
					&$\sigma_I = 1/\sigma_{fg}$	\\
					\hline	
				\end{tabular}
			\end{center}
			\caption{\label{table2} Properties of the width of functions $H$ and $I$ for convolution with an addition complex phase in terms of their origin functions $f$ and $g$, giving rise to an additional term $I(p)$.}
		\end{table}	
		
	\end{itemize}
	%%%%% A-2 %%%%%%%%%%%%%%%%%%%%%%%%%%%%%%%%%%%%%%%%%%%%
	
	\section{Calculation of the Neutrino Flavor Transition Amplitude}  \label{Sec.AppB}
	
	\subsection{Neutrinos represented by Entanglement States}
\begin{figure}
		\centering
		\includegraphics[width=0.8 \textwidth]{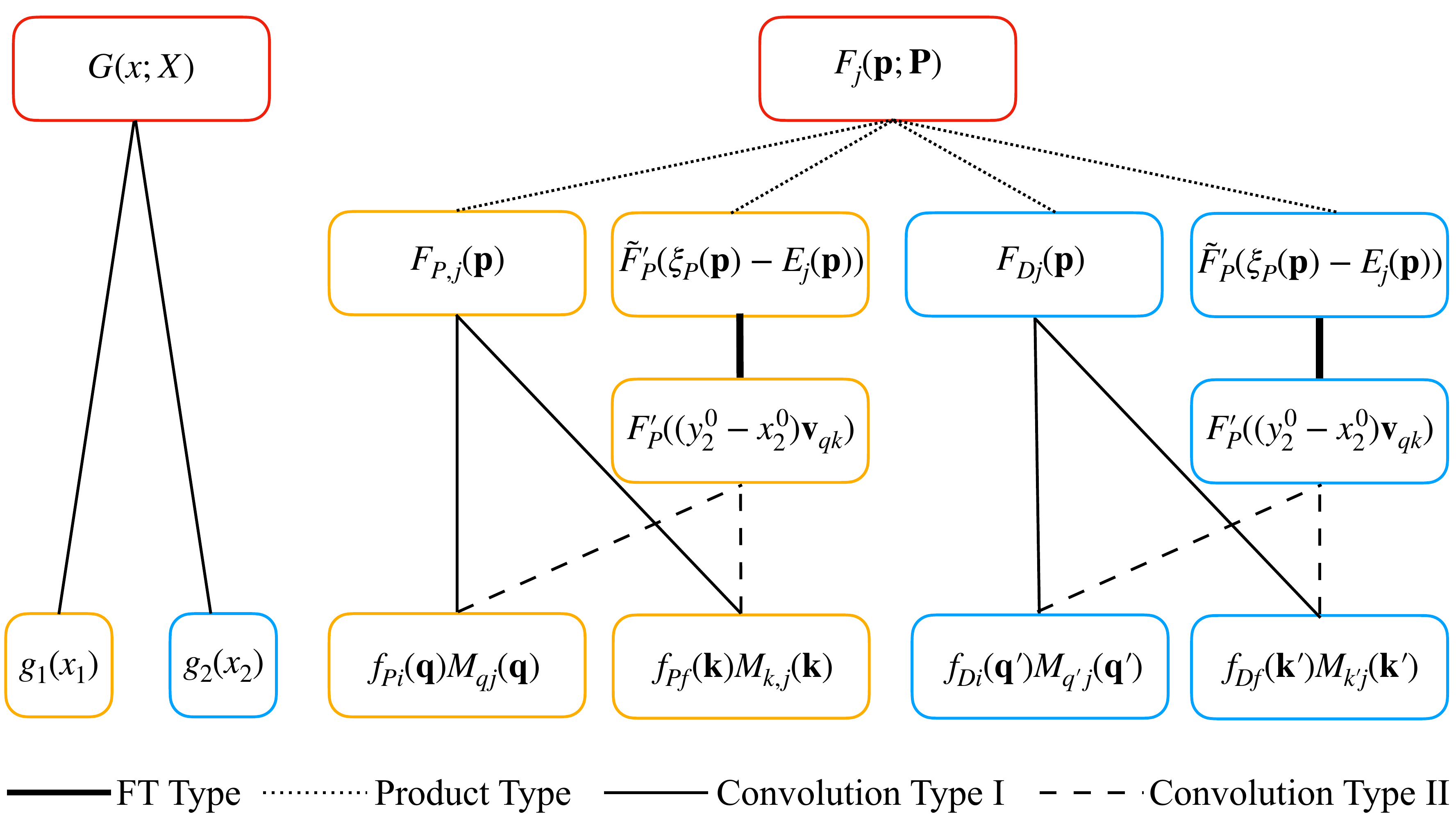}
		\caption{\label{fig:CompDist} The final (width of the) weighting functions ($G(x;X)$ and $F_j(\vp,\vP)$) in terms of the wavepacket (size) of each of the external particles, and the spatial uncertainty (size) at the vertices. Referring to Table~\ref{table1} and Table~\ref{table2}, this diagram is useful for finding how the widths are related, and how they contribute to the width of the weighting functions, which are $\sigma_x$ and $\sigma_p$ in the main text.}
	\end{figure}

	In this subsection, we work out the details in Eq.~\eqref{Aj_full}, where the phase space representing layer 1 is composed by $x = x_2 - x_1$ and $\vp = \bold{q}-\bold{k} = \bold{k}'-\bold{q}'$, where $x_1/x_2$ are the space-time coordinates of the production/detection vertices and $\bold{q},\bold{k},\bold{q}',\bold{k}'$ are the momenta of the initial and final states of the production and detection sites. Therefore, $x$ and $\vp$ represent the traveling distance and the momentum of the neutrino decided by the external particles and the position of the vertices, i.e.\ the states entangled to the neutrino. 
	The process to reach Eq.~\eqref{Aj_full} includes a series of Fourier transformations and convolutions, which is illustrated in Fig.~\ref{fig:CompDist}, from which we can clearly see how each of the uncertainties carried by each of the external states and the vertices affect  the weighting functions with the help of Table~\ref{table1} and Table~\ref{table2}. 	
	Below, we will show the derivation from Eq.~\eqref{tot_amp} to Eq.~\eqref{Aj_full}, and also the relations in Fig.~\ref{fig:CompDist}. 
	We first include all the uncertainties following Eq.~\eqref{tot_amp}:       
	\begin{align}
	&A_{2,j}(T,\vL,\vP) =
	\int[dq]f_{Pi}(\bold{q})\int[dk] \, f^*_{Pf}(\bold{k})\int[dq'] \, f_{Di}(\bold{q}')\int[dk']f^*_{Df}(\bold{k}')  \nonumber\\ 
	& \times \int d^4 x_1 \, g_{P}(x_1) \int d^4 x_2 \, g_{D}(x_2)
	\int d^4 \, y_2 \, M_{Dj}(q',k')e^{-i(q'-k')(y_2-x_2)} \nonumber\\ %
	&\times \int \frac{d^4 p_\nu}{(2\pi)^4} \, \frac{\fsl{p_\nu}+m_j}{p_\nu^2-m_j^2+i\epsilon}e^{-ip_\nu(y_1-y_2)} 
	\int d^4y_1 \, M_{Pj}(q,k)e^{-i(q-k)(y_1-x_1)}. \label{A1}
	\end{align}	
	For all $h =\{q,k,q',k'\}$, since the external states are on the mass-shell, $h^0=E_h(\bold{h})=\sqrt{\bold{h}^2-m_h^2}$. 
	Furthermore, if the wavepackets are sharply peaked at the expectation value $\langle h \rangle$, then by the saddle point approximation, we can write $E_h(\bold{h})\simeq E_h + \bold{v}_h(\bold{h}-\bold{h}_0)$, where $E_h=E_h(\langle h \rangle) $.
	Then, for the production site, by doing a change of variables: $\{q,k\} \rightarrow \{p,k\} $, where $p = k - q$, the integration over $k$ performs the convolution between the initial state and the final state including the plane wave amplitudes $M_{Pj}(q,k)$, i.e.
	\begin{align}
	& \int \frac{d^3 k}{(2\pi)^3}
	f^*_{Pf}(\bold{k}) f_{Pi}(\bold{k}-\bold{p})
	M'_{Pj}(\bold{p},\bold{k})
	e^{-i(y_1^0-x_1^0)(E_q(\bold{k}-\vp)-E_k(\bold{k}))}
	\nonumber \\ &
	= F_{Pj}(\vp)F_P'(y_1^0-x_1^0)
	e^{i\xi(\vp)(y_1^0-x_1^0)}, \label{F_P}  
	\end{align}
	where we write $M'_{Pj}(\vp,\bold{k})=M_{Pj}(k-p,k)|_{p^0=E_q(\bold{k}-\vp)-E_k(\bold{k}),\,k^0=E_k(\bold{k})}$ for convenience.
	For $f_{Pi}$ and $f_{Pj}$ being Gaussian functions with width $\sigma_q$ and $\sigma_k$ respectively, and applying the saddle point approximation,  the momentum uncertainties from the external states at the production site are 
	\begin{equation}
	F_{Pj} =  M'_{Pj}
	(\langle\bold{p}\rangle,\langle\bold{k}\rangle)
	\exp \left[\frac{-(\vp-\vP)^2}{4(\sigma_q^2+\sigma_k^2)}\right],
	\end{equation}
	where $\vP = \langle \bold{q} \rangle - \langle \bold{k} \rangle $. The other terms are
	\begin{align}
	&F_P'(y_1^0-x_1^0)=\exp \left[ - (y_1^0-x_1^0)^2
	\sigma_{qk}^2 \, \bold{v}_{qk}^2 \right],\\
	& \xi_P(\vp)=
	E_q-E_k-\bold{v}_q \bold{q}_0+\bold{v}_k \bold{k}_0
	+\frac{\bold{v}_{qk}^2 }{\Delta_{qk}} \vP 
	+\vp \left(\bold{v}_q+\frac{\bold{v}_{qk}^2 }{\Delta_{qk}}\right),
	\end{align}
	where $\bold{v}_{qk}=\bold{v}_q-\bold{v}_k$, $\Delta_{qk} = (\sigma_q^2+\sigma_k^2)/\sigma_k^2$, and $\sigma_{kq}$ refers to the notation in Table~\ref{table1}. Then analogously for the detection site, with $\vp'=\bold{k}'-\bold{q}'$ and also  $\vP' = \langle\bold{k}'\rangle-\langle\bold{q}'\rangle$,  Eq.~\eqref{A1} becomes: 
	\begin{align}
	&A_{2,j} = 
	\int  d^3 p \, F_{Pj}(\vp)\int d^3 p'\, F_{Dj}(\vp') \int d^4 x_1 \, g_P(x_1) \int d^4 x_2 \, g_D(x_2) \nonumber\\
	&\times 
	\int d^4 y_1 e^{-i(y_1^0-x_1^0)\xi_P(\vp)+i(\bold{y}_1-\vx_1)\vp}F'_P(y_1^0-x_1^0) 
	\int d^4 y_2 e^{-i(y_2^0-x_2^0)\xi_D(\vp')+i(\bold{y}_2-\vx_2)\vp'}F'_D(y_2^0-x_2^0) \nonumber\\
	&\times \int \frac{d^4 p_{\nu}}{(2\pi)^4} \frac{\fsl{p_{\nu}}+m_j}{p_{\nu}^2-m_j^2+i\epsilon}
	\, e^{-ip_\nu(y_1-y_2)}. \label{A2}
	\end{align}	
	The integration over $\int d^3y_1$ and $\int d^3y_2$ gives rise to $\delta^3(\vp-\vp_\nu)$ and $\delta^3(\vp'-\vp_\nu)$, respectively, while the integration over $\int d y_1^0$ returns 
	\begin{equation}
	\int d y_1^0 e^{-iy_1^0(\xi(\vp)-p_\nu^0)}F_P'(y_1^0-x_1^0)
	= e^{-ix_1^0(\xi(\vp)-p_\nu^0)}\tilde{F}'_P(\xi(\vp)-p_\nu^0),
	\end{equation}
	where $\tilde{F}'_P$ is the Fourier transformation of $F'_P$.
	After analogous calculations for the detection site, Eq.~\eqref{A2} finally takes the form of the layer moving operator, with weighting function $F_j(\vp;\vP)G(x,X)$:
	\begin{equation}\label{A_lmo_1to2}
	A_{2,j} = 
	\int  d^3 p \, 
	\int d^4 x \, F_j(\vp;\vP) G(x;X) \, 
	A_{1,j}(x,\vp),
	\end{equation}
	where $x=x_2-x_1$, $F_j(\vp,\vP)=F_{Pj}(\vp)F_{Dj}(\vp)$,
	\begin{equation} 
	G(x)=\int d^4x_2 g_P(x_2-x)g_D(x_2), \label{G}
	\end{equation}
	and the first layer transition amplitude is
	\begin{equation} \label{A_1}
	A_{1,j}(x,\vp)=
	e^{i\vp\vx} \,
	\int d p_\nu^0 e^{-i p_\nu^0x^0}
	\tilde{F}'_P(\xi_P(\vp)-p_\nu^0)
	\tilde{F}'_D(\xi_D(\vp)-p_\nu^0)
	\frac{\fsl{p_{\nu}}+m_j}{p_{\nu}^2-m_j^2+i\epsilon}\bigg|_{\vp_\nu=\vp}	. 
	\end{equation}
	This appears as the collection of configuration of all the energetically allowed states for some $\vp$ given by the external particles.
	Moreover, since the neutrino propagates a macroscopic distance, it can be approximated as traveling on the mass-shell. Hence, $p_\nu^0=E_j(\vp)\equiv \sqrt{\vp^2+m_j^2}$, then we can replace $A_{1,j}(x,\vp)\rightarrow e^{-itE_j(\vp)+i\vx\vp }$ and 
	\begin{equation}
	F_j(\vp) \rightarrow F_j(\vp) 
	\tilde{F}'_P(\xi(\vp)-E_j(\vp))
	\tilde{F}'_D(\xi'(\vp)-E_j(\vp)),
	\end{equation}
	in Eq.~\eqref{A_lmo_1to2}. 
	Therefore, either from the calculation we have shown, or more efficiently from Fig.~\ref{fig:CompDist}, the width of $F_j$, $\sigma_p$ is 
	\begin{equation} \label{sig_p}
	\frac{1}{\sigma_p^2}=\frac{1}{\sigma_q^2+\sigma_k^2}+\frac{1}{\sigma_{q'}^2+\sigma_{k'}^2}
	+\frac{|\bold{v}_q-\bold{v}_j+\bold{v}_{qk}^2/\Delta_{qk}|}{\sigma_{qk}^2 \bold{v}_{qk}^2}
	+\frac{|\bold{v}_{q'}-\bold{v}_j+\bold{v}_{q'k'}^2/\Delta_{q'k'}|}{\sigma_{q'k'}^2 \bold{v}_{q'k'}^2}.
	\end{equation}
	Next, to move $A_{1,j}$ onto the next layer, we first integrate out the coordinate space $\int d^4 x$, for $X=(T,\vL)$, and have
	\begin{equation}
	A_{2,j} = 
	\int  d^3 p \,  e^{-iTE_j(\vp)+i\vL\vp }
	F_j(\vp;\vP) \tilde{G}(\vp). 
	\end{equation}  
	Then as long as one of the functions $F_P$, $F_D$, $\tilde{F}_P'$, $\tilde{F}_D'$ or $\tilde{G}$ is sharply peaked, we can apply the saddle point approximation at $\vP_j$, such that 
	\begin{equation}\label{sadd}
	\frac{d}{d\vp}F_j(\vp;\vP)\tilde{G}(\vp)\bigg|_{\vp=\vP_j}=0,
	\end{equation}
	then the approximation gives $E_j(\vp) \simeq E_j + \vj(\vp-\vP_j)$, resulting in the final form of the second layer transition amplitude as
	\begin{equation} \label{A_2j}
	A_{2,j}=e^{-iE_jT + i\vP_j\vL} 
	\hat{\Phi}_{j}(\vL_j,\vP_j),
	\end{equation}
	where $\vL_j=\vL-\vj T$. 
	In fact, with such approximation, and taking  $G(x;X)$ as
	\begin{equation}
	G_x(x,X)=\exp\left[-\frac{(t-T)^2}{4\sigma_t^2}
	-\frac{(\vx-\vL)^2}{4\sigma_\vx^2}\right],
	\end{equation}
	the function after the integration of $x$ (a Fourier transformation to the momentum space) is:
	\begin{align}
	& \int dt  \int d^3x
	e^{-it(E_j+\vj\vp-\vj\vP_j)+i\vx\vp}
	G_x(x,X) \\
	& =e^{-i(E_j-\vj\vPj)T+i\vp(\vL-\vj T)}
	\exp \left[ -(\sigma_t^2 \vj^2+\sigma_\vx^2)(\vp-\tilde{m}_j)^2\right],
	\end{align}
	where $\tilde{m}_j=m_j\vj\sigma_t^2/(\sigma_t^2 \vj^2+\sigma_\vx^2)$.
	Therefore, Eq.~\eqref{A_lmo_1to2} can be written as
	\begin{equation}\label{forWig}
	e^{-iE_j T+\vj \vP_j T}
	\int dp^3 \int dx^3 \, e^{i\vx\vp}
	G(\vx;\vL_j)F_j(\vp;\vP_j),
	\end{equation}
	where $G$ has width $\sigma_x^2=\sigma_t^2\vj^2+\sigma_\vx^2$ and is centred at $\vL_j=\vL-\vj T$; $F_j$ has width $\sigma_p$ and centred at $\vP'_j=\Delta \,\vP_j$, where $\Delta = 1+4 \sigma_x^2\sigma_p^2$, such that the saddle point from Eq.~\eqref{sadd} is at $\vPj$.
	In general, if all the input distributions are Gaussian distributed and with the saddle point approximation, $\vx$ and $\vp$  both linear dependent, $F_j$ and $G$ will also be Gaussian distributed with some width $\sigma_p$ and $\sigma_x$, respectively. In this case, we obtain the formalism in Eq.~\eqref{Phi}.

	\subsection{Neutrinos represented directly} \label{Sec.AppB-2}
	Instead of representing the first layer phase space for the neutrinos by its entangled states, we represent it by the neutrinos directly. Therefore, we need to leave $y=y_2-y_2$ and $p_\nu$ non-integrated. Nonetheless, as we will see later, with this representation, we cannot have uncertainties representing the external particles and the internal vertex explicitly in the weighting function, but as an effective one with either just space-time or energy-momentum uncertainties. Therefore, we still apply the other representation in the main text for the sake of investigating neutrino decoherence in terms of these two sorts of uncertainties. Nonetheless, the representation presented in this subsection could also be used in our structure, resulting in the same effects. In particular, with this representation we can derive the total spatial uncertainty mentioned in Sec.~\ref{Sec.2-1}, since the width of the function of $y$ would indicate the total coordinate uncertainty, when it is the only non-integrated variable. Hence following Eq.~\eqref{A2}, 
	\begin{align}\label{B14}
	& A_{2,j}(T,\vL,\vP) =
	\int d^4 y_1 \int d^4 y_2 \int \frac{d^4 p_\nu}{(2\pi)^4} \Delta(p_\nu) e^{-ip_\nu(y_1-y_2)} 
	\nonumber \\
	& \times \int d^4 x_1 \, g_P(x_1) \tilde{F}_P(y_1-x_1) e^{iP_P(y_1-x_1)}
	\int d^4 x_2 \, g_D(x_2) \tilde{F}_D(y_2-x_2) e^{-iP_D(y_2-x_2)},
	\end{align}	
	where $\Delta(p_\nu)$ is the neutrino propagator in the momentum space.
	Here we write the wavepackets at the production and detection site in coordinate space as:
	\begin{align}
	& \tilde{F}_{Pj}(y_1-x_1) \, e^{iP_P(y_1-x_1)}
	\simeq \int d^3 p F_{Pj}(\vp) 
	e^{-i(y_1^0-x_1^0)\xi_P(\vp)+i(\bold{y}_1-\vx_1)\vp} ,\\ 
	& \tilde{F}_{Dj}(y_2-x_2) \, 
	e^{-iP_D(y_2-x_2)}
	\simeq \int d^3 p' F_{Dj}(\vp') 
	e^{-i(y_2^0-x_2^0)\xi_D(\vp')+i(\bold{y}_2-\vx_2)\vp'} , 
	\end{align}
	where $P_P$ and $P_D$ are the saddle point of $F_{Pj}(\vp) $ and $F_{Dj}(\vp') $, respectively. If there is negligible energy loss during the neutrino propagation, then $P_P=P_D \simeq P$. Next, after the integration over $x_1$ and $x_2$, Eq.~\eqref{B14} becomes
	\begin{equation} \label{B17}
	\int d^4 y 
	\left\{\int d^4 y_1 I_P(y_1;P_P) I_D(y_1-y;P_D) e^{-[x_P(y_1)-x_D(y_1-y)]}\right\}
	e^{iPy}\tilde{\Delta}(y),
	\end{equation}
	where $y = y_1-y_2$, and $\tilde{\Delta(y)}$ is the Fourier transformation of the propagator, i.e.\ the two point function with distance $y$ of the neutrino.
	Here $I_P$ and $I_D$ represent the total coordinate uncertainties for the production and detection site respectively, which are the convolutions between the coordinate uncertainties of the external states and the vertices, i.e.
	\begin{align}\label{B18}
	&  I_P(y_1;P) e^{iP[y_1-x_P(y_1)]}
	= \int d^4 x_1 \, g_P(x_1) \tilde{F}_{Pj}(y_1-x_1) F_P'(y_1^0-x_1^0) e^{iP(y_1-x_1)}\\
	&  I_D(y_2;P) e^{-iP[y_2-x_D(y_2)]}
	= \int d^4 x_2 \, g_D(x_2) \tilde{F}_{Dj}(y_2-x_2) F_D'(y_2^0-x_2^0) e^{-iP(y_2-x_2)}.
	\end{align}	
	This can be related to Property 5 in Appendix \ref{Sec.AppA}.
	Finally, the large bracket in Eq.~\eqref{B17}  represents the total coordinate uncertainty, which turns out to be the width of the convolution function $(I_P*I_D)(y)$, or 	$((g_P*\tilde{F}_{Pj}F_P')*(g_D*\tilde{F}_{Dj}F_D'))(y)$.
	Moreover, with the association and commutation property for convolution, we can rewrite the total coordinate width as  $(G*\tilde{F}_{Pj}^{\rm tot}*\tilde{F}_{Dj}^{\rm tot})(y)$, where $G$ is in Eq.~\eqref{G}, $\tilde{F}_{Pj}^{\rm tot}=\tilde{F}_{Pj}F_P'$ and $\tilde{F}_{Dj}^{\rm tot}=\tilde{F}_{Dj}F_D'$.

	%A-3%%%%%%%%%%%%%%%%%%%%%%%%%%%%%%%%%%%%%%%%%%%%%%%%%%%%%%%%%%%%%
	
	\section{Factorization Condition} \label{Sec.AppC}
	In this appendix, we derive the conditions under which a function can be factorized out of some integration, i.e. 
	\begin{equation} \label{factorize}
	\int_{-\infty}^{\infty} dX_2 S(X_2) Y(X_2;X_3) 
	\simeq S(X_3) \int_{-\infty}^{\infty} dX_2 Y(X_2 ;X_3) ,
	\end{equation}
	where $S(X_2)$ and $Y(X_2;X_3)$ are both localized, i.e.\ $S(X_2)=Y(X_2)=0$ as $X_2 \rightarrow \pm \infty$. 
	This is useful for describing state decoherence as a phase wash-out effect on the Wigner-PS in Eq.~\eqref{SD_factor}, and disentangling  $\sigma_{E}$ and $\sigma_{L}$ in Eq.~\eqref{PD_factor}.
	Intuitively, the condition where $S(X_2)$ can be factorized out of the integral as Eq.~\eqref{factorize} is when the width of $S(X_2)$ is much larger then $Y(X_2)$, since the product of the two functions would be dominated by the function which is more localized. Nevertheless, in order to see if this condition is sufficient and to have a more concrete idea, we derive Eq.~\eqref{factorize} as follows:   
	\begin{align} 
	LHS = &
	-\int_{-\infty}^{\infty}dX_2 \frac{d S(X_2)}{dX_2} \int_{-\infty}^{X_2} dX_2' Y(X_2';X_3) \label{C1}\\
	\simeq &	-\int_{X_3-\Lambda}^{\infty}dX_2 \frac{d S(X_2)}{dX_2} \int_{X_3-\Lambda}^{X_2} dX_2' Y(X_2';X_3) \label{C2}\\
	\simeq & -\int^{\infty}_{X_3-\Lambda}dX_2 \frac{d S(X_2)}{dX_2} 
	\int_{-\infty}^{\infty} dX_2' Y(X_2';X_3) \nonumber\\
	&+ \int^{X_3+\Lambda}_{X_3-\Lambda}dX_2 
	\frac{d S(X_2)}{dX_2} \int_{X_2}^{X_3+\Lambda} dX_2' Y(X_2';X_3) 
	= RHS. \label{C3}
	\end{align}
	Eq.~\eqref{C1} is achieved by doing integration by parts with the boundary terms vanishing due to the localization property of $S(X_2)$.  
	\begin{figure} 
		\centering
		\includegraphics[width=.65\textwidth]{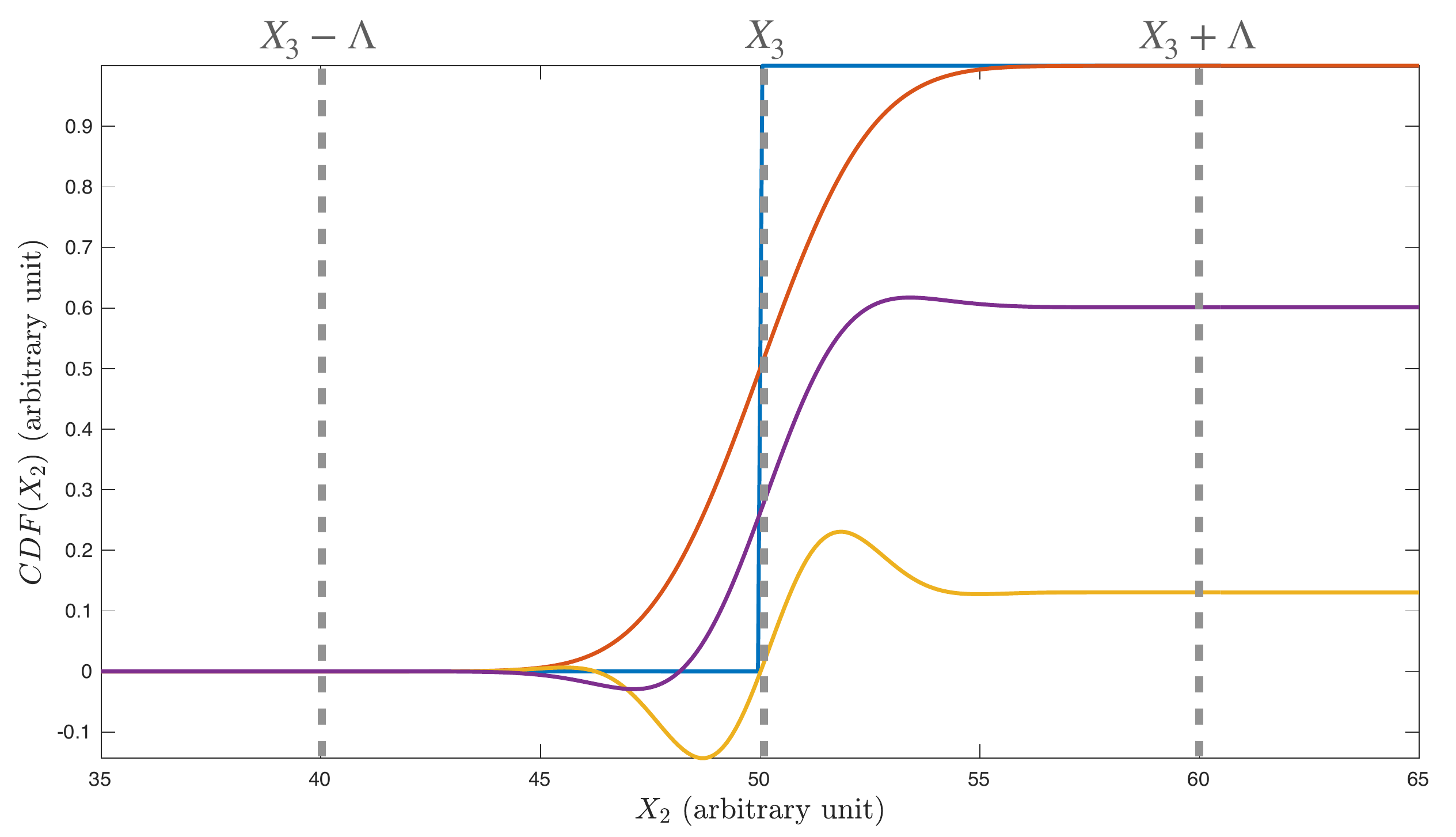}
		\caption{\label{fig:CDF} The cumulative distribution function (CDF) of $Y(X_2;X_3 = 50)$ as a Gaussian PDF with width = 2 AU (arbitrary units of $X_2$), centered at 50 AU: $N(X_2)$ for the red line; a delta function centered at 50 AU for the blue line; and $N(X_2)\cos(X_2)$/$N(X_2)\cos(X_2/2)$ for the yellow/purple line; $\Lambda$ is the cutoff value in Eq.~\eqref{C2}.}
	\end{figure}	
	In Eq.~\eqref{C2} and Eq.~\eqref{C3}, we use the localization property of $Y(X_2)$ to make a cut at $\Lambda$ w.r.t.\ $X_3$ such that the cumulative distribution function $\int_{-\infty}^{X_3 - \Lambda} dX_2' Y(X_2';X_3)=0$ and $\int^{X_3 + \Lambda}_{-\infty} dX_2' Y(X_2';X_3)$ converges to a constant, so that $\int^{\infty}_{X_3 + \Lambda} dX_2' Y(X_2';X_3)=0$, as shown in Fig.~\ref{fig:CDF}. Finally, if $S(X_2)$ varies slowly in the interval $(X_3-\Lambda,X_3+\Lambda)$, i.e.\ the width of $S(X_2)$($\sigma_S$), is much larger than that of $Y(X_2)$($\Lambda$), then the latter part in Eq.~\eqref{C3} can be neglected and we arrive at the $RHS$ of Eq.~\eqref{factorize}, where $S(X_3-\Lambda)\rightarrow S(X_3)$ can be taken out of the integral if  $X_3 \gg \Lambda$. Furthermore, $Y(X_2;X_3)$ could be any distribution as long as it is localized, even if it includes a non-zero phase term, as we can see from the orange and purple lines in Fig.~\ref{fig:CDF}.

	\section{Phase Decoherence for Discrete Neutrino Sources} \label{Sec.AppD}
	In this section, we show the formalism of phase decoherence effect including a damping term and a phase shift term for neutrino detection coming from multiple sources. In other words, we formulate the case where the weighting function on layer 2 for the coordinate uncertainty is composed of multiple delta functions discretely scattered. We start with the simple case where there are only two point-like sources located at $x_1$ and $x_2$, contributing neutrino flux $A$ and $B$, then the phase decoherence term in Eq.~\eqref{PhaseDecoh_def} is simply 
	\begin{equation}
	\Phi_{jk}= e^{-i\ajk L_3}
	\left( a\, e^{i \ajk x_1}+ b \, e^{i \ajk x_2} \right)
	\equiv 
	\phi^{(1)}\, e^{i \ajk (x_{\rm eff}^{(1)}-L_3)} ,
	\end{equation}    
	where $a = A/(A+B)$, $b = B/(A+B)$  and $0\leq c \leq 1$, are real. This requires
	\begin{equation} \label{D1}
	A \sin \left( \ajk \delta_1 \right)+B \sin \left( \ajk \delta_2 \right)=0,
	\end{equation}
	where $x_1 = x_{\rm eff}^{(1)}+\delta_1$ and $x_2 = x_{\rm eff}^{(1)}+\delta_2$. 
	Hence, by solving Eq.~\eqref{D1} for $\Delta x^{(1)} = x_1- x_2 = \delta_1 - \delta_2$, we have
	\begin{equation}
	\delta_1 \equiv f_{jk}(\Delta x,\frac{a}{b})
	= \frac{1}{\ajk}\tan ^{-1}
	\left[ \frac{-\sin(\ajk \Delta x)}
	{\frac{a}{b}+\cos(\ajk \Delta x)}\right],
	\end{equation}
	and thus
	\begin{align} \label{recrusion}
	&x_{\rm eff}^{(1)} = x_1 +f_{jk}(\Delta x^{(1)},\frac{a}{b}), \nonumber \\
	&\phi^{(1)} = a \cos\left[ \ajk f_{jk}(\Delta x^{(1)},\frac{a}{b})\right]
	+b \cos\left[ \ajk (f_{jk}(\Delta x^{(1)},\frac{a}{b})+\Delta x^{(1)})\right],
	\end{align}
	indicating that the damping term $\phi\leq 1$ as excepted since $a+b = 1$.
	In the case where the detector is placed far from all the sources, $x_1, x_2 \gg \Delta x$, then $x_{\rm eff}=x_2$, and $c = a+b\cos(\Delta x)$.
	Similarly, if there are three point like neutrino sources, 
	\begin{equation}
	\Phi_{jk}= e^{-i\ajk L_3}
	\left( a\, e^{i \ajk x_1}+ b \, e^{i \ajk x_2} 
	+ c \, e^{i \ajk x_3}\right)
	\equiv 
	\phi^{(2)}\, e^{i \ajk (x_{\rm eff}^{(2)}-L_3)} ,
	\end{equation} 
	the damping term ($\phi^{(2)} $) and the phase term ($x_{\rm eff}^{(2)}$) are obtained by  replacing $x_1 \rightarrow  x_{\rm eff}^{(1)}$, $\Delta x \rightarrow  x_{\rm eff}^{(1)}-x_3$, $a \rightarrow \phi^{(1)} $ and $b \rightarrow c$ in Eq.~\eqref{recrusion}, and so on for more point-like sources.

	% References 	
	\bibliographystyle{utphys} 
	\bibliography{NuDecoh.bib}

\providecommand{\href}[2]{#2}\begingroup\raggedright\begin{thebibliography}{10}

\bibitem{SajjadAthar:2021prg}
M.~Sajjad~Athar {\em et~al.}, ``{Status and perspectives of neutrino
  physics},'' \href{http://dx.doi.org/10.1016/j.ppnp.2022.103947}{{\em Prog.
  Part. Nucl. Phys.} {\bfseries 124} (2022) 103947},
  \href{http://arxiv.org/abs/2111.07586}{{\ttfamily arXiv:2111.07586
  [hep-ph]}}.

\bibitem{Gribov:1968kq}
V.~N. Gribov and B.~Pontecorvo, ``{Neutrino astronomy and lepton charge},''
  \href{http://dx.doi.org/10.1016/0370-2693(69)90525-5}{{\em Phys. Lett. B}
  {\bfseries 28} (1969) 493}.

\bibitem{Maki:1962mu}
Z.~Maki, M.~Nakagawa, and S.~Sakata, ``{Remarks on the unified model of
  elementary particles},'' \href{http://dx.doi.org/10.1143/PTP.28.870}{{\em
  Prog. Theor. Phys.} {\bfseries 28} (1962) 870--880}.

\bibitem{Pontecorvo:1967fh}
B.~Pontecorvo, ``{Neutrino Experiments and the Problem of Conservation of
  Leptonic Charge},'' {\em Zh. Eksp. Teor. Fiz.} {\bfseries 53} (1967)
  1717--1725.

\bibitem{Schlosshauer:2003zy}
M.~Schlosshauer, ``{Decoherence, the Measurement Problem, and Interpretations
  of Quantum Mechanics},''
  \href{http://dx.doi.org/10.1103/RevModPhys.76.1267}{{\em Rev. Mod. Phys.}
  {\bfseries 76} (2004) 1267--1305},
  \href{http://arxiv.org/abs/quant-ph/0312059}{{\ttfamily
  arXiv:quant-ph/0312059}}.

\bibitem{Zeh}
H.~D. Zeh, ``{On the interpretation of measurement in quantum theory},''
  \href{http://dx.doi.org/10.1007/BF00708656}{{\em Foundations of Physics}
  {\bfseries 1} (1970) 69--76}.

\bibitem{PhysRevD.48.4318}
J.~Rich, ``Quantum mechanics of neutrino oscillations,''
  \href{http://dx.doi.org/10.1103/PhysRevD.48.4318}{{\em Phys. Rev. D}
  {\bfseries 48} (Nov, 1993) 4318--4325}.
  \url{https://link.aps.org/doi/10.1103/PhysRevD.48.4318}.

\bibitem{Giunti:1997wq}
C.~Giunti and C.~W. Kim, ``{Coherence of neutrino oscillations in the wave
  packet approach},'' \href{http://dx.doi.org/10.1103/PhysRevD.58.017301}{{\em
  Phys. Rev. D} {\bfseries 58} (1998) 017301},
  \href{http://arxiv.org/abs/hep-ph/9711363}{{\ttfamily arXiv:hep-ph/9711363}}.

\bibitem{Akhmedov:2010ms}
E.~K. Akhmedov and J.~Kopp, ``{Neutrino Oscillations: Quantum Mechanics vs.
  Quantum Field Theory},''
  \href{http://dx.doi.org/10.1007/JHEP04(2010)008}{{\em JHEP} {\bfseries 04}
  (2010) 008}, \href{http://arxiv.org/abs/1001.4815}{{\ttfamily arXiv:1001.4815
  [hep-ph]}}. [Erratum: JHEP 10, 052 (2013)].

\bibitem{Kiers:1995zj}
K.~Kiers, S.~Nussinov, and N.~Weiss, ``{Coherence effects in neutrino
  oscillations},'' \href{http://dx.doi.org/10.1103/PhysRevD.53.537}{{\em Phys.
  Rev. D} {\bfseries 53} (1996) 537--547},
  \href{http://arxiv.org/abs/hep-ph/9506271}{{\ttfamily arXiv:hep-ph/9506271}}.

\bibitem{Kayser:1981ye}
B.~Kayser, ``{On the Quantum Mechanics of Neutrino Oscillation},''
  \href{http://dx.doi.org/10.1103/PhysRevD.24.110}{{\em Phys. Rev. D}
  {\bfseries 24} (1981) 110}.

\bibitem{Beuthe:2001rc}
M.~Beuthe, ``{Oscillations of neutrinos and mesons in quantum field theory},''
  \href{http://dx.doi.org/10.1016/S0370-1573(02)00538-0}{{\em Phys. Rept.}
  {\bfseries 375} (2003) 105--218},
  \href{http://arxiv.org/abs/hep-ph/0109119}{{\ttfamily arXiv:hep-ph/0109119}}.

\bibitem{Giunti:2002xg}
C.~Giunti, ``{Neutrino wave packets in quantum field theory},''
  \href{http://dx.doi.org/10.1088/1126-6708/2002/11/017}{{\em JHEP} {\bfseries
  11} (2002) 017}, \href{http://arxiv.org/abs/hep-ph/0205014}{{\ttfamily
  arXiv:hep-ph/0205014}}.

\bibitem{Akhmedov:2010ua}
E.~K. Akhmedov and A.~Y. Smirnov, ``{Neutrino oscillations: Entanglement,
  energy-momentum conservation and QFT},''
  \href{http://dx.doi.org/10.1007/s10701-011-9545-4}{{\em Found. Phys.}
  {\bfseries 41} (2011) 1279--1306},
  \href{http://arxiv.org/abs/1008.2077}{{\ttfamily arXiv:1008.2077 [hep-ph]}}.

\bibitem{Naumov}
N.~V. Naumov~D.V., ``{Quantum Field Theory of Neutrino Oscillations},''
  \href{http://dx.doi.org/https://doi.org/10.1134/S1063779620010050}{{\em Phys.
  Part. Nuclei} {\bfseries 51} (2020) 1--106}.

\bibitem{Grimus:2019hlq}
W.~Grimus, ``{Revisiting the quantum field theory of neutrino oscillations in
  vacuum},'' \href{http://dx.doi.org/10.1088/1361-6471/ab716f}{{\em J. Phys. G}
  {\bfseries 47} no.~8, (2020) 085004},
  \href{http://arxiv.org/abs/1910.13776}{{\ttfamily arXiv:1910.13776
  [hep-ph]}}.

\bibitem{Lind}
G.~Lindblad, ``{On the generators of quantum dynamical semigroups},''
  \href{http://dx.doi.org/10.1007/BF01608499}{{\em Commun.\ Math.\ Phys.}
  {\bfseries 48} (1976) 119}.

\bibitem{Lisi:2000zt}
E.~Lisi, A.~Marrone, and D.~Montanino, ``{Probing possible decoherence effects
  in atmospheric neutrino oscillations},''
  \href{http://dx.doi.org/10.1103/PhysRevLett.85.1166}{{\em Phys. Rev. Lett.}
  {\bfseries 85} (2000) 1166--1169},
  \href{http://arxiv.org/abs/hep-ph/0002053}{{\ttfamily arXiv:hep-ph/0002053}}.

\bibitem{Hansen:2016klk}
R.~S.~L. Hansen and A.~Y. Smirnov, ``{The Liouville equation for flavour
  evolution of neutrinos and neutrino wave packets},''
  \href{http://dx.doi.org/10.1088/1475-7516/2016/12/019}{{\em JCAP} {\bfseries
  12} (2016) 019}, \href{http://arxiv.org/abs/1610.00910}{{\ttfamily
  arXiv:1610.00910 [hep-ph]}}.

\bibitem{Benatti:2000ph}
F.~Benatti and R.~Floreanini, ``{Open system approach to neutrino
  oscillations},'' \href{http://dx.doi.org/10.1088/1126-6708/2000/02/032}{{\em
  JHEP} {\bfseries 02} (2000) 032},
  \href{http://arxiv.org/abs/hep-ph/0002221}{{\ttfamily arXiv:hep-ph/0002221}}.

\bibitem{Stirner:2018ojk}
T.~Stirner, G.~Sigl, and G.~Raffelt, ``{Liouville term for neutrinos: Flavor
  structure and wave interpretation},''
  \href{http://dx.doi.org/10.1088/1475-7516/2018/05/016}{{\em JCAP} {\bfseries
  05} (2018) 016}, \href{http://arxiv.org/abs/1803.04693}{{\ttfamily
  arXiv:1803.04693 [hep-ph]}}.

\bibitem{Coelho:2017byq}
J.~a. A.~B. Coelho and W.~A. Mann, ``{Decoherence, matter effect, and neutrino
  hierarchy signature in long baseline experiments},''
  \href{http://dx.doi.org/10.1103/PhysRevD.96.093009}{{\em Phys. Rev. D}
  {\bfseries 96} no.~9, (2017) 093009},
  \href{http://arxiv.org/abs/1708.05495}{{\ttfamily arXiv:1708.05495
  [hep-ph]}}.

\bibitem{Gomes:2020muc}
A.~L.~G. Gomes, R.~A. Gomes, and O.~L.~G. Peres, ``{Quantum decoherence and
  relaxation in neutrinos using long-baseline data},''
  \href{http://arxiv.org/abs/2001.09250}{{\ttfamily arXiv:2001.09250
  [hep-ph]}}.

\bibitem{Farzan:2008zv}
Y.~Farzan, T.~Schwetz, and A.~Y. Smirnov, ``{Reconciling results of LSND,
  MiniBooNE and other experiments with soft decoherence},''
  \href{http://dx.doi.org/10.1088/1126-6708/2008/07/067}{{\em JHEP} {\bfseries
  07} (2008) 067}, \href{http://arxiv.org/abs/0805.2098}{{\ttfamily
  arXiv:0805.2098 [hep-ph]}}.

\bibitem{Jones:2014sfa}
B.~J.~P. Jones, ``{Dynamical pion collapse and the coherence of conventional
  neutrino beams},'' \href{http://dx.doi.org/10.1103/PhysRevD.91.053002}{{\em
  Phys. Rev. D} {\bfseries 91} no.~5, (2015) 053002},
  \href{http://arxiv.org/abs/1412.2264}{{\ttfamily arXiv:1412.2264 [hep-ph]}}.

\bibitem{Wigner:1932eb}
E.~P. Wigner, ``{On the quantum correction for thermodynamic equilibrium},''
  \href{http://dx.doi.org/10.1103/PhysRev.40.749}{{\em Phys. Rev.} {\bfseries
  40} (1932) 749}.

\bibitem{Vlasenko:2013fja}
A.~Vlasenko, G.~M. Fuller, and V.~Cirigliano, ``{Neutrino Quantum Kinetics},''
  \href{http://dx.doi.org/10.1103/PhysRevD.89.105004}{{\em Phys. Rev. D}
  {\bfseries 89} no.~10, (2014) 105004},
  \href{http://arxiv.org/abs/1309.2628}{{\ttfamily arXiv:1309.2628 [hep-ph]}}.

\bibitem{Blasone:1995zc}
M.~Blasone and G.~Vitiello, ``{Quantum field theory of fermion mixing},''
  \href{http://dx.doi.org/10.1006/aphy.1995.1115}{{\em Annals Phys.} {\bfseries
  244} (1995) 283--311}, \href{http://arxiv.org/abs/hep-ph/9501263}{{\ttfamily
  arXiv:hep-ph/9501263}}. [Erratum: Annals Phys. 249, 363--364 (1996)].

\bibitem{PhysRevD.45.2414}
C.~Giunti, C.~W. Kim, and U.~W. Lee, ``Remarks on the weak states of
  neutrinos,'' \href{http://dx.doi.org/10.1103/PhysRevD.45.2414}{{\em Phys.
  Rev. D} {\bfseries 45} (Apr, 1992) 2414--2420}.
  \url{https://link.aps.org/doi/10.1103/PhysRevD.45.2414}.

\bibitem{Giunti:2003dg}
C.~Giunti, ``{Fock states of flavor neutrinos are unphysical},''
  \href{http://dx.doi.org/10.1140/epjc/s2004-02100-4}{{\em Eur. Phys. J. C}
  {\bfseries 39} (2005) 377--382},
  \href{http://arxiv.org/abs/hep-ph/0312256}{{\ttfamily arXiv:hep-ph/0312256}}.

\bibitem{Tureanu:2020odo}
A.~Tureanu, ``{Comment on the Comment on the paper ''Can oscillating neutrino
  states be formulated universally?''},''
  \href{http://arxiv.org/abs/2005.02219}{{\ttfamily arXiv:2005.02219
  [hep-ph]}}.

\bibitem{Torres:2020gzm}
B.~d. S.~L. Torres, T.~R. Perche, A.~G.~S. Landulfo, and G.~E.~A. Matsas,
  ``{Neutrino flavor oscillations without flavor states},''
  \href{http://dx.doi.org/10.1103/PhysRevD.102.093003}{{\em Phys. Rev. D}
  {\bfseries 102} no.~9, (2020) 093003},
  \href{http://arxiv.org/abs/2009.10165}{{\ttfamily arXiv:2009.10165
  [hep-ph]}}.

\bibitem{Smaldone:2021mii}
L.~Smaldone and G.~Vitiello, ``{Neutrino Mixing and Oscillations in Quantum
  Field Theory: A Comprehensive Introduction},''
  \href{http://dx.doi.org/10.3390/universe7120504}{{\em Universe} {\bfseries 7}
  no.~12, (2021) 504}, \href{http://arxiv.org/abs/2111.11809}{{\ttfamily
  arXiv:2111.11809 [hep-th]}}.

\bibitem{Blasone:2020wer}
M.~Blasone and L.~Smaldone, ``{A note on oscillating neutrino states in quantum
  field theory},'' \href{http://dx.doi.org/10.1142/S0217732320503137}{{\em Mod.
  Phys. Lett. A} {\bfseries 35} no.~38, (2020) 2050313},
  \href{http://arxiv.org/abs/2004.04739}{{\ttfamily arXiv:2004.04739
  [hep-ph]}}.

\bibitem{Akhmedov:2020vua}
E.~Akhmedov, ``{Neutrino oscillations in matter: from microscopic to
  macroscopic description},''
  \href{http://dx.doi.org/10.1007/JHEP02(2021)107}{{\em JHEP} {\bfseries 02}
  (2021) 107}, \href{http://arxiv.org/abs/2010.07847}{{\ttfamily
  arXiv:2010.07847 [hep-ph]}}.

\bibitem{Esteban:2020cvm}
I.~Esteban, M.~C. Gonzalez-Garcia, M.~Maltoni, T.~Schwetz, and A.~Zhou, ``{The
  fate of hints: updated global analysis of three-flavor neutrino
  oscillations},'' \href{http://dx.doi.org/10.1007/JHEP09(2020)178}{{\em JHEP}
  {\bfseries 09} (2020) 178}, \href{http://arxiv.org/abs/2007.14792}{{\ttfamily
  arXiv:2007.14792 [hep-ph]}}.

\bibitem{deGouvea:2020hfl}
A.~de~Gouvea, V.~de~Romeri, and C.~A. Ternes, ``{Probing neutrino quantum
  decoherence at reactor experiments},''
  \href{http://dx.doi.org/10.1007/JHEP08(2020)049}{{\em JHEP} {\bfseries 08}
  (2020) 018}, \href{http://arxiv.org/abs/2005.03022}{{\ttfamily
  arXiv:2005.03022 [hep-ph]}}.

\bibitem{deGouvea:2021uvg}
A.~de~Gouv\^ea, V.~De~Romeri, and C.~A. Ternes, ``{Combined analysis of
  neutrino decoherence at reactor experiments},''
  \href{http://dx.doi.org/10.1007/JHEP06(2021)042}{{\em JHEP} {\bfseries 06}
  (2021) 042}, \href{http://arxiv.org/abs/2104.05806}{{\ttfamily
  arXiv:2104.05806 [hep-ph]}}.

\bibitem{JUNO:2021ydg}
{\bfseries JUNO} Collaboration, J.~Wang {\em et~al.}, ``{Damping signatures at
  JUNO, a medium-baseline reactor neutrino oscillation experiment},''
  \href{http://arxiv.org/abs/2112.14450}{{\ttfamily arXiv:2112.14450
  [hep-ex]}}.

\bibitem{Stuttard:2020qfv}
T.~Stuttard and M.~Jensen, ``{Neutrino decoherence from quantum gravitational
  stochastic perturbations},''
  \href{http://dx.doi.org/10.1103/PhysRevD.102.115003}{{\em Phys. Rev. D}
  {\bfseries 102} no.~11, (2020) 115003},
  \href{http://arxiv.org/abs/2007.00068}{{\ttfamily arXiv:2007.00068
  [hep-ph]}}.

\bibitem{Boriero:2017tkh}
D.~Boriero, D.~J. Schwarz, and H.~Velten, ``{Flavour composition and entropy
  increase of cosmological neutrinos after decoherence},''
  \href{http://dx.doi.org/10.3390/universe5100203}{{\em Universe} {\bfseries 5}
  no.~10, (2019) 203}, \href{http://arxiv.org/abs/1704.06139}{{\ttfamily
  arXiv:1704.06139 [astro-ph.CO]}}.

\bibitem{DeGouvea:2020ang}
A.~De~Gouv\^ea, I.~Martinez-Soler, Y.~F. Perez-Gonzalez, and M.~Sen,
  ``{Fundamental physics with the diffuse supernova background neutrinos},''
  \href{http://dx.doi.org/10.1103/PhysRevD.102.123012}{{\em Phys. Rev. D}
  {\bfseries 102} (2020) 123012},
  \href{http://arxiv.org/abs/2007.13748}{{\ttfamily arXiv:2007.13748
  [hep-ph]}}.

\bibitem{Carpio:2017nui}
J.~Carpio, E.~Massoni, and A.~M. Gago, ``{Revisiting quantum decoherence for
  neutrino oscillations in matter with constant density},''
  \href{http://dx.doi.org/10.1103/PhysRevD.97.115017}{{\em Phys. Rev. D}
  {\bfseries 97} no.~11, (2018) 115017},
  \href{http://arxiv.org/abs/1711.03680}{{\ttfamily arXiv:1711.03680
  [hep-ph]}}.

\bibitem{Carpio:2018gum}
J.~A. Carpio, E.~Massoni, and A.~M. Gago, ``{Testing quantum decoherence at
  DUNE},'' \href{http://dx.doi.org/10.1103/PhysRevD.100.015035}{{\em Phys. Rev.
  D} {\bfseries 100} no.~1, (2019) 015035},
  \href{http://arxiv.org/abs/1811.07923}{{\ttfamily arXiv:1811.07923
  [hep-ph]}}.

\bibitem{BalieiroGomes:2018gtd}
G.~Balieiro~Gomes, D.~V. Forero, M.~M. Guzzo, P.~C. De~Holanda, and R.~L.~N.
  Oliveira, ``{Quantum Decoherence Effects in Neutrino Oscillations at DUNE},''
  \href{http://dx.doi.org/10.1103/PhysRevD.100.055023}{{\em Phys. Rev. D}
  {\bfseries 100} no.~5, (2019) 055023},
  \href{http://arxiv.org/abs/1805.09818}{{\ttfamily arXiv:1805.09818
  [hep-ph]}}.

\bibitem{Coloma:2018idr}
P.~Coloma, J.~Lopez-Pavon, I.~Martinez-Soler, and H.~Nunokawa, ``{Decoherence
  in Neutrino Propagation Through Matter, and Bounds from IceCube/DeepCore},''
  \href{http://dx.doi.org/10.1140/epjc/s10052-018-6092-6}{{\em Eur. Phys. J. C}
  {\bfseries 78} no.~8, (2018) 614},
  \href{http://arxiv.org/abs/1803.04438}{{\ttfamily arXiv:1803.04438
  [hep-ph]}}.

\bibitem{Sen}
G.~L.~S. R.~N.~Sen, ``{Fiber bundles in quantum physics},''
  \href{http://dx.doi.org/https://doi.org/10.1063/1.1447309}{{\em J. Math.
  Phys.} {\bfseries 43} (2002) 1323--1339}.

\bibitem{Case}
W.~B. Case, ``{Wigner functions and Weyl transforms for pedestrians},''
  \href{http://dx.doi.org/https://doi.org/10.1063/1.1447309}{{\em Am.J.Phys.}
  {\bfseries 76(10)} (2008) 937}.

\bibitem{Jacob}
R.~Jacob and R.~G. Sachs, ``{Mass and Lifetime of Unstable Particles},''
  \href{http://dx.doi.org/https://doi.org/10.1063/1.1447309}{{\em Phys. Rev.}
  {\bfseries 121} (1961) 350}.

\bibitem{Akhmedov:2012uu}
E.~Akhmedov, D.~Hernandez, and A.~Smirnov, ``{Neutrino production coherence and
  oscillation experiments},''
  \href{http://dx.doi.org/10.1007/JHEP04(2012)052}{{\em JHEP} {\bfseries 04}
  (2012) 052}, \href{http://arxiv.org/abs/1201.4128}{{\ttfamily arXiv:1201.4128
  [hep-ph]}}.

\bibitem{Kersten:2015kio}
J.~Kersten and A.~Y. Smirnov, ``{Decoherence and oscillations of supernova
  neutrinos},'' \href{http://dx.doi.org/10.1140/epjc/s10052-016-4187-5}{{\em
  Eur. Phys. J. C} {\bfseries 76} no.~6, (2016) 339},
  \href{http://arxiv.org/abs/1512.09068}{{\ttfamily arXiv:1512.09068
  [hep-ph]}}.

\bibitem{Martini:2012uc}
M.~Martini, M.~Ericson, and G.~Chanfray, ``{Energy reconstruction effects in
  neutrino oscillation experiments and implications for the analysis},''
  \href{http://dx.doi.org/10.1103/PhysRevD.87.013009}{{\em Phys. Rev. D}
  {\bfseries 87} no.~1, (2013) 013009},
  \href{http://arxiv.org/abs/1211.1523}{{\ttfamily arXiv:1211.1523 [hep-ph]}}.

\bibitem{DeRomeri:2016qwo}
V.~De~Romeri, E.~Fernandez-Martinez, and M.~Sorel, ``{Neutrino oscillations at
  DUNE with improved energy reconstruction},''
  \href{http://dx.doi.org/10.1007/JHEP09(2016)030}{{\em JHEP} {\bfseries 09}
  (2016) 030}, \href{http://arxiv.org/abs/1607.00293}{{\ttfamily
  arXiv:1607.00293 [hep-ph]}}.

\bibitem{Akhmedov:2009rb}
E.~K. Akhmedov and A.~Y. Smirnov, ``{Paradoxes of neutrino oscillations},''
  \href{http://dx.doi.org/10.1134/S1063778809080122}{{\em Phys. Atom. Nucl.}
  {\bfseries 72} (2009) 1363--1381},
  \href{http://arxiv.org/abs/0905.1903}{{\ttfamily arXiv:0905.1903 [hep-ph]}}.

\bibitem{Giunti:2000kw}
C.~Giunti and C.~W. Kim, ``{Quantum mechanics of neutrino oscillations},''
  \href{http://dx.doi.org/10.1023/A:1012230026160}{{\em Found. Phys. Lett.}
  {\bfseries 14} no.~3, (2001) 213--229},
  \href{http://arxiv.org/abs/hep-ph/0011074}{{\ttfamily arXiv:hep-ph/0011074}}.

\bibitem{Farzan:2008eg}
Y.~Farzan and A.~Y. Smirnov, ``{Coherence and oscillations of cosmic
  neutrinos},'' \href{http://dx.doi.org/10.1016/j.nuclphysb.2008.07.028}{{\em
  Nucl. Phys. B} {\bfseries 805} (2008) 356--376},
  \href{http://arxiv.org/abs/0803.0495}{{\ttfamily arXiv:0803.0495 [hep-ph]}}.

\bibitem{Esteban-Pretel:2007jwl}
A.~Esteban-Pretel, S.~Pastor, R.~Tomas, G.~G. Raffelt, and G.~Sigl,
  ``{Decoherence in supernova neutrino transformations suppressed by
  deleptonization},'' \href{http://dx.doi.org/10.1103/PhysRevD.76.125018}{{\em
  Phys. Rev. D} {\bfseries 76} (2007) 125018},
  \href{http://arxiv.org/abs/0706.2498}{{\ttfamily arXiv:0706.2498
  [astro-ph]}}.

\bibitem{Akhmedov:2017mcc}
E.~Akhmedov, J.~Kopp, and M.~Lindner, ``{Collective neutrino oscillations and
  neutrino wave packets},''
  \href{http://dx.doi.org/10.1088/1475-7516/2017/09/017}{{\em JCAP} {\bfseries
  09} (2017) 017}, \href{http://arxiv.org/abs/1702.08338}{{\ttfamily
  arXiv:1702.08338 [hep-ph]}}.

\bibitem{Hooper:2004xr}
D.~Hooper, D.~Morgan, and E.~Winstanley, ``{Probing quantum decoherence with
  high-energy neutrinos},''
  \href{http://dx.doi.org/10.1016/j.physletb.2005.01.034}{{\em Phys. Lett. B}
  {\bfseries 609} (2005) 206--211},
  \href{http://arxiv.org/abs/hep-ph/0410094}{{\ttfamily arXiv:hep-ph/0410094}}.

\bibitem{NuSTEC:2017hzk}
{\bfseries NuSTEC} Collaboration, L.~Alvarez-Ruso {\em et~al.}, ``{NuSTEC White
  Paper: Status and challenges of neutrino\textendash{}nucleus scattering},''
  \href{http://dx.doi.org/10.1016/j.ppnp.2018.01.006}{{\em Prog. Part. Nucl.
  Phys.} {\bfseries 100} (2018) 1--68},
  \href{http://arxiv.org/abs/1706.03621}{{\ttfamily arXiv:1706.03621
  [hep-ph]}}.

\bibitem{Kayser:2010pr}
B.~Kayser and J.~Kopp, ``{Testing the Wave Packet Approach to Neutrino
  Oscillations in Future Experiments},''
  \href{http://arxiv.org/abs/1005.4081}{{\ttfamily arXiv:1005.4081 [hep-ph]}}.

\bibitem{Blennow:2005yk}
M.~Blennow, T.~Ohlsson, and W.~Winter, ``{Damping signatures in future neutrino
  oscillation experiments},''
  \href{http://dx.doi.org/10.1088/1126-6708/2005/06/049}{{\em JHEP} {\bfseries
  06} (2005) 049}, \href{http://arxiv.org/abs/hep-ph/0502147}{{\ttfamily
  arXiv:hep-ph/0502147}}.

\bibitem{ParticleDataGroup:2020ssz}
{\bfseries Particle Data Group} Collaboration, P.~A. Zyla {\em et~al.},
  ``{Review of Particle Physics},''
  \href{http://dx.doi.org/10.1093/ptep/ptaa104}{{\em PTEP} {\bfseries 2020}
  no.~8, (2020) 083C01}.

\bibitem{RENO:2018dro}
{\bfseries RENO} Collaboration, G.~Bak {\em et~al.}, ``{Measurement of Reactor
  Antineutrino Oscillation Amplitude and Frequency at RENO},''
  \href{http://dx.doi.org/10.1103/PhysRevLett.121.201801}{{\em Phys. Rev.
  Lett.} {\bfseries 121} no.~20, (2018) 201801},
  \href{http://arxiv.org/abs/1806.00248}{{\ttfamily arXiv:1806.00248
  [hep-ex]}}.

\bibitem{RENO:2016}
{\bfseries RENO} Collaboration, S.~H. Seo {\em et~al.}, ``{Spectral Measurement
  of the Electron Antineutrino Oscillation Amplitude and Frequency using 500
  Live Days of RENO Data},''
  \href{http://dx.doi.org/10.1103/PhysRevD.98.012002}{{\em Phys. Rev. D}
  {\bfseries 98} no.~1, (2018) 012002},
  \href{http://arxiv.org/abs/1610.04326}{{\ttfamily arXiv:1610.04326
  [hep-ex]}}.

\bibitem{RENO:2010vlj}
{\bfseries RENO} Collaboration, J.~K. Ahn {\em et~al.}, ``{RENO: An Experiment
  for Neutrino Oscillation Parameter $\theta_{13}$ Using Reactor Neutrinos at
  Yonggwang},'' \href{http://arxiv.org/abs/1003.1391}{{\ttfamily
  arXiv:1003.1391 [hep-ex]}}.

\bibitem{Abe:2011ts}
K.~Abe {\em et~al.}, ``{Letter of Intent: The Hyper-Kamiokande Experiment ---
  Detector Design and Physics Potential ---},''
  \href{http://arxiv.org/abs/1109.3262}{{\ttfamily arXiv:1109.3262 [hep-ex]}}.

\bibitem{DUNE:2020ypp}
{\bfseries DUNE} Collaboration, B.~Abi {\em et~al.}, ``{Deep Underground
  Neutrino Experiment (DUNE), Far Detector Technical Design Report, Volume II:
  DUNE Physics},'' \href{http://arxiv.org/abs/2002.03005}{{\ttfamily
  arXiv:2002.03005 [hep-ex]}}.

\bibitem{Ajimura:2017fld}
S.~Ajimura {\em et~al.}, ``{Technical Design Report (TDR): Searching for a
  Sterile Neutrino at J-PARC MLF (E56, JSNS2)},''
  \href{http://arxiv.org/abs/1705.08629}{{\ttfamily arXiv:1705.08629
  [physics.ins-det]}}.

\bibitem{Baxter:2019mcx}
D.~Baxter {\em et~al.}, ``{Coherent Elastic Neutrino-Nucleus Scattering at the
  European Spallation Source},''
  \href{http://dx.doi.org/10.1007/JHEP02(2020)123}{{\em JHEP} {\bfseries 02}
  (2020) 123}, \href{http://arxiv.org/abs/1911.00762}{{\ttfamily
  arXiv:1911.00762 [physics.ins-det]}}.

\bibitem{Alonso:2010fs}
J.~Alonso {\em et~al.}, ``{Expression of Interest for a Novel Search for CP
  Violation in the Neutrino Sector: DAEdALUS},''
  \href{http://arxiv.org/abs/1006.0260}{{\ttfamily arXiv:1006.0260
  [physics.ins-det]}}.

\bibitem{Harnik:2019iwv}
R.~Harnik, K.~J. Kelly, and P.~A.~N. Machado, ``{Prospects of Measuring
  Oscillated Decay-at-Rest Neutrinos at Long Baselines},''
  \href{http://dx.doi.org/10.1103/PhysRevD.101.033008}{{\em Phys. Rev. D}
  {\bfseries 101} no.~3, (2020) 033008},
  \href{http://arxiv.org/abs/1911.05088}{{\ttfamily arXiv:1911.05088
  [hep-ph]}}.

\bibitem{Strumia:2003zx}
A.~Strumia and F.~Vissani, ``{Precise quasielastic neutrino/nucleon
  cross-section},'' \href{http://dx.doi.org/10.1016/S0370-2693(03)00616-6}{{\em
  Phys. Lett. B} {\bfseries 564} (2003) 42--54},
  \href{http://arxiv.org/abs/astro-ph/0302055}{{\ttfamily
  arXiv:astro-ph/0302055}}.

\bibitem{Ajimura:2020qni}
S.~Ajimura {\em et~al.}, ``{Proposal: JSNS$^2$-II},''
  \href{http://arxiv.org/abs/2012.10807}{{\ttfamily arXiv:2012.10807
  [hep-ex]}}.

\bibitem{COHERENT:2020iec}
{\bfseries COHERENT} Collaboration, D.~Akimov {\em et~al.}, ``{First
  Measurement of Coherent Elastic Neutrino-Nucleus Scattering on Argon},''
  \href{http://dx.doi.org/10.1103/PhysRevLett.126.012002}{{\em Phys. Rev.
  Lett.} {\bfseries 126} no.~1, (2021) 012002},
  \href{http://arxiv.org/abs/2003.10630}{{\ttfamily arXiv:2003.10630
  [nucl-ex]}}.

\bibitem{ESSnuSB:2021azq}
{\bfseries ESSnuSB} Collaboration, A.~Alekou {\em et~al.}, ``{Updated physics
  performance of the ESSnuSB experiment: ESSnuSB collaboration},''
  \href{http://dx.doi.org/10.1140/epjc/s10052-021-09845-8}{{\em Eur. Phys. J.
  C} {\bfseries 81} no.~12, (2021) 1130},
  \href{http://arxiv.org/abs/2107.07585}{{\ttfamily arXiv:2107.07585
  [hep-ex]}}.

\bibitem{Kolbe:2002gk}
E.~Kolbe, K.~Langanke, and P.~Vogel, ``{Estimates of weak and electromagnetic
  nuclear decay signatures for neutrino reactions in Super-Kamiokande},''
  \href{http://dx.doi.org/10.1103/PhysRevD.66.013007}{{\em Phys. Rev. D}
  {\bfseries 66} (2002) 013007}.

\bibitem{Bodek:1980ar}
A.~Bodek and J.~L. Ritchie, ``{Fermi Motion Effects in Deep Inelastic Lepton
  Scattering from Nuclear Targets},''
  \href{http://dx.doi.org/10.1103/PhysRevD.23.1070}{{\em Phys. Rev. D}
  {\bfseries 23} (1981) 1070}.

\end{thebibliography}\endgroup
	
\end{document}